\documentclass[11pt]{report}
\usepackage{graphicx,url,anysize,rotating,hyperref}
\usepackage[T1]{fontenc}
\usepackage{amsmath}
\usepackage{verbatim}
\usepackage{color}
\usepackage{ulem}
\usepackage{multirow}
\usepackage{amsthm}
\usepackage{caption}
\usepackage{algorithm}
\usepackage{algpseudocode}
\usepackage{blindtext}
\usepackage{url}
\usepackage{booktabs}
\newtheorem*{DI*}{Design Insight}

\marginsize{3.2cm}{2cm}{2.5cm}{3.2cm}  % left right top bottom?

\newtheorem{theorem}{Theorem}

\newtheorem{example}[theorem]{{\bf Example}}

\newtheorem{definition}[theorem]{Definition}

\algnewcommand{\LineComment}[1]{\State \(\triangleright\) #1}

\def\@onedot{\ifx\@let@token.\else.\null\fi\xspace}

\def\prog{{\mathcal{P}}}
\def\cP{{\prog}}

\def\cA{{\mathcal{A}}}

\def\cQ{{\mathcal{Q}}}
\def\pref{{\mbox{\sc Pref}}}

\def\wl{{\mbox{\sc WaitList}}}
\def\mc{{\mbox{\sc Min-Cutoff}}}

\def\ip{{\mbox{\sc Is-Processed}}}
\def\merit{{\mbox{\sc Merit}}}
\def\rank{{\mbox{\sc Rank}}}
\def\LP{{i}}
\def\length{{\mbox{\sc Length}}}

\title{
Joint Seat Allocation 2018: An algorithmic perspective \\
{\large Technical Report}
}
\date{October 2018}
\author{
  Surender Baswana\thanks{Indian Institute of Technology Kanpur, Kanpur, India}\\
  \and
  Partha Pratim Chakrabarti\thanks{Indian Institute of Technology Kharagpur, Kharagpur, India}\\
  \and
  Sharat Chandran\thanks{Indian Institute of Technology Bombay, Mumbai, India}\\
  \and
  Yashodhan Kanoria\thanks{Columbia Business School, New York, New York, USA}\\
  \and
  Utkarsh Patange\footnotemark[\value{footnote}]\\
}

\date{November 2019}

\begin{document}
\maketitle
\pagenumbering{roman}
\begin{abstract}
\small

Until 2014, admissions to the Indian Institutes of Technology
(IITs) were conducted under one umbrella, whereas the admissions to
the non-IIT Centrally Funded Government Institutes (CFTIs) were
conducted under a different umbrella, the Central Seat Allocation
Board (CSAB). The same set of candidates were eligible to apply for a
seat in each of the two sets, and several hundred
candidates would indeed receive two seats from the two different
sets. Each such candidate could use at most one of the seats, leaving
a vacancy in the other seat; this would be noticed much later, in many
cases after classes began. Such seats would either remain vacant or
would be reallocated at a later stage, leading to logistical hardship
for candidates, or inefficiency in seat allocation in the form of
unnecessary vacancies, and also misallocation of seats (e.g., a
particular CSAB seat could be offered to a candidate A, despite
denying the same seat earlier to a candidate B with better rank, who
had meanwhile taken some IIT seat).  The two sets also operated
under different time windows, with the net result that classes
would begin later in the academic year, compared to, say, colleges
offering only the sciences.

Since 2015, a new joint seat allocation process has been implemented
improving the efficiency and productivity of concerned stakeholders.
The process brings all CFTIs under one umbrella for admissions:
%It
%began in 2015 with 86 institutes, and approximately 34000 available
%seats. The numbers increased steadily in subsequent years with
100
institutes and approximately 39000 seats in 2018.  In this
scheme, each candidate submits a single choice list over
all available programs, and receives no more than a single seat from
the system, based on the choices and the ranks in the relevant merit
lists. Significantly, \emph{overbooking} of seats is forbidden.

Earlier~\cite{TechReport:2015}, we described the Multi-Round Multi-Run
Deferred Acceptance scheme that was first used for the joint seat
allocation in 2015.  Crucially, unlike the 2014 and earlier admissions
processes, the scheme seamlessly handles \emph{multiplicity of merit
  lists} across different institutes and programs; indeed every
program may have a separate merit list, and these lists need not have
any relation with each other. In addition, the scheme has several
other desired objectives. The scheme makes it safe and optimal for
candidates to report their \emph{true} preferences over programs. The
seat allocation produced does not waste seats and is \emph{fair} in
the sense that it does not give a seat to a lower-ranked candidate
when it was denied to a higher ranked candidate. Further, the
allocation is \emph{optimal} in a formal sense, providing each
candidate with the best possible seat subject to fairness.  Without
compromising on these tenets, the scheme factors in various business
rules, including reservations for different birth categories,
reservations for home state candidates, and rules regarding
\emph{dereservation} when sufficient candidates are not available.
The scheme also factors in changes that are inevitable when it is
discovered, for instance, that candidates have inadvertently or
otherwise, incorrectly declared their birth category, or when it is
discovered that the qualifying criteria have been incorrectly recorded
by state education authorities.

% The joint seat allocation scheme is inspired by the single run
% Deferred Acceptance algorithm attributed to Gale and Shapley.  The
% reader may refer to \cite{TechReport:2015} for the 1st technical
% report of joint seat allocation that was used in 2015.

Over a period of the last four years, several business rules have
changed, added, or dropped. Principal among them is a rule requiring
the allocation of supernumerary seats to females, provided the program
did not have a sufficient desired percentage, while, at the same time,
not reducing the amount of seats available to non-females.  The
current report incorporates all old and new business rules that were
used in 2018, thereby displaying the versatility
of the original method.  Instead of providing only changes, this
report provides more details of other process improvements made since
2015, and up to 2018.  We hope that this self contained report will be
helpful for the reader, and will also serve as a valuable reference
for joint seat allocations in future years as well.

%   %At first glance, a common allocation scheme in which all CFTI seats
% are declared open, and all candidates express their choices
% truthfully in advance, is an obvious solution to the varied problems
% painted in the scenario above.  It should take care of candidate's
% preferences after their intense hard work.  It should take care of
% ripple effects in state engineering college to a certain effect. And
% it would soothe the tension for a number of people.
% %
% Implementing a common allocation scheme with a unified choice list
% from the candidates and with a (single) unique rank list is theoretically
% straightforward.  However, a unique rank list may be unavailable,
% and the problem takes the hue of a scientific problem.
%

Looking forward, we posit first that it is inevitable that different
colleges will prefer different mechanisms of judging merit, and
assigning relative rank.  For example, programs such as Architecture
and those in the Humanities require different skills. The IITs do not use
higher secondary marks in their admission process.  States may impose
home quotas, and the Government of India recognizes quotas (and
preferential allotment) among certain birth categories.  We believe
the ability of our algorithm to gracefully handle multiple merit lists
gives us hope to express optimism that \emph{all} undergraduate admissions in
the country, beyond the CFTIs, can beneficially use the suggested
scheme.

%

% In this report, we analyze the problem of seat allocation with
% \textbf{multiple rank lists}.  We show that retro fitting prior
% unique rank algorithms may not be desirable from the candidates
% point of view.  Instead we offer a different algorithm that is
% \textbf{provably optimal} in a formal sense.  Thus, in the
% foreseeable future, regardless of the number of seats, the number of
% colleges, and the choices of the individuals, we assert that the
% algorithm is correct. We also incorporate the business rules of the
% IITs and NITs such as quotas for categories and the physically
% challenged. We believe the same process is suitable for
% implementation for all CFTI admissions.

  %In this report, and based on the experience of running the scheme in
  %July 2015. we also present several important implementation
  %considerations, and some recommendations for the future.
  %We strike
  %two notes of caution that is necessary to maximally improve the
  %efficiency of the seat allocation: (i) the \emph{calendar} should be suitably
  %constructed so that a common allocation is implementable both in
  %theory and practice. (ii) educating candidates to \emph{fill choices properly}
  %is outside the scope of this report but an equally crucial, and
  %difficult task that must be given attention.

   \normalsize
\end{abstract}

\tableofcontents
% chapter 1
%% Introduction, Background and Scope

\chapter{Introduction}
\pagenumbering{arabic}

Allocation (earlier, {\em counseling}) of students to seats in
colleges is a key step in the admission process from an institutional
perspective. It decides the final fruit of all the hardships taken by
a candidate over years of schooling and special preparations for
competitive examinations. Importantly, this decision shapes the
candidate's future career.  For better or for worse, currently in the
Indian system, seats are allocated based on marks obtained in
competitive exams such as the Joint Entrance Examination (JEE).  A lot
of care needs to be taken to ensure that the {\em appropriate} seat is
delivered to the candidate.

Until 2014, two different entrance examination-based methods were
available for a candidate to obtain a seat in undergraduate
engineering disciplines offered by the Centrally Funded Technical
Institutions (CFTI) including the Indian Institute of Technologies
(IITs).  The admissions to the IITs
were conducted under one umbrella, whereas the admissions to the
non-IIT (CFTIs) were conducted under a different umbrella, the Central
Seat Allocation Board (CSAB). This system resulted in an inefficient
allocation.

\begin{example}
  Consider an applicant Rohini who was assigned B.Sc.\ Chemistry in IIT
  Bombay through her rank in the IIT JEE (Advanced) exam.  She is
  unable to obtain her dream choice of a B.Tech.\ program in Chemical
  Engineering at the National Institute of Technology (NIT), Nagpur in the
  first round of counseling administered by CSAB (due to her
  relatively poorer rank in her JEE Mains exam). While Rohini was
  offered Civil Engineering in NIT Silchar, she rejected this offer
  from CSAB, and settled for the B.Sc.\ program at IIT Bombay.

  For another candidate Bhavesh, it is the opposite.  He is successful
  in getting B.Tech., Chemical Engg.\ at VNIT but, having a lower rank
  than Rohini in the JEE (Advanced), does not get a seat in IIT
  Bombay. He decides to stay back in his home state of Gujarat and
  rejects the VNIT seat and also elbows out a local candidate, Gurjar,
  in L.D.\ Engg.\ College, Ahmedabad.

In the subsequent rounds of CSAB allocation, Rohini is annoyed to find
a candidate (with marks lower than her own) allotted her favored
Chemical Engg.\ seat at VNIT.  Rohini realizes that if she had accepted
the Civil Engineering seat at NIT Silchar she would have obtained a
seat in VNIT in the final reckoning. In this process, she would have
left a seat vacant in IIT Bombay.

Meanwhile, recent high school graduate Gurjar, and most likely his
parents, were trying to get a train ticket to Assam in the hope of
securing a civil engineering seat in Silchar in the so-called spot round.
\end{example}

Taking a step back, we observe that two different points of view need
to be considered to correctly allocate seats --- the point of view of
the candidates, and the point of view of the participating
institutions. All candidates should get the best available choice, and
every institution should fill all seats in the varying courses they
offer. Consistent efforts have been taken by different organizations
in charge of counseling to satisfy both viewpoints.
%It is very interesting to note that even in prestigious
%examinations like the JEE (Advanced), the cut-off marks for the Common
%Merit List is fixed based on the {\em number of seats} available
%across all participating IITs.
%The major drawback has been on
%capturing the choice of the candidate, and we explore the reasons.

Many institutions have grown to a stature of national importance. Each
of them have their own entrance examination for admission and also
have a separate rank list and a separate allocation
process. Previously, the candidate filled a {\em choice sheet} for
each of these institutions, or sometimes, group of institutions. Thus,
there was no provision in the past for the candidate to list her
choices in a {\em single} choice sheet across all institutions. Every
choice sheet he filled could indicate only the preferences of the
candidate among available choices in the institution he applies to.
As a result, the candidate who filled $K$ choice sheets may receive
admission to up to $K$ programs from which he has to select
one. However, as counseling dates are not synchronized, and there is
no legal provision for overbooking seats, the candidate may rationally
{\em block} more than one seat by paying required fees for safety
reasons. At most one of these seats is occupied by the candidate,
whereas the remaining seats go vacant. As different allocation
processes have no mechanism to track this, many seats in {\em
  unpopular} (in the eyes of the young candidates) courses end up
being unfilled.  At the same time, many candidates in the waiting
lists could not be admitted as the decision of the deserting
candidates who have been allotted a course is known only after the
courses start and that becomes too late.

For better or for worse, overbooking of seats is not allowed by the
authorities.

Different institutions and admission boards have tried different
mechanisms to alleviate this problem, which in our opinion are not
very effective, leading to allocation inefficiency as well as
logistical difficulties for candidates and institutes. Some of these
mechanisms are:

\begin{enumerate}
\item {\em Spot admissions after classes begin:} Many institutions do
  this admission within a very short window of time. It becomes
  practically impossible for the candidates to move around thousands
  of kilometers attending one spot round after another.  Not only
  there are severe logistical costs, there is also inappropriate
  allocation of seats.\footnote{In fact, in  2014
    candidates holding a CSAB seat
    were not permitted to participate in the CSAB spot round causing
    so-called rank violations.}

\item {\em Penalty:} When a candidate blocks a seat in a course the
  amount he pays may not be refunded if he does not accept the
  admission offer; further the candidate may not be allowed to write
  the entrance examination for the corresponding institution the next
  year. Such penalties may be infeasible\footnote{The courts have
    typically insisted on a full refund of fees.} or may not be
  effective.
  % \footnote{Monetary penalties may not be effective. Being
  %   debarred from a particular entrance exam next year is immaterial
  %   if the candidate is not planning to write the exam anyway, which
  %   is true for many candidates who block seats.}
\end{enumerate}

\subsection*{Solution}
So what is the solution? Rather than passing the buck to the
candidate, one can ask ``Why don't the different admission bodies join
together and solve the problem?'' With this view this document
describes the details of a Joint Seat Allocation process. It has
been successfully implemented since 2015. The participating institutes
are IITs, NITs, and Other-GFTIs (Government funded
technical institutes.)
%
%a method to join two of the major admission process (for the
%National Institutes of Technology (the NITs) and the Indian Institutes
%of Technology (the IITs), and conduct a common allocation procedure. \emph{The
%common procedure was implemented in July 2015 based on an earlier
%version of this document published in Feb 2015.}
%

The process which we formulated conducts a {\em single window
  multi-round} counseling for admissions to the Centrally Funded
Technical Institutions (CFTI) based on the JEE (Main) and JEE
(Advanced) examination. Note that there are {\em multiple} sets of
merit lists.\footnote{Different merit lists need not have any
  relationship with each other. They may or may not be based on the
  same set of examinations.} We note that each set comprises merit
lists for the General (GEN) category, Other Backward Castes Non-Creamy
Layer (OBC-NCL) category, Scheduled Caste (SC) category, Scheduled
Tribe (ST) category, and Persons With Disability (PwD)
category.\footnote{While the existence of these categories is
  significant, the definitions (or rationale) of these categories are
  outside the scope of this document.} The candidate fills {\em only
  one} choice list indicating her choices from programs offered across
all CFTIs.  Thus, the admission bodies get a {\em global} view of the
choices and, in principle and practice, a better
allocation~\cite{Interfaces-paper:2018} is
possible that incorporates both the candidate viewpoint and the
institutional viewpoint.

We believe that the problem of a single window Joint Seat Allocation
is non-trivial. In the current context, we require a careful
adaptation of existing methods based on the celebrated Deferred
Acceptance algorithm by Gale and Shapley
\cite{gale1962college,PeransonRothNRMP,abdulkadirouglu2005new}. Our
scheme is \emph{truthful}, namely the scheme makes it safe/optimal for
each candidate to report her true preferences over programs in her
submitted choice list. The seat allocation produced does not waste
seats and is \emph{fair/stable} in the sense that it does not give a
seat to a lower ranked candidate that was denied to a higher ranked
candidate. Further, the allocation is \emph{optimal} in a formal
sense, providing each candidate with the best possible seat subject to
fairness. See Section~\ref{sec:problem_statement} for formal
definitions.

With the implementation of this scheme in the last 4 years, the joint
seat allocation has led to a substantial reduction in vacancies in the
IITs quantified in \cite{Interfaces-paper:2018}. The reduction in
vacancies in CSAB institutes have been more modest.  We discuss
methods to minimize these vacancies in %our recent article
\cite{Interfaces-paper:2018}.
%{\tt http://www.columbia.edu/~yk2577/submissionIndian-engg-admissions.pdf.}

\section{Organization of the report}

Chapter~2 presents the problem formally along with introducing some
basic notations and terminologies to be used in the rest of the
report.  Chapter~3 first presents the deferred acceptance (DA)
algorithm in its simplest form. Later it describes the version that
handles two important aspects of multiple rounds, namely, \mc\ and
multiple candidates with the same rank.  Chapter~4 describes some
fundamental business rules for admissions into three types of
institutes, and how they are incorporated using virtual programs and
virtual preference tables. Chapter~5 presents the complete description
of the multiple round DA algorithm. Chapter~6 presents the way
de-reservation is handled in each round by executing multiple runs of
our algorithm. In view of this consideration, we term our scheme as
MRDA, a multi-run deferred acceptance scheme. Chapter~7 presents the
adaptation of the rule of supernumerary seats for females. In
Chapter~8 and Chapter~9, we conclude with a brief summary of the
impact of joint seat allocation in the last 4 years along with the
recommendations for future years.

Finally, in Chapter~10, which is the Appendix, we provide the details
of the following aspects of the MRDA algorithm and Joint Seat
Allocation 2018:
\begin{enumerate}
%\item Proof of correctness of the DA algorithm.
\item A key requirement, that of increasing the ratio of females to
  males, cropped up in 2018.  While Chapter~7 introduces the
  constraints, and provides our algorithm, first look alternative
  algorithms that have significant drawbacks are listed in the
  appendix for a better understanding of the nuances.
\item A summary of the activities at the reporting center during a round
  that influences the seat allocation of the next round. 
\item Survey questions that are asked when a candidate refuses a seat
\item Validation modules required to ensure that the seat allocation
  produced meets all business rules. 
\item Implementation details of MRDA.
\item The changes in the business rules of the joint seat allocation since 2015.
\item Implementation details of a 2015 business rule that we believe may be interesting to practitioners elsewhere.
\item History and background of the admission process to engineering colleges in India.
\item The challenge and opportunity of overbooking seats to reduce vacancies
\end{enumerate}

% Its simplest
% version consists of trying to evolve a seat allotment process that
% takes into account multiple merit lists of multiple entrance
% examinations using a single common choice policy. This is much more
% intricate than seat allocation from a single merit list that is done
% for IIT seat allocation today.

% chapter 2
\chapter{Preliminaries}
\section{Background and Challenges}
In the case of a single merit list, candidates are sequentially
ordered, and the allotment of seats is done by processing candidates
in the same order and allowing each candidate to choose from among the
seats that are still available.\footnote{This is known as \emph{serial
    dictatorship} in the literature \cite{roth1992two}.} For example,
consider three candidates, A, B and C.  If in the NIT merit list, they
are ranked as A, B and C, then, given the choices of A, B and C for
different branches, we give preference to A over B and C. Likewise, we
give preference to B over C. The allotment process first allocates a
seat to A, then to B, and finally to C.

In practice, the process is much more intricate as one has to take care
of various categories like GEN, OBC-NCL, SC, ST, PwD, their cut-offs,
and then recouping of unallocated OBC-NCL and PwD seats. Nevertheless, a
solution can, and has been devised.

Unfortunately, if we have multiple merit lists, the issues are much
more complex. Consider a simple scenario of three merit lists, namely
those of the IITs, the NITs and a merit list for the Architecture
(ARCH) program in which an additional examination needs to be cleared.
Again consider our 3 candidates, A, B and C. Let the three merit lists
be as follows:
\begin{verbatim}
NIT merit List: (1, A), (2, B) and (3, C) (as mentioned earlier)

IIT Merit List: (1, B), (2, C), and (3, A)

ARCH merit List: (1, C), (2, A) and (3, B)
\end{verbatim}

Note that now, unlike in the case of a single merit list, there
is no overall strict ordering among A, B and C as their relative
performance is different in the three examinations. In practice, we may
have several merit lists covering tens of examinations and lakhs of
candidates.

Now consider a joint seat allocation process where each candidate
fills up a single choice sheet. An example is shown below where we
assume, for illustration, that there is only a single seat available
in each program.

\begin{verbatim}
A: 1: IIT, 2: ARCH, 3: NIT

B: 1: ARCH, 2: NIT, 3: IIT

C: 1: NIT, 2: IIT, 3: ARCH

\end{verbatim}

Note that the choices of the candidates are their personal preferences
and may not have any linkages to their relative ranks in different
merit lists.

How do we allocate seats? The first condition that comes to one's mind
is \emph{fairness}.

% There are two conditions which we one may
% like to satisfy, namely {\em fairness} and {\em optimality}.

Fairness may appear straightforward: A candidate's choice for a
program should be honored in the order of merit for that program.
% This example contradicts what has been given above and has been
% pruned and modified.
% In the
% example, if candidate B prefers to go to NIT over ARCH and IIT,
% then it would not be fair if candidate C is allotted NIT in preference
% to B, while the allocation to A is fair. Similarly if candidate C
% prefers ARCH to IIT and NIT then candidates B or A cannot be allotted
% ARCH.
In the example, since candidate B does express a preference to go to
NIT, it should never be the case that any candidate in the NIT merit
list worse than B displaces B, and denies B a seat; B may however be
given a seat which she prefers to NIT, say ARCH.

Given multiple merit lists, there may be several solutions for the
same choices that satisfy fairness. Consider three
different allocations:
\begin{verbatim}
Allocation 1: A (NIT), B (IIT), C (ARCH)

Allocation 2: A (ARCH), B(NIT), C (IIT)

Allocation 3: A (IIT), B (ARCH), C (NIT)

\end{verbatim}

Allocation~1 is fair because A is ranked first in NIT merit list, B is
ranked first in IIT merit list and C is ranked first in ARCH merit
list. But Allocation~1 gives each candidate their worst
choice. In fact, should A and B meet each other, they would be willing
to swap their allocation! Allocation~2 is also fair, and better from the candidate's
point of view than Allocation~1 because it gives everyone their second
choice. Finally, Allocation~3 is also fair, and better than both
Allocations 1 and 2 from the candidate's point of view because
everyone gets their first choice.  This brings about the following
scientific question, given that fairness is not good enough.

\begin{quote}
Is there a ``best'' choice among the different available fair allocations, and how can one compute it?
\end{quote}

We have highlighted these questions through a simple scenario. The
solution gets complicated when we have the practical situations of
lakhs of candidates having to fill a common choice list over multiple
merit lists where each institution have their own business rules in
the handling of state and central SC/ST/OBC-NCL/PwD categories,
multi-session allocation, and spot allocations.\footnote{In the simple case with no categories, \cite{gale1962college} shows that there is a unique
\emph{candidate optimal} fair allocation, which simultaneously gives
each candidate their best possible fair allocation. Further, this allocation can be computed
using the candidate-proposing deferred acceptance algorithm.}

% It has been observed that a straightforward extension to the single
% merit list algorithm that mimics the process of sequential allotment
% does not guarantee optimal solutions even when none of the above
% complications are considered.\footnote{It is unclear what sequence to
%   consider the candidates in, considering that there are multiple
%   merit lists.}

This document presents details of design, analysis and implementation
of a scalable solution to the problem mentioned above.

%The rest of this document is organized as follows. 

% Problem statement and Challenges

\section{Formal Problem Statement}
\label{sec:problem_statement}

We start with a simplified problem with multiple merit lists for the
different programs\footnote{One may view the word `program', `course',
  `branch' and `programme' interchangeably in the rest of the
 document. For ease of understanding at this stage, one may think of
a program as a
college having a single course, say, Electrical Engineering in IIT Kharagpur.}.
Differing programs may have the same merit list, or may have different
merit lists. As discussed earlier, the problem is complex, but we term
it `simplified' for now because (for example) there are no quotas or
ties in the presentation of this section.

%\noindent {\bf Definition}:
%Let $E$ be the number of examinations conducted. Let $P(i)$ be the total number of programs for which admission is made based on
%exam $i, 1 \leq i \leq E$.

Let $\cP$ be the set of programs
%, with $P=|\cP|$ being the number of
%programs. 
For a program $p \in \cP$, let $c(p)$ denote the number of
seats in $p$. Let $\cA$ denote the set of applicants (or candidates),
with $A=|\cA|$ being the number of candidates. Each candidate is
allowed only one seat in the system. Candidates are asked to submit a
choice (or preference) list over programs; the choice list is a
strictly ordered list containing any subset of the programs in $\cP$.

We denote the preference list of candidate $x$ by
\begin{equation}
\label{eq:choiceList}
\pref(x) = p_{x,1}, p_{x,2}, \, \ldots \,, p_{x,n(x)}\
\end{equation}
which means that candidate $x$ has listed $n(x)$ programs, with
program $p_{x,1}$ being her top choice, program $p_{x,2}$ being her
next choice and so on. The candidate is asked to list only programs
she is interested
in. %We write $p >_a p'$ if (as per $\pref(x)$), candidate $x$ prefers program $p$ to program $p'$.

Each program must submit its capacity $c(p)$, as well as its merit
list of candidates, which is a strictly ordered list containing a
subset of the candidates in $\cA$. We denote the merit list of program
$p$ by
\begin{equation}
\label{eq:meritlist}
\merit(p) = x_{p,1}, x_{p,2}, \, \ldots \, , x_{p,m(p)} \
\end{equation}
which means that program $p$ has ranked $m(p)$ candidates in its merit
list, with candidate $x_{p,1}$ ranked $1$, candidate $x_{p,2}$ being
ranked 2,  and so on.

% {\sf Warning: $ x$ is overloaded, can we use $y$ instead?

% Y: In what sense is $x$ overloaded? Did we use $x$ for something other than a candidate?}

Let $\mu(x) \in \cP$ be the program allotted to candidate $x$ by some
mechanism, with some candidates possibly not getting any seat, in
which case we write $\mu(x) = \phi$. Denote the overall allocation by
$\bar{\mu} = (\mu(x))_{x \in \cA}$. We want a mechanism with the
following properties:
\begin{enumerate}

\item {\em Fairness}: Suppose candidate $x$ is allotted program
  $p$. Then for any other candidate $y$ such that $y$ has a better
  (smaller) rank than candidate $x$ in the merit list of $p$, the
  allocation of $y$ should be $p$ or some other program that $y$
  prefers to $p$.\footnote{This property is called \emph{stability} in
    the literature \cite{gale1962college}.} %Mathematically, $\mu(x)=p$ and $x<_p y \ \Rightarrow \mu(y) \geq_{y} p$

  Further, the mechanism must ensure that a candidate is not allotted
  a program that she did not list, and that no program is allotted to a
  candidate that was not a part of the merit list of the
  program.\footnote{This property is called \emph{individual
      rationality} in the literature \cite{roth1992two}.}

\item {\em Optimality}: There does not exist any other allocation
  $\bar{\mu}'$ that satisfies the \emph{Fairness} property, and
  provides \emph{any} candidate $x$ with an allocation she prefers to
  $\mu(x)$ based on her preference list.

% {\sf COMMENT: What does ``like'' mean?

% Y: Please check now.}

\item {\em Truthfulness}: The mechanism must make it optimal for candidates to report their true preferences.
\end{enumerate}

A priori it is unclear that a mechanism satisfying all these
properties exists. However, it turns out that such a mechanism \emph{does} exist,
and was constructed by Gale and Shapley.
\cite{gale1962college}).

\section{Previous Work}
\label{sec:prev-work}
An initial solution to the simplified problem was proposed by Gale and
Shapley in 1962 \cite{gale1962college}  by formulating
it as a ``Stable marriage problem''. The proposed solution was shown to
have a multitude of desirable properties including fairness and
candidate optimality \cite{gale1962college}, and truthfulness
\cite{DubinsFreedman} (no candidate or group of candidates can benefit
from misreporting their preferences).
The Gale and Shapley mechanism has been adapted and implemented
successfully in a multitude of real world settings, e.g., the National
Residency Matching Program (NRMP) (running in the USA since 1951, redesigned in
1999 \cite{PeransonRothNRMP}), New York City high school admissions
since 2003 \cite{abdulkadirouglu2005new}, and school and college
admissions in Hungary \cite{biro2008student}. 

However, our problem involves a variety of business rules governing
flows of different systems that need to be streamlined for evolving a
sound process of common allocation. We need to suitably adapt the
Gale-Shapley mechanism to incorporate these business rules while
retaining all these desirable properties.  In the sequel, we
demonstrate these and come up with a practical algorithm.

% chapter 3
% The proposed scheme

\chapter{A Combined Seat Allocation Scheme}
\label{sec:DA}
%\subsection{Assumptions}

%We begin by solving the problem stated in Section
%\ref{sec:problem_statement} which has multiple merit list, but no
%quotas.

Our proposed mechanism first collects information from the
participating programs and candidates. It then uses this information
to produce an allocation of seats.

\section{Initial Information collection}
\label{subsec:initial_information}
Each participating program  provides two pieces of information:
\begin{itemize}
  \item The capacity $c(p)$, i.e., the number of available seats
  \item A merit list of eligible candidates (Equation~\ref{eq:meritlist}).
        The purpose of a merit list of program is to compare two candidates.
        For implementation of the algorithm, it is quite possible that
        this merit list is not explicitly computed.
        Instead, the relative order of any two
        candidates can be determined from the information associated with each
        of them (marks, birth category, PwD status).
\end{itemize}

Each candidate (who is eligible for at least one program) enters a
preference list (Equation~\ref{eq:choiceList}) of programs for which she is
eligible, with the first entry being her most preferred program, the next
entry being her next most preferred program, and so on.

%Logically the whole data can be viewed as a bipartite graph between programs and candidates.
We now describe the algorithm for allocating programs to candidates
which incorporates all the business rules (see
Chapter~\ref{chapter:businessrules}), 
%CFTIs 
and has three properties, namely, fairness, candidate-optimality, and
truthfulness.

The Deferred Acceptance (DA) algorithm that forms the core of the
joint allocation is described in the following section.
\section{Basic DA algorithm}
We stress that in the DA algorithm, ``applications'', insertions into
``waitlists'' and ``rejections'' mentioned are all merely part of the
algorithm, and do not actually involve any participation from the candidates
and programs in the real world. That is, the algorithm internally
generates applications, using the information that the candidates and
programs have provided. We now state the algorithm in words.
See Section~\ref{subsec:psedocode-of-DA}
for the complete pseudocodes.

\noindent Input:
\begin{itemize}
\item For each program, its capacity and the rank list of eligible candidates.
\item For each candidate, a preference list of programs.
\end{itemize}
\noindent Algorithm:
\begin{enumerate}
\item All candidates apply  (in any order) to the
  first program in their preference list.

  \item Each program $p$ considers the applications it has received.
    Applications from candidates who are not eligible are immediately
    dropped. Let the capacity of the program be $c(p) > 0$.  If the
    program has received $c(p)$ or fewer eligible applications, then
    it retains all candidates on a waitlist.  Otherwise, it ranks the
    candidates
    %\footnote{assuming for the moment that equal ranks do not exist; ties are handled in a later section.} 
    making these requests (as per the merit list of
    the program) and retains only the $c(p)$ best candidates on its
    waitlist, and rejects other candidates.

    If no rejections are made by any program, the algorithm
    terminates. %The final waitlists are the admitted candidates.

  \item Only rejected candidates apply (in any order) to the next
    program on their list, if any, and the algorithm returns to Step~2.

    If not even a single application is generated, then the
    algorithm terminates.

% {\sf How does the algorithm terminate if all rejected candidates have
%   no program to apply to?

%   Y: Thanks, please check.}

\end{enumerate}

\noindent Output: When the algorithm terminates, the (final) ``wait
list'' for each program $p$ constitutes candidates admitted to
program $p$.

%The output of the algorithm is the seat allocation produced by our
%mechanism.
We present complete details of the DA algorithm through pseudocode in the
following sections.
%A formal proof of correctness and other computational considerations of
%the algorithm are described 
%%respectively 
%in 
%Section~\ref{subsec:proofs} 
%%and
%\ref{sec:computational-considerations-of-DA} 
%in Appendix.

\section{Pseudocode}
\label{subsec:psedocode-of-DA}
Denote rank of candidate $x$ with respect to $\merit(p)$ by
$\rank(x,p)$. For a list $l$, denote the number of entries in the list
by $\length(l)$.

We narrate two versions of the pseudocode.  In the first version, the
assumption is that the entire seat allocation happens in a single
round and all candidates in a merit list have distinct ranks.  It is
presented to understand the general flow of the algorithm. The second
version gets rid of this assumption, and is for multi-round scenarios
as described in Chapter~\ref{subsec:multi_round}.  In particular, it
handles two non-trivial issues (i) there may be multiple candidates
with the same rank, (ii) the credentials of candidates, and thus
relative ordering, may change between rounds.

\begin{algorithm}
\caption{Deferred Acceptance Simple Version}
\label{alg:DA-simple}

INPUT:\\
Candidates $\cA$, Programs $\cP$\\
Preference list $\pref(x)$ for each $x \in \cA$\\
Capacity $c(p)$ and merit list $\merit(p)$ for each $p \in \cP$\\

OUTPUT:\\
For each candidate $x\in \cA$, the allocation $\mu(x) \in \cP \cup \{ \emptyset \}$ \\
Also for each program $p \in \cP$, the list of admitted candidates $\wl(p)$\\

\begin{algorithmic}[1]
\ForAll{$p \in \cP$}%\label{line:startAlgo}
    \State Create an empty ordered list $\wl(p)$ that will consist of
    \State candidates ordered by their rank in $\merit(p)$
\EndFor
\State Create an empty queue $\cQ$
\ForAll{$x \in \cA$} %\label{line:enqueuefirsttime}
    \State $\LP(x)\gets 1$ \Comment{Initialize list position to 1.}
    \If{$\length(\pref(x)) > 0$}
        \State {\sc Enqueue}($x$,$\cQ)$  \Comment{$x$ enters queue $\cQ$}
    \EndIf
\EndFor  %\label{line:beforeProcessQ}
\While{$\cQ$ is non-empty}
    \State $x \gets \mbox{\sc Dequeue}(\cQ)$ \Comment{$x$ is any candidate
removed from queue $\cQ$}
    \State $p \gets p_{x, \LP(x)}$ \Comment{$x$ applies to program $p_{x,\LP(x)}$}
    \If{$x$ is not eligible for $p$}
        \State \Call{Reject}{$x$}
        \State {\bf continue}
    \EndIf
    %\State
\algstore{myalg}
\end{algorithmic}
\end{algorithm}

\begin{algorithm}
\begin{algorithmic}[1]
\algrestore{myalg}
    \If{$\length(\wl(p))=c(p)$} \Comment{The waitlist is full}
        \State $y \gets \mbox{Last candidate in }\wl(p)$
        \If{$\rank(x,p)<\rank(y,p)$}
            \State Remove $y$ from $\wl(p)$
            \State \Call{Reject}{$y$}
            \State Insert $x$ into ordered list $\wl(p)$ at correct location
        \Else
            \State \Call{Reject}{$x$}
        \EndIf
    \Else
        \State Insert $x$ into ordered list $\wl(p)$ at correct location
    \EndIf
\EndWhile
\ForAll{$x \in \cA$}
    \If{$p_{x,i(x)}$ exists in $\pref(x)$}
        \State $\mu(x) \gets p_{x,i(x)}$
    \Else
        \State $\mu(x) \gets \emptyset$
    \EndIf
\EndFor
\State \Return $\mu(x)$ for all $x \in \cA$ and $\wl(p)$ for all $p \in \cP$
\State
\Function{Reject}{$x$}
    \State Increment $\LP(x)$
    \If{$p_{x,\LP(x)}$ exists in $\pref(x)$} \Comment{$x$ wants to apply further}
        \State Enqueue($x$,$\cQ$) \Comment{$x$ enters queue $\cQ$ again}
    \EndIf
\EndFunction

\end{algorithmic}
\end{algorithm}

\newpage
\subsection{Multi-Round Scenario}
\label{sec:multi-round-intro}
Many candidates who obtain seats in the first round of seat allocation
will surrender or reject their respective seats at a later stage. In
order to utilize these surrendered seats, the business rules allow
multiple rounds of seat allocation.  We now present Algorithm~2 that
incorporates the following two issues --- the first issue arises due to
multiple round and the second issue arises due to multiple candidates
with same rank competing for a program.\footnote{In 2015, similar
  issues had more complicated business rules;
  see~\cite{TechReport:2015} severely complicating the algorithm.}

\begin{itemize}
\item {\em Seat guarantee in future rounds}.\\
  Credentials (such as birth category, or qualifying marks) of
  candidates may change (due to faulty reporting) during future
  rounds. In some cases, fresh candidates become eligible to
  participate between rounds. In such situations, we need to provide
  seat guarantee to candidates who were allotted a seat in an earlier round.
  %but their credentials did not change.  
  We introduce the idea of 
  $\mc(p)$ for each $p \in \cP$: this quantity is used in second and
  later rounds of allocation (see Chapter~\ref{subsec:multi_round} for
  details). Intuitively, a candidate with rank better than, or as good
  as $\mc(p)$ will never be rejected by $p$ regardless of the capacity
  of $p$. Candidates offered a seat in the first round will thus be no
  worse off in subsequent rounds. As alluded above, there might be new
  candidates who become eligible for seat allocation in subsequent
  rounds\footnote{For example, a candidate might be ineligible for joint seat
  allocation earlier due to less board marks, but may become eligible
  in subsequent round if her board marks increase after
  re-evaluation. In such cases, the allocated seat shall be what the
  candidate would have got on the basis of the revised rank in the
  first round. This seat can be a supernumerary seat.}.
  %We use the notation $x | y$ to indicate that $\rank(x,p)$ is as good
  %as, or better than $\mc(p)$, and $\rank(y,p)$ is
  %worse.
%
%---------------Commented on 22 May 2018--------------------------
%\item Another optional input is the list \cc\ which consists of
%  candidates who might have had their category changed. For example, a
%  person presumed to be with an OBC-NCL in an earlier round may be
%  reclassified as a general category candidate.  As per the business
%  rules, in this case, the candidate will be allocated a seat in the
%  best (in terms of the filled-in choices) possible choice of academic
%  program that has unfilled seats and supernumerary seats
%  are NOT created.

\item  
{\em Multiple candidates with equal rank}.\\
Suppose multiple candidates obtain exactly the same rank. In the
situation when we are forced to reject some candidate due to the
possibility that the program is oversubscribed, we have to also reject
(and remove) all other candidates in the waiting list who
also have the same rank.  This may not be always possible.

We carry out suitable modifications as stated in Algorithm \ref{alg:DA} to
allow such candidates to obtain seats in the program on a supernumerary basis.
We keep $\wl(p)$ fully ordered at each step, by
choosing an arbitrary order between candidates in $\wl(p)$ who have
the exact same rank as per $\merit(p)$.  

There are two concepts involved here.  First, when an attempt is made
to remove a candidate $x$, a removal of all candidates with the same
rank of $x$ may cause the waitlist which is initially overflowing, to
now underflow.  To identify this situation, we compute the length $G$ of
the waitlist, and the length $L$ of the list of candidates all of whom
have the same rank as $x$.  If the difference between the two exceeds
or matches the capacity, we are free to remove all such
candidates. Second, if this condition is not satisfied, it is not
possible to remove all candidates with rank of $x$.

%\item {\color{red} The starting position on the preference list $\LP(x)$ for each
%  $x \in \cA$ and the current queue $\cQ$ of candidates to be
%  processed. These starting positions inputs are meant as an optimization
%  mechanism and can be ignored in the first reading of the pseudo-code.}
  
  %incorporation of certain unusual
  %business  rule for defense service (DS) candidates, see
  %Section~\ref{subsec:DS}.
\end{itemize}
The pseudocode of Algorithm~2 allows for additional optional inputs in order to implement second and later
rounds of seat allocation. In Chapter~\ref{chapter:businessrules}, we show how various business rules of IITs
and NITs (such as handling quotas, supernumerary seats for DS candidates, etc.) can be
incorporated seamlessly in this algorithm. The full description of how to use
Algorithm~2 to implement the seat allocation in each round is provided
in Chapter~\ref{subsec:multi_round}. 

\begin{algorithm}
\caption{Deferred Acceptance (Full version allowing multiple rounds and multiple candidates with the same rank)}
\label{alg:DA}

INPUTS:\\
Candidates $\cA$, Programs $\cP$\\
For each $x \in \cA$:\\
\phantom{xx} Preference list $\pref(x)$\\
\phantom{xx} Optional input: integer $\LP(x)$. Default value $\LP(x) = 1$.
(Start from beginning of preference list by default.)\\
For each $p \in \cP$:\\
\phantom{xx} Capacity $c(p)$ and merit list $\merit(p)$.\\
\phantom{xx} Optional input: $\wl(p)$.\\
\phantom{xx} Optional input: {\mc($p$)}. $0$ by default.\\
\phantom{xx} Optional input: $\cQ$ a queue of candidates. By default
contains all %Indian 
candidates $x$ with $\length(\pref(x)) > 0$.\\ 
%\phantom{xx} Optional input: $\cc$. A list of candidates whose
%category has changed (empty by default)

OUTPUTS:\\
For each candidate $x\in \cA$, the allocation $\mu(x) \in \cP \cup \{ \emptyset \}$ and $\LP(x)$.\\
Also for each program $p \in \cP$, the list of admitted candidates $\wl(p)$\\

\begin{algorithmic}[1]
%%%%%%%%%%%%%%%%---COMMENTED ON 14th October 2018-------------
%\State \ldots Everything up to Line~\ref{line:beforeProcessQ} in
%  Algorithm~\ref{alg:DA-simple}
%%%%%%%%%%%%%%%-----------------------------------------------
%
%
%
%
%%%%%%%%%%%---INSERTED on 14th October------------
\ForAll{$p \in \cP$}\label{line:startAlgo}
    \State Create an empty ordered list $\wl(p)$ that will consist of
    \State candidates ordered by their rank in $\merit(p)$
\EndFor
\State Create an empty queue $\cQ$
\ForAll{$x \in \cA$} \label{line:enqueuefirsttime}
    \State $\LP(x)\gets 1$ \Comment{Initialize list position to 1.}
    \If{$\length(\pref(x)) > 0$}
        \State {\sc Enqueue}($x$,$\cQ)$  \Comment{$x$ enters queue $\cQ$}
    \EndIf
\EndFor  \label{line:beforeProcessQ}
%%%%%%%%%%%---INSERTED on 14th October------------

\While{$\cQ$ is non-empty}
    \State $x \gets \mbox{\sc Dequeue}(\cQ)$ \Comment{$x$ is any candidate
removed from queue $\cQ$}
    \State $p \gets p_{x, \LP(x)}$ \Comment{$x$ applies to program $p_{x,\LP(x)}$}
    \If{$x$ is not eligible for $p$ OR $c(p)=0$}
        \State \Call{Reject}{$x$}
        \State {\bf continue}  \Comment move to next person in $\cQ$
    \EndIf
\State Insert $x$ into ordered list $\wl(p)$ at correct location
\If{$\length(\wl(p)) \leq c(p)$}{\bf ~continue}
    \Comment space in program
\EndIf
\State $y \gets \mbox{Last candidate in } \wl(p)$ %\Comment{y can be null}
\If{rank($y$)$>$MinCutOff($p$)} 
\Call{RemoveAndReject}{$y,p$}
\EndIf
\EndWhile
%
%%%%%%%%%%%%%%%%ADDED on 14th October---------------
%\algrestore{myalg}
\algstore{myalg}
\end{algorithmic}
\end{algorithm}

\begin{algorithm}
\begin{algorithmic}[1]
\algrestore{myalg}
\ForAll{$x \in \cA$} \label{line:assign_output}
    \If{$\LP(x)\leq \length(\pref(x))$}
        \State $\mu(x) \gets p_{x,\LP(x)}$
    \Else \Comment{$x$ reached the end of her list}
        \State $\mu(x) \gets \emptyset$
    \EndIf
\EndFor
\State \Return $\mu(x), \LP(x)$ for all $x \in \cA$ and $\wl(p)$ for all $p \in \cP$
\Function{Reject}{$x$}
 \Comment{The function is assumed to have access to $\LP(x)$, $\pref(x)$, $\cQ$}
  \State Increment $\LP(x)$
  \If{$\LP(x)\leq \length(\pref(x))$} \Comment{$x$ wants to apply further}
        \State Enqueue($x$,$\cQ$) \Comment{$x$ enters queue $\cQ$ again}
    \EndIf
\EndFunction
%%%%%%%%%%%%%%%%%14th October%%%%%%%%%%%%%%%%%%%%%%%%%%%%%%%%%%
%
%
\Function{RemoveAndReject}{$w,p$}
\label{line:apply_noties}
  \State $G \gets |\wl(p)|$
  \State $L \gets \mbox{number of people with rank same as $w$}$
  \Comment L is at least 1
  \If{$(G - L) \geq c(p)$}
%      \ForAll{$x \in \wl(p)$ with rank same as rank($w$)}    
      \ForAll{$x \in \wl(p)$ with $\rank(x,p) = \rank(w,p)$}
          \State Remove $x$ from $\wl(p)$
          \State \Call{Reject}{$x$}
      \EndFor    
  \EndIf  
\EndFunction
%\algstore{myalg}
\end{algorithmic}
\end{algorithm}

\chapter{Business Rules}
\label{chapter:businessrules}

As discussed earlier, the Government of India recognizes quotas for
certain birth categories.  In general, a candidate applies for a program
rather than for a seat in a particular category; she may be eligible
for multiple seat categories.  The so-called business rules~\cite{Business-rules:2018} describe how to allocate seats in such scenarios. In this
and subsequent sections, we describe how these rules are to be
incorporated into the proposed algorithm. The following is a broad
classification of the variety in allocation.
\begin{enumerate}
\item Allocations based on reservations based on  birth-categories.
\item Allocations for persons with disabilities (PwD).
%\footnote{We
%    use PwD and PD interchangeably in this document.}.
\item Allocations for candidates who are not nationals of India
  (International students). This particular section applies only to the IITs.
\item Allocations for certain children of military personnel killed in
  military operations (DS candidates). This particular section applies
  only to the IITs.
%\item Allocation for students who cannot be distinguished on merit
%  (multiple candidates with the same rank).
\item Allocations for candidates who fail to clear certain minimum
  thresholds but can be groomed for admission a year later (preparatory
  course (PC) candidates).  This particular section applies only to the IITs.
\item Allocations based on reservations on the residency state of a
  candidate. This particular section does not apply to the IITs.
\item Allocations based on de-reservation of seats.
\item Allocations incorporating the rule of supernumerary seats for female candidates.
\end{enumerate}

Allocations based on de-reservation of seats (Item 7), is presented in
full details in Chapter~\ref{chapter:deReserve}. Allocation
incorporating the rule of supernumerary seats for female candidates
(Item 8) is presented in
Chapter~\ref{chapter:SupernumeraryFemales}. In the current chapter we
describe how our DA algorithm can incorporate Items 1-6.

\section{Incorporating Quotas}

We first consider the important cases of 1-2 above.
%Later we consider other cases.

\subsection{Virtual programs}
All candidates, irrespective of their respective categories, 
declare programs in the decreasing order of preference. However,
internally we introduce virtual programs based on the quota for each
category.  Each virtual program will be associated with a merit list
constructed out of rank lists (which in turn is based on marks
obtained in the exam).  Each candidate will now be
associated with a virtual preference list.  The concept of virtual
programs, together with virtual merit lists and virtual preference
lists for candidates is the suggested way for handling the majority of
business rules.

There are 8 virtual programs for each actual program as per Table \ref{table:categoryTag}. The DS virtual program for the IITs is considered later (Section~\ref{subsec:DS}).

%\footnote{In addition, there may be PC virtual programs defined (PC-OP-PD, PC-OBC-PD, PC-SC-PD, PC-SC, PC-ST-PD and PC-ST) if a preparatory course is available and we choose to implement de-reservations using multiple runs of DA, cf. Section \ref{sec:multiPass}. Alternatively, if we use single pass DA with de-reservations, cf. Section \ref{sec:singlePass}, then we do not define PC virtual programs.}
\begin{table}[h]
\begin{center}
\begin{tabular}{|c|c|c|c|}
\hline
OPEN & OBC-NCL & SC & ST \\ \hline
OPEN-PwD & OBC-NCL-PwD & SC-PwD & ST-PwD \\
\hline
\end{tabular}
\caption{\label{table:categoryTag}
Candidates are partitioned into categories and assigned a
  tag.  A candidate can possess only one of these 8 tags.
  Further, for each actual program, there is a separate virtual program corresponding to each of these 8 categories.}
\end{center}
\end{table}

\subsection{Virtual Preference List}

\begin{figure}[!ht]
\centering
\includegraphics[width=\textwidth]{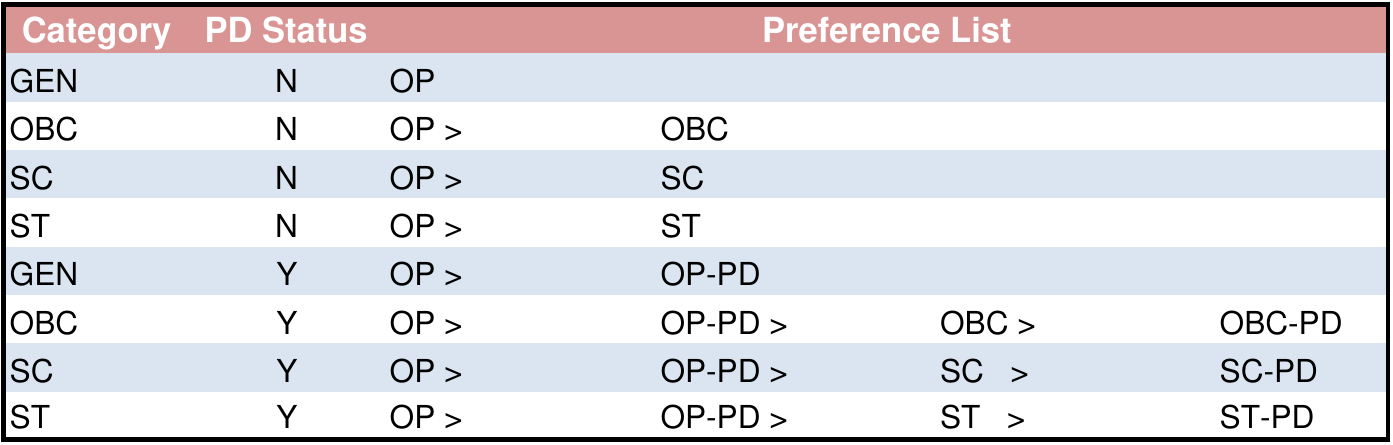}
\caption{\label{Figure:NoDereserve}Virtual preference list 
%\protect\footnotemark
for candidates when quotas are
 involved.}
\end{figure}

The sequence of seat allocation mentioned in Rule VII of 
%\cite{BJan3}
\cite{Business-rules:2018}
are very important.  These are given in Figure~\ref{Figure:NoDereserve} and merit
careful consideration.\footnote{In this and all further preference tables, OP, OBC and PD are used instead of OPEN, OBC-NCL and PwD respectively. That is, OP refers to OPEN, OBC refers to OBC-NCL, OBC-PD refers to OBC-NCL-PwD, etc.}  For example, the sixth row states that for an
OBC-NCL candidate with PwD tag, we try to fill the OPEN seats before any
other seat, failing which we consider the OPEN-PwD seat before venturing
into OBC-NCL seats.

\subsection{Virtual Merit Lists}
\label{sec:standardRankList}

\textbf{Standard Rank Lists}
How does one decide on which candidate to award a seat if there are
competing candidates for a program?  Associated with each virtual
program is a merit list constructed from ranks provided by various
examinations.

Quoting rule VI from \cite{Business-rules:2018}
%\cite{BJan3}
 the following TYPES of rank lists
will be prepared based on pre-defined cut-offs:

\begin{enumerate}
\item Common rank list (CRL): It includes candidates who are assigned the
tag GEN, GEN-PwD, OBC-NCL, OBC-NCL-PwD, SC, SC-PwD, ST or ST-
PwD.
Foreign nationals are also included in the CRL prepared based on JEE
(Advanced) 2018.

\item OBC-NCL rank list: It includes candidates who are assigned the tag
OBC-NCL or OBC-NCL-PwD.

\item SC rank list: It includes candidates who are assigned the tag SC or SC-PwD.

\item ST rank list: It includes candidates who are assigned the tag ST or ST-PwD.

\item CRL-PwD rank list: It includes candidates who are assigned the tag
GEN-PwD, OBC-NCL-PwD, SC-PwD or ST-PwD.

\item OBC-NCL-PwD rank list: It includes candidates who are assigned the
tag OBC-NCL-PwD.

\item SC-PwD rank list: It includes candidates who are assigned the tag SC-PwD.

\item ST-PwD rank list: It includes candidates who are assigned the tag
ST-PwD.
\end{enumerate}

%We also assume that rank lists \emph{will} be prepared for preparatory
%candidates (PC).

Once rank lists are available, the rules for allocation for various
seat categories are specified in Rule VII of \cite{Business-rules:2018}.
%\cite{BJan3}.

% \footnotetext{In this and all further preference tables, OP, OBC and PD are used instead of OPEN, OBC-NCL and PwD respectively. That is, OP refers to OPEN, OBC refers to OBC-NCL, OBC-PD refers to OBC-NCL-PwD, etc.}

In each virtual queue, candidates with differing tags are eligible to
participate. Thus in the OPEN virtual programs, we can find candidates
with different categories from Table~\ref{table:categoryTag}. The
virtual merit list encodes the order of consideration for seats in
each program.  
%
%Although the virtual merit list appears to be exactly
%the same as the standard rank list at this juncture, de-reservation
%may be implemented by constructing virtual merit lists that suitably
%combine standard rank lists, cf. Section \ref{sec:DA}.
%
\subsection{Preparatory courses allocation}
\label{subsec:preparatory}

The notion of PC (Item 5 at the beginning of Chapter~\ref{chapter:businessrules}) for the IITs, will, however, uncover why
only the standard merit lists are not sufficient in allocating seats.
Quoting Rule XI of \cite{Business-rules:2018} we have the rules for seat allocation
to preparatory courses applicable for every round. 
%{\color{red} \footnote{
%In the case of IITs and ISM, seat allocation in the 4th round
%will be made only for preparatory courses.  This consideration should
%be kept in mind for multi-round (Chapter~\ref{subsec:multi_round})
%of seat allotment.}}.

\begin{enumerate}

\item Unfilled OPEN-PwD seats will be allocated to candidates in the
  CRL-PwD preparatory rank list subject to no more candidates in the CRL-PwD
  rank list having opted for them.

\item Unfilled OBC-NCL-PwD seats will be allocated to candidates in
  the OBC-NCL-PwD preparatory rank list subject to no more candidates
  in the OBC-NCL-PwD rank list having opted for them.

\item Unfilled SC-PwD seats will be allocated to candidates in the
  SC-PwD preparatory rank list subject to no more candidates in the
  SC-PwD rank list having opted for them.

\item Unfilled ST-PwD seats will be allocated to candidates in the
  ST-PwD preparatory rank list subject to no more candidates in the
  ST-PwD rank list having opted for them.

\item Unfilled SC seats will be allocated to candidates in the SC
  preparatory rank list subject to no more candidates in the SC rank
  list have opted for them.

\item Unfilled ST seats will be allocated to candidates in the ST
  preparatory rank list subject to no more candidates in the ST rank
  list have opted for them.
\end{enumerate}

In order to allot seats to PC candidates, preparatory rank lists may
be prepared. This leads us to the following definition.

\begin{definition}
\label{def:primary_merit_list}
We construct the \emph{extended merit list} for any virtual program,
cf. Table \ref{table:categoryTag}, by taking the standard
rank list followed by the corresponding preparatory course (PC) rank
list if
\begin{itemize}
\item such a list has been prepared, and
\item the corresponding preparatory course exists.
\end{itemize}
\end{definition}

Up to six lists are possible: PC-CRL-PwD, PC-OBC-NCL-PwD, PC-SC,
PC-SC-PwD, PC-ST, PC-ST-PwD.  In each case, if the PC list is
prepared, it is included as part of the extended merit list for the
parent category: e.g., if the PC-SC rank list is prepared, the virtual
merit list for SC courses (in institutes that have PC courses) will
include the SC rank list followed by the PC-SC rank list. This will
automatically lead to unfilled seats being offered to PC
candidates. Note that
% Moreover, this does not lead to any kind of complication since
the PC rank lists do not have any commonality with the standard
rank lists, i.e., a candidate in PC rank list will not appear in any standard rank list.

In summary, at this juncture, in the case no PC exists for a program (as
is the case of the NITs), the virtual merit list of a virtual program
is identical to the corresponding standard rank list. On the other
hand, if PC courses are present, the virtual merit list of a virtual
program is identical to the extended merit list.

Later on, we will see more complex virtual merit lists.

\subsection{Updated Algorithm: DA With Quotas}
\label{sec:daWithQuotas}

Armed with virtual programs, virtual merit lists corresponding to
virtual programs, and virtual preferences, we now explicitly construct
the internal preference list for each candidate.

\begin{example}
Consider a candidate with category tag SC-PwD, who is
eligible for both the SC virtual program, and the SC-PwD virtual
program. This candidate has not cleared the cutoff specified for OPEN
seat. Then for each IIT program $p$ in the preference list of the
candidate, we instantiate the virtual preference from
Figure~\ref{Figure:NoDereserve} by excluding the OPEN option, and considering
only the three other options corresponding to row 7.
\end{example}

Algorithm~\ref{alg:DA} is run for each round based on the construction
of the virtual preference list for all candidates, and extended merit list
%(see Definition~\ref{def:primary_merit_list}) 
for each virtual program.

It is important to note that by modifying the virtual preference
tables, the standard DA with quotas can be applied based on variations
in business rules on how quotas should be administered when, for
example, unfilled seats are found.

%\section{DA with multiple candidates having same rank}
%\label{sec:DAwithTies}

\section{DA with International Students}
\label{sec:foreign}
%%%%%%%COMMENTED On 14th October
%The business rules for the IITs specify that foreign nationals should be given the
%best program for which they qualify, on a supernumerary basis. 
%However, if the number of Indian candidates in a program is below
%capacity, then every foreign candidate, who requests for the program,
%will be accepted in the program.
%
%
The business rules for the IITs specify that international candidates
can be admitted to programs in IITs through supernumerary seats
subject to the following constraints.
\begin{enumerate}
\item Eligibility: The candidate (with international student
  credentials) must have a valid rank in CRL and satisfy all the
  eligibility requirements meant for a GEN category candidate.
  Furthermore, such a  candidate is eligible for a program if her
  rank is not worse than the rank of the worst rank Indian candidate
  in OPEN category getting admitted to that program.  However, if
  there is any OPEN seat lying vacant in a program, then every international
  candidate is eligible for that program.
\item As per rules,
%the number of supernumerary seats for foreign candidates must not exceed 10\% of the total seats in an academic program. 
  the total number of international allocations must not exceed 10\% of the
  total allocations in an academic program.  However, each IIT may
  choose to restrict the seats for such candidates to any number
  below 10\% in some, or all programs. There is a separate seat matrix
  provided by IITs for international candidates each year. This seat matrix
  is provided as input along with the usual seat matrix which is meant
  for Indian (i.e., non-international) students.
\end{enumerate}

In order to incorporate this business rule, we proceed as
follows. First, corresponding to each OPEN program, we introduce a
virtual program for international candidates, with a capacity decided
by each institute. Next, we compute the seat allocation
only for Indian candidates using Algorithm 2.  Then we process all
international candidates. In order to satisfy the eligibility criteria
mentioned in Item~1 above, we replace lines 15 to lines 18 in
Algorithm~2 by the following code. Here $\mbox{OPEN}(p)$ is the OPEN
virtual program for Indian candidates corresponding to the international
virtual program $p$.

\begin{algorithmic}[1]
              \If{$c(p)=0$}
                  \State \Call{Reject}{$x$}
                  \State {\bf continue}
              \EndIf
              %\State $k \gets$ number of Indian candidates admitted in $\mbox{OPEN}(p)$    
              \If{$\length(\wl(\mbox{OPEN}(p))) \ge c(\mbox{OPEN}(p))$} \Comment $\mbox{OPEN}(p)$ had no vacancy
                  \State $y \gets \mbox{Last \textsc{Indian} candidate in }\wl(\mbox{OPEN}(p))$
                  \If{$\rank(x,p) > \rank(y,p)$} \Comment{$x$ has inferior rank than $y$} 
                        \State \Call{Reject}{$x$}
                        \State {\bf continue}                 
                  \EndIf
              \EndIf       
\end{algorithmic}

\noindent

\section{DA with candidates from defense service quota}
 \label{subsec:DS}
 The business rule specifies that DS (defense services) candidates
 should be given the best program they prefer in an IIT subject to the
 constraint that there are only 2 seats for DS candidates per
 IIT. Moreover, these seats will be supernumerary seats.  
 Note that for a DS candidate to be eligible for virtual DS program, she must have a valid
rank in CRL.
 
\subsection{Incorporating the business rule for DS candidates}
To implement the rule, a new virtual DS program with capacity 2 is
created for each IIT, namely IITK-DS, IITB-DS, IITR-DS, and so on.
Only DS candidates are eligible for this virtual program. Furthermore,
the preference list of each DS candidate is modified as follows.  If
the preference list of the DS candidate is
$\langle p_1,p_2,p_3\rangle$, then her preference list will first be
modified as $\langle p_1$, Institute($p_1$), $p_2$,
Institute($p_2$),$p_3$, Institute($p_3$)$\rangle$, where
Institute($p_i$) is the DS virtual program created for the IIT
corresponding to $p_i$.  For example, if $p_i$ is a program in IIT
Roorkee, then Institute($p_i$) is IITR-DS.  Then $p_1$ (likewise
$p_2,p_3$) will be replaced as usual by the list of virtual programs
of $p_1$ for which the DS candidate is eligible based on the birth
category and PwD status. This ensures that each DS candidate first
competes to get a program of some IIT based on birth category and PwD
status, and only after that, she competes (based on rank in CRL) to
get that program through DS rule. DS candidates admitted to, for
example, the IITR-DS virtual program, are allotted a supernumerary seat in
their most preferred program at IIT Roorkee.

\section{DA for Admission into IITs, NITs, and other GFTIs}
\label{sec:coreDA}

In Section~\ref{sec:daWithQuotas} we have seen the standard DA with
quotas which is applicable to all central government funded
institutions.  Further, we have seen various business rules, and how
the core algorithm is modified to take care of various business
rules.  In this section we summarize the process of allocation.

\subsection{Virtual Preference Table for IITs}
\label{sec:IIT}
\begin{figure}[!ht]
\centering
\includegraphics[width=\textwidth]{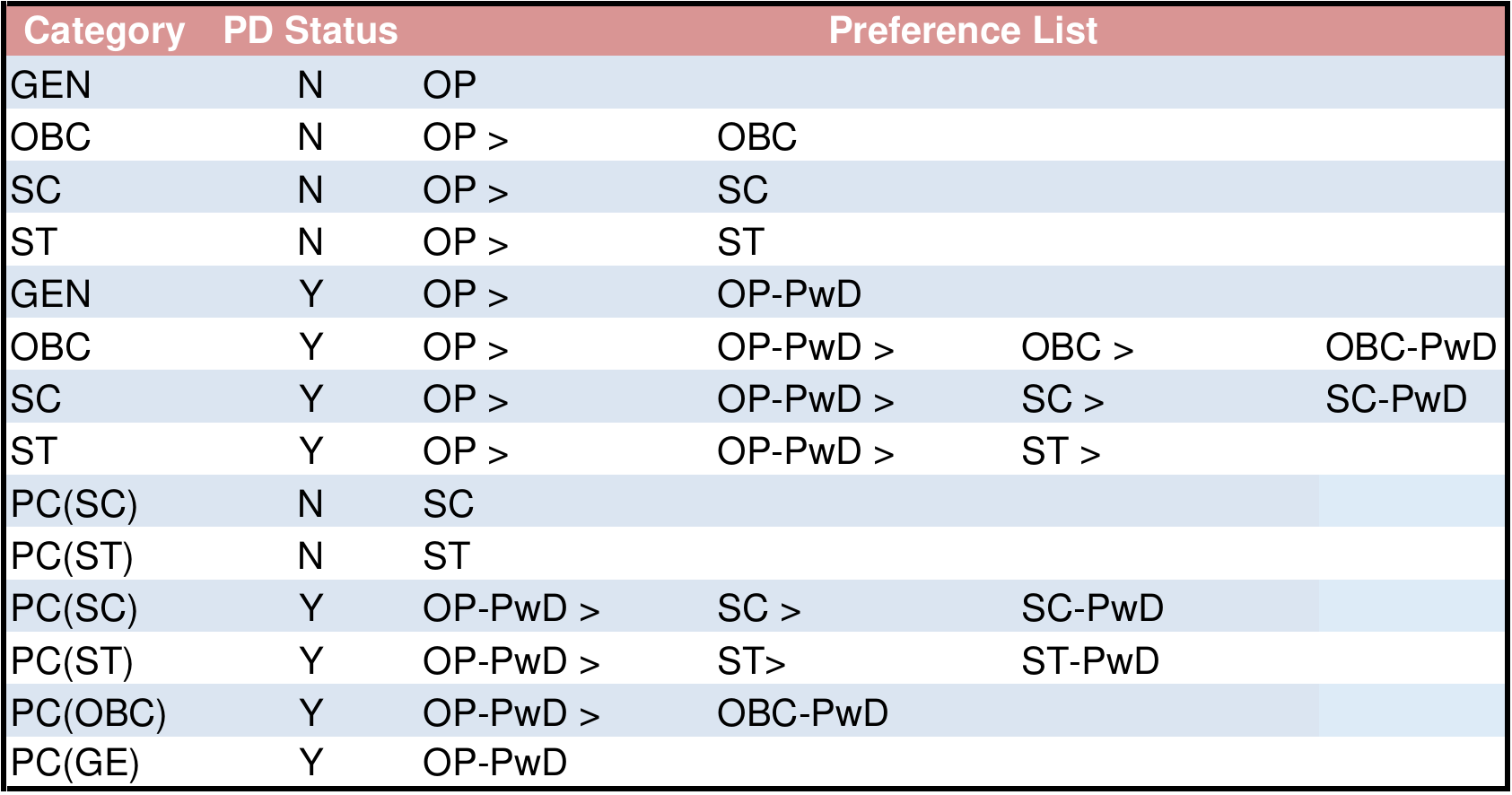}
\caption{Virtual preference list for candidates when PC candidates are
  involved.}
\label{Figure:NoDereserveIIT}
\end{figure}

Business Rule items 1-5 and items 7 \& 8 from the list at the beginning of
Chapter~\ref{chapter:businessrules} apply. To include PC candidates,
Figure~\ref{Figure:NoDereserve} is modified slightly and is presented
in Figure~\ref{Figure:NoDereserveIIT}.

Notice that, as compared to Fig.~\ref{Figure:NoDereserve}, there are
new types of categories in the last six rows, and the extended merit
list comes into play.  We may use the Standard DA algorithm with
quotas (Section~\ref{sec:daWithQuotas}) with the discussions so far in
%Sec.~\ref{sec:DAwithTies}, 
Sec.~\ref{sec:foreign}, and Sec.~\ref{subsec:DS} if only seats in the
IITs are offered. 

In reality, a candidate may be eligible to apply for many programs
across multiple institutions beyond the IITs. The virtual preference
lists need to be considered in an appropriate fashion, depending on
whether a program is in the IITs or not.  The next two sections
describe the process for NITs and other GFTIs.

\subsection{Virtual Preference Table for NITs}
\label{sec:NIT}

\begin{sidewaysfigure}[p!]
\centering
%\hspace*{-2cm}
\includegraphics[width=\textwidth]{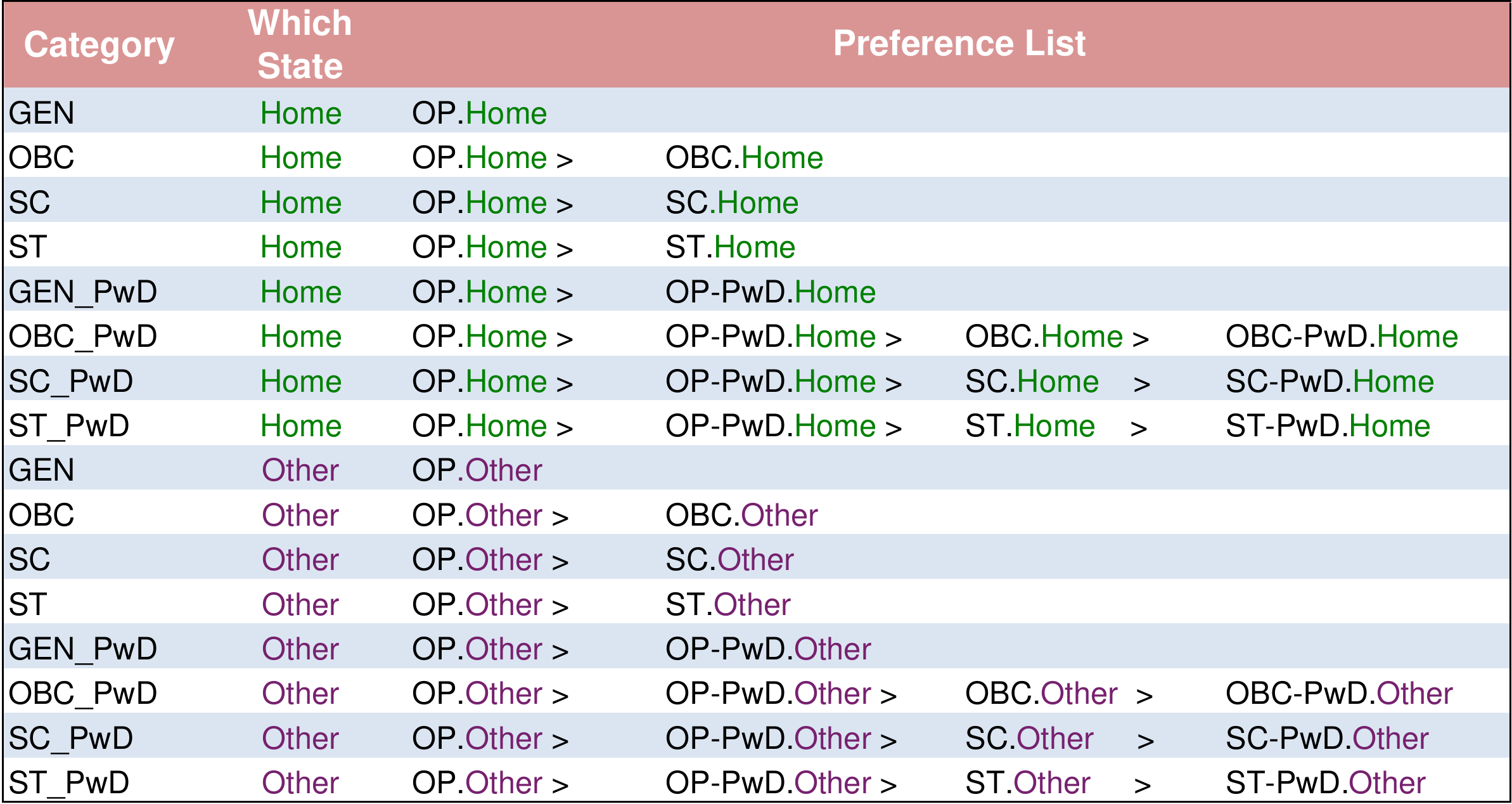}
\caption{Virtual preference list for candidates for seats in the NITs.
  The first 8 rows represent preferences of candidates applying to the
  NIT in their ``Home State'' quota; seats
  in this quota are available only to state residents.
  Candidates applying to a NITs in $\nu$ which is not their home state
  quota (the last
  8 rows) will compete with
  all residents of India, excluding candidates whose home state is
  $\nu$, the state under consideration.}
\label{Figure:NIT-no-dereserve}
\end{sidewaysfigure}

All items in the beginning of Chapter~\ref{chapter:businessrules} apply,
except Items 3,4, and 5. There are, however, additional qualifiers
corresponding to Item 6. We focus on Item 6. Seats for academic
programs offered by NITs are divided into Home State quota and Other
States quota.  These additional quotas are reflected in the virtual
preference table as shown in Figure~\ref{Figure:NIT-no-dereserve}, and
do not change the core of the algorithm.

\begin{example}
As an example, consider the virtual preference list for an OBC-NCL-PwD
person whose ``Home State'' is Maharashtra.  He may express a
preference to apply to the electrical engineering program of  NIT in this
state, viz., the NIT in Nagpur.  In this case, her virtual preference
will be represented by the sixth row in the table.  If her next preference
is  the electrical engineering program of the  NIT in Tamil Nadu,
then the virtual preference will be read from the 14th row of the
table and he will not be eligible for the Home State quota of the NIT
in Tamil Nadu.  Finally, if her next preference is the electrical
engineering program in IIT Roorkee, Item 6 does not apply; we refer back to 
the sixth row in Figure~\ref{Figure:NoDereserveIIT}.
\end{example}

\subsection{Virtual Preference Table for Other GFTIs}
\label{sec:GFTI}

\begin{sidewaysfigure}[p!]
  \begin{center}
\includegraphics[width=1.1\textwidth]{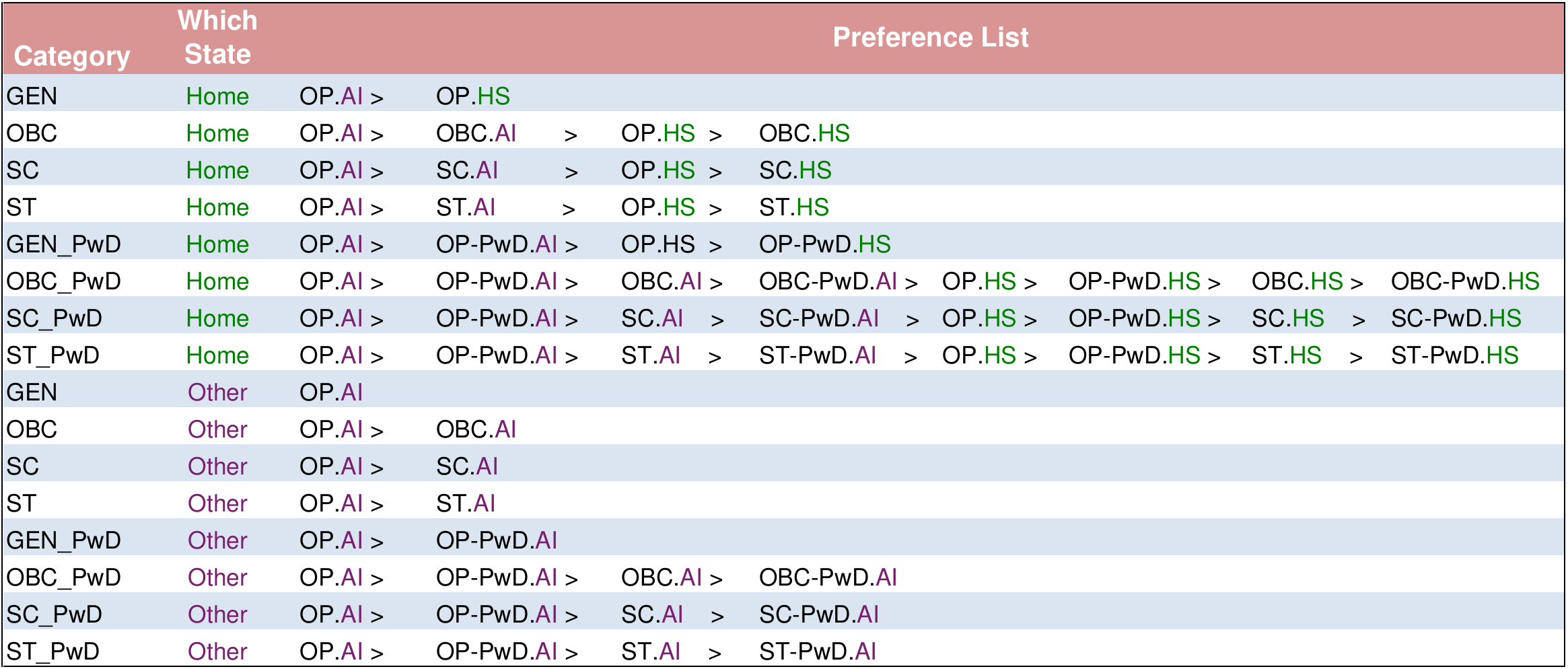}
\caption{Virtual preference list for candidates for seats in certain
  other GFTIs.  The last 8 rows represent preferences of candidates
  applying to the GFTI not in their ``Home State'' (HS); seats in
  this quota are available to all Indian nationals.  Candidates applying
  to a GFTI in their home state (the first 8 rows)
  will first be considered for seats in the All India (AI) quota
  failing which they will be considered for seats in the Home State
  quota.
}
\label{Figure:GFTI-no-dereserve}
\end{center}
\end{sidewaysfigure}

For Government Funded Technical Institutions (GFTIs) other than the
IITs and the NITs, all items in the beginning of
Chapter~\ref{chapter:businessrules} except for Items 3,4, and 5 apply. On
the basis of the existence of quotas based on the residency state of a
candidate, there are the following three types of other GFTIs and their
respective virtual preference tables.
\begin{enumerate}
\item All India (AI) quota:~ If a GFTI has only AI quota, then the
  virtual preference table is the same as the one shown in
  Figure~\ref{Figure:NoDereserve}. All Indian Institutes of Information
  Technology (IIITs) have AI quota only. Other examples of
  GFTIs that have only AI quota are School of Planning \&
  Architecture, Bhopal and Gurukula Kangri Vishwavidyalaya, Haridwar.

\item Home State (HS) and Other State (OS) quota:~If a GFTI has HS and
  OS quota, then the virtual preference table is the same as that for
  NITs (shown in   Figure~\ref{Figure:NIT-no-dereserve}). BITS Mesra is one
  such GFTI that follows the HS/OS quota model.

\item Home State (HS) and All India (AI) quota:~ The virtual
  preference table for such GFTIs is shown in
  Figure~\ref{Figure:GFTI-no-dereserve}.  Currently Assam university,
  Silchar is the only GFTI institute having HS and AI quota.
\end{enumerate}

%\sbas{The earlier version of this para is commented in the tex file. There was an example which was wrong since it referred to NIT Hamirpur
%as GFTI}
%Business Rule Items 1-2 apply and items 3-6 do
%not apply. There are however additional qualifiers corresponding to
%Item 7.
%
%As in the case of the NITs, all we need to do is to modify
%the virtual preference table which, for this case, appears in Figure
%\ref{Figure:GFTI-no-dereserve}.
%\sbas{The following example is wrong since NIT Hamirpur is an NIT and does not follow this rule.
%The example should be modified as follows.
%As an example, consider the virtual preference list for a OBC-PwD
%person whose ``Home State'' is Himachal Pradesh.  She may express a
%preference to apply to the mechanical engineering program of NIT in
%this state, viz., the NIT in Hamirpur.  In this case, her virtual
%preference will be represented by the sixth row in the table.  If her
%next preference is the electrical engineering program of the NIT in
%Rajasthan, then the virtual preference will be read from the 14th row
%of the table.  Note that not only will she be ineligible for the seats
%in the Home State quota of the NIT in Rajasthan, she will also be
%competing for seats with candidates whose home state is Rajasthan.

As an example, consider the virtual preference list for a ST-PwD
person whose ``Home State'' is Assam.  She may express a preference to
apply to the mechanical engineering program of Assam University
Silchar.  In this case, her virtual preference list for this program
will be represented by the ST-PwD row (the eighth row) in the table
shown in Figure~\ref{Figure:GFTI-no-dereserve}. Let us consider an
OBC-NCL-PwD candidate whose Home State is not Assam. If one of her
preference is a program in the Assam University Silchar, then the
virtual preference list for this program will be read from the fourteenth
row of the table shown in Figure~\ref{Figure:GFTI-no-dereserve}.  Note
that not only will she be ineligible for the seats in the Home State
quota of this university, she will also be competing for seats from
All India quota with candidates whose home state is Assam.

\subsection{Summary}

The core algorithm is the standard DA algorithm with quotas.  However,
the virtual preference lists must be correctly constructed based on the
programs a candidate applies to, and the tag she has.  Once virtual
programs, virtual preference lists, and virtual merit lists are constructed, the
algorithm is applied for all candidates in the pool, including the
PC candidates, DS candidates, international students, as well as
the remaining bulk of candidates who do not belong to these special cases.

Note that the business rules involve complex de-reservation.  As such,
in order to complete the allocation, de-reservation as per 
business rules must be employed on top of the process described so
far.  Chapter~\ref{chapter:deReserve} is devoted to handling
de-reservation in a holistic way. Starting from 2018, the rule of
supernumerary seats for females has also been introduced. We discuss
it in Chapter~\ref{chapter:SupernumeraryFemales}.

% chapter 5
\chapter{Multi-round Implementation}
\label{subsec:multi_round}

Allocation of seats is conducted over multiple rounds for a variety of
reasons, including the reality that offered seats may not be accepted,
and overbooking of seats is not permitted.  As introduced in
Section~\ref{sec:multi-round-intro}, the notion of fairness becomes
complicated since new candidates may join the seat requisition
process. Second and later rounds facilitate the utilization of
surrendered seats, including the possibility of awarding a surrendered
open category seat to a person earlier denied such an open seat, and
earlier awarded a seat in a particular restricted quota.  To maintain
truthfulness, fairness, and optimality across rounds requires careful
algorithmic design.

After initial information collection
(Section~\ref{subsec:initial_information}) our algorithm proceeds in
rounds with the following activities taking place in each round.

\begin{enumerate}

\item We execute the algorithm described in Section~\ref{sec:coreDA}
  using Algorithm~\ref{alg:DA} with the candidates initiating
  applications (proposing to the programs in the decreasing order of
  their preferences). Such an execution is termed as a \emph{run} of
  the algorithm. The seat allocation computed by the algorithm is
  announced to the candidates.
  %Before executing it the first time, the optional
  %inputs are not provided. 
  %\footnote{As an implementation note, the \cc\
  %  flag is arbitrarily set to 2 to indicate the default situation
  %  that the credentials of none of the candidates have changed.}.
  %In subsequent rounds, optional inputs are
  %expected to be provided. Once the seat allocation is computed,
  %candidates are informed about the programs offered to them.
  %
  %Specifically, the $\mc(p)$ for each virtual
  %program $p \in \cP$ is chosen appropriately as described below in
  %Section~\ref{subsec:changes_to_other_DA} based on the output of the
  %prior round.  
%  {\color{red} The list $\cc$ of candidates whose category has
%  changed at any time after the first round is also provided as an
%  input.  The outcome of DA algorithm is an allocation of programs
%  to candidates.}

\item A candidate who is allotted a program is required to go
  physically to a reporting center for document verification and seat
  confirmation in the program. There, the candidate may also exercise
  the options `freeze', `float', or `slide' for future rounds. These options
  are described in more details later.
%If the candidate does not report at the reporting center, she will be out of Joint Seat Allocation
%for that year.% and hence will not be considered for seat allocation in subsequent rounds. 
  A candidate who accepts a program may also surrender her seat and
  withdraw from the joint seat allocation later in the same round or
  any subsequent round before the  withdrawal deadline
  \footnote{Withdrawal was allowed till the sixth round in 2018.}.  It
  is also possible that after document verification, the 
  credentials of the candidate change which may potentially lead to
  the cancellation of the seat of the candidate. A candidate may
  reject the offered seat; rejection is implicit if the candidate does
  not report.  Candidates who reject a seat do not participate in the
  process thereafter.
% may get cancelled if she fails to produce suitable document(s). 
%It may also happen that she is found to be ineligible for future rounds of seat allocation. 
%Please read Section \ref{sec:Reporting-center-activities} in Appendix to know about the complete details of the
%activities that take place at reporting center.
%on the summary of the reporting center activities.
%  {\color{red} This is the case also before the closure round (i.e.,
%  after the fourth round).  The details of these options are described
%  in Section~\ref{subsec:multiround_virtualprefs}.}
\item Based on the activities at the reporting center, inputs for
  the next round of seat allocation is updated.
%At the end of the round, the status of each allotted seat is also required as an input for computing the seat allocation for the next round. 
%
%Some candidates leave the system. Some candidates are still part of the system but their credentials might have changed.
%Based on the last round allotment and the reporting center activities, we need to compute the min-cut-off of each virtual program.
%We also have to modify the preference list of the candidates based on the option they took at the reporting center.
%These modified preference lists and min-cut-off of each virtual program are taken into account while computing the
%seat allocation for the next round by the DA algorithm. 
 %{\color{red}\footnote{A
 %  special note is made about the pre-processing the preference lists
 % appropriately in Section~\ref{subsec:preprocessing_ofpref} to
 %implement the different rules for the fourth and closure rounds.}}.
%
\end{enumerate}

\section{MRDA}
\begin{center}
\begin{figure}[h]
\centering
\includegraphics[height=0.5\textheight]{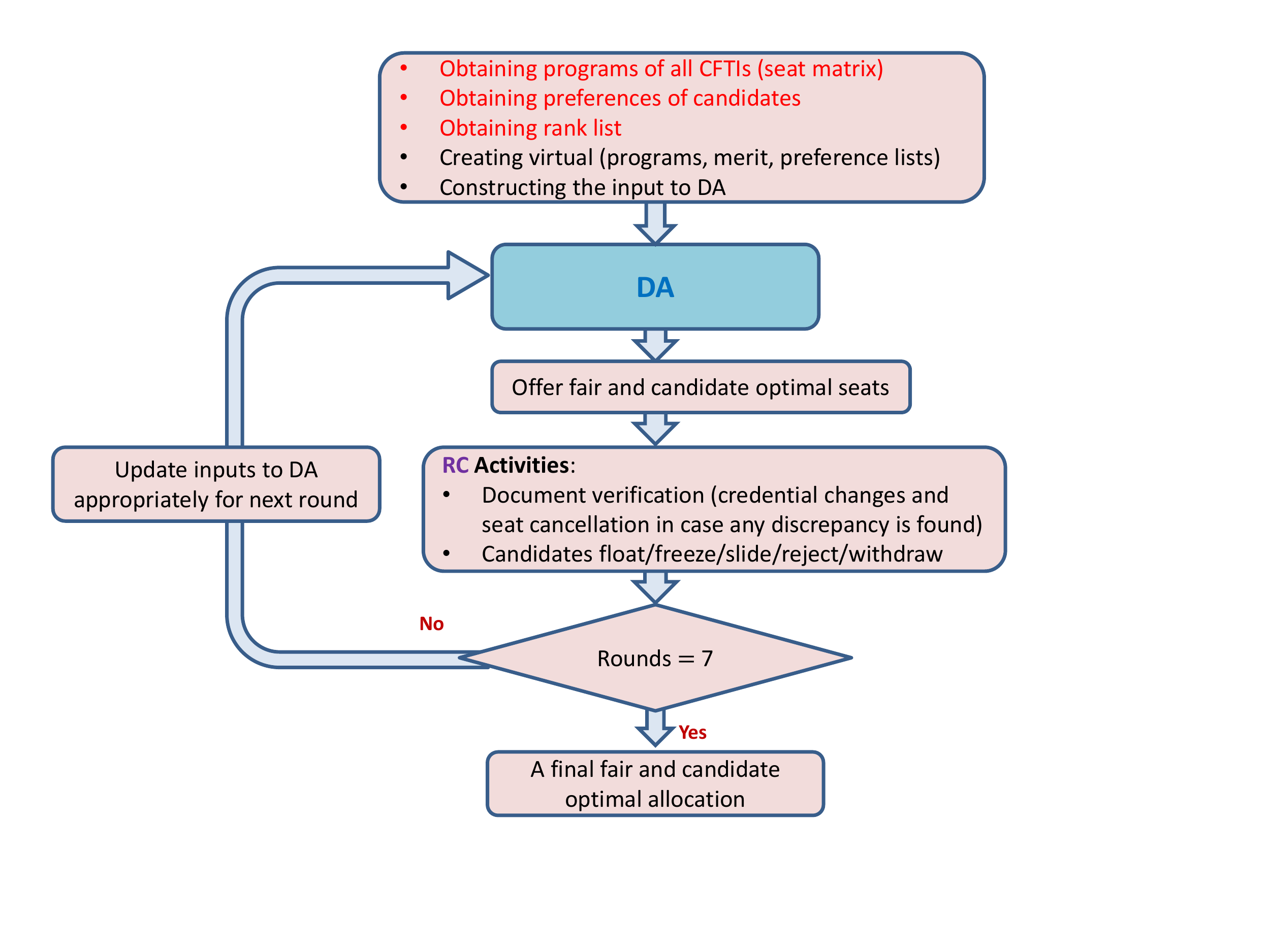}
\caption{Multi-round deferred acceptance  allocation.}
\label{Figure:overview-allocation}
\end{figure}
\end{center}

The process described in Steps 1--3 is summarized in
Figure~\ref{Figure:overview-allocation} and it constitutes the
Multi Round Deferred Acceptance scheme. The process mentioned here
does not take care of de-reservation which is addressed in
Chapter~\ref{chapter:deReserve}. In brief, in de-reservation, multiple
\emph{runs} of the algorithm in Step~1 are performed, and we term
the modified version as the \emph{Multi Run DA
  algorithm}, {\sc MRDA}.

``Multi-Round'' is a term that refers to the overall mechanism that
occurs over days and possibly weeks, with the repeated involvement and
inputs of humans, especially the candidates.  In the context of a
mechanism, the acronym {\sc MR} in {\sc MRDA} may actually be
considered to refer to Multi-Round Deferred Acceptance scheme. These
two concepts -- run and round -- are orthogonal: one can perform
multiple runs of the DA in situations where only one round is planned,
and de-reservation is necessary; that would be a multi-run deferred
acceptance algorithm.  One could also perform a single run per round,
and have multiple rounds, and run the algorithm as in
Figure~\ref{Figure:overview-allocation}; this would be a multi-round
scheme. Or one could do both, as performed in the joint seat
allocation every year since 2015.  In terms of nomenclature, we choose
not to distinguish between these cases as the intent is clear from the
context.

%\yk{Please check the flow chart.}

\section{Input for second and subsequent rounds}
Apart from seat matrices which remain unchanged throughout all the
rounds of seat allocation, the following input files are updated due
to the reporting center activities of each round. These updated tables
are to be used for seat allocation in subsequent rounds.
\begin{enumerate}
\item {\em Candidate Table}\footnote{This table stores the credentials
    of the eligible candidates participating the joint seat
    allocation.}.\\
  The credentials of the candidates may change during the document
  verification at the reporting center during a round. The credentials
  are updated accordingly in Candidate table for the next round.  In
  addition, a candidate who is offered a seat may opt for
  Freeze/Slide/Float/Reject for the seat allocation in subsequent
  rounds.  This information is updated in the Candidate table.
%The choice list of these candidate has to be pruned accordingly before carrying out seat allocation for the next round. 
  Candidates who become ineligible for seat allocation after document
  verification are removed from the Candidate table in subsequent
  rounds.
%This information is also provided in the Candidate table.
%
\item {\em Preference list}:
Based on the changes in the credentials of a candidate, the preference
list of the candidate may have to be updated. In particular, some
choices may become invalid after document verification. (The rules for
these are tabulated in the business rules, but suffice to say that
this step is necessary).
% The
                                % ``Validity'' field of such choices
                                % is marked N.  
%
\item {\em Previous Round Allotment}.\\
  The Allotment Table is one of the output files of the seat
  allocation algorithm after each round. Details about the program
  allotted to a candidate appear here. However, a candidate who is
  allotted a seat for the first time in the previous round might not
  report at the reporting center (thereby implicitly rejecting the
  offer). Another possibility is that a candidate after accepting a
  seat may surrender it and withdraw from the joint seat allocation.
  This additional information is appended to the Allotment table of
  the round. The Allotment table at the end of the previous round, is
  very crucial for computing the seat allocation for the next
  round. Hence, for the second and subsequent round, this table is also
  provided as an additional input file.
\end{enumerate}

The reader is advised to study
Section~\ref{sec:Reporting-center-activities} in the Appendix to know
more details of the activities that take place at reporting center and
how the input files as mentioned above for the second (or subsequent)
round are computed.

\section{Preprocessing of input for second and subsequent rounds}
\label{subsec:updating_DA_input}
Once the input for second (or any subsequent) round is received, it needs
to be preprocessed before we carry out the seat allocation of the
round. In particular, the following tasks need to be carried out (1)
Editing preference list as mentioned above (2) Computing \mc, the minimum
cut-off rank of each virtual program to ensure (as mentioned in
Section~\ref{sec:multi-round-intro}) that no candidate abruptly loses an
earlier awarded seat in a later round. We now describe how this
preprocessing is carried out.

\subsection{Preference list editing}
\label{subsec:multiround_virtualprefs}
Each candidate who is allocated a program in the previous round (call
it $N$) chooses one of the following options before the next round
($N+1$) is conducted. In each case we describe how the candidate's
preference list is
%(after, if required, pre-processing as in Section~\ref{subsec:preprocessing_ofpref}) 
appropriately edited before round $N+1$. This edited preference list
is then used to construct virtual preferences as usual.
%; precise details are described in footnotes.

\begin{itemize}
\item {\em Reject}:~The candidate rejects the program. In this case,
  the choice list of the candidate is set to empty to remove the
  candidate from the process.  (Note that this reference to
  \emph{Reject} is different from the {\sc Reject} in the algorithm
  pseudocode.)
%can be eliminated from the seat allocation process (or   the preference list can be made completely empty).
\item {\em Freeze}:~The candidate confirms the acceptance of the
  program allocated. In this case, we leave the allotted
  program\footnote{In other words, the virtual preference list for the
    candidate now contains only virtual programs corresponding to the
    allotted program and programs ranked below it by the candidate, as
    per the relevant table
    % in Figure   \ref{Figure:preference-lists-for-de-reservation-for-IITs}
    (using the complete row corresponding to the candidate category).}
  on her choice list along with the entries below the allotted
  program,
  % (in   case marks are revised downwards), 
  % Sharat: marks change has not been discussed so far
  and eliminate all other entries.
\item {\em Float}:~The candidate expresses that the program is
  acceptable but would like to be considered for future rounds to get
  a more desirable program, if possible. In this case, we leave the
  candidate choice list unchanged.
\item {\em Slide}:~The candidate wishes to be considered for future
  rounds but would prefer programs in the same institute where he is
  offered the current program. In this case, we remove all programs
  from his list that are above the current allotment and belong to
  the institute different from the institute of the program offered to
  him\footnote{Virtual preferences over virtual programs remain
    unchanged for programs that are not removed.}. The allotted program
  and the entries below the allotted program are left unchanged.
  %(in   case marks are revised downwards).
\item {\em Withdraw}:~ The candidate withdraws from the joint seat
  allocation. In this case, the choice list of the candidate is set to
  empty. Although `Reject' and `Withdraw' ultimately have the same
  effect (the candidate will no longer be part of the process), the
  difference lies in the timing. Once a candidate is offered a
  program, and she accepts, she can
  later withdraw till the withdrawal deadline.
\end{itemize}

%\section{Other Inputs Needed After First Round}
%\label{subsec:changes_to_other_DA}
%
%After the allocation of seats, a number of changes in the assumptions
%of the \emph{input} materialize. The variety of these changes are
%large, and unpredictable; some of them are due to the inadequacies of the
%candidates, and some are due to the inadequacies of the institutional
%framework.  We believe, that at least in the near term, some of these
%are hard to avoid.  The MRDA algorithm makes provision for the
%following kinds of possibilities. 
%
%
%
%% \sbas{This section has been extended with more details on CatChange
%% field and Min-cut-off computation and Min-cut-off benefits. If these
%% details obstruct the flow, please consider shifting it to Appendix
%% suitably.}  \yk{Seems ok to me.}
%
% \begin{itemize}

\subsection{Computing \mc\ }%and setting \cc\ }
\label{sec:mincutoff}
The seat guarantee across multiple rounds is needed because, as
discussed above, the rules provide for a freeze, float, and slide
option. The seat guarantee is implemented by the notion of \mc: A
candidate is allowed to retain a seat in round $J+1$ if her rank is
better off, or equal to the rank corresponding to the minimum cut-off
rank in a virtual program at the end of $J$th round.  This rank might
be based on her rank, or, more subtly, \emph{the rank of any other
  candidate previously allotted a seat in the virtual program under
  consideration.}  Further, to avoid merit violation, the candidate
may get something even better (from her perspective) in yet another
program based on the minimum cut-off of that program, and this is
permitted even on a supernumerary (SN) basis.

%For computing \mc\ for a virtual program $p$ for round $i+1$, we proceed as follows. If there was any vacancy in virtual program $p$ in 
%the allotment of round $i$, we assign $\mc(p)$ as $\infty$. Otherwise, we compute set $S(p)$
%of candidates allocated to $p$ in the seat allocation of round $i$ and proceed as follows. From $S(p)$, we remove the following
%candidates
%\begin{enumerate}
%\item Those who rejected or withdrew
%\item Those whose seat got canceled
%\end{enumerate} 
%$\mc(p)$ is defined as the rank of the worst rank candidate in the resulting $S(p)$. If $S(p)$ is empty, we define $\mc(p)$ as 0.
%
\noindent {\bf Computing \mc: } The word minimum in
min-cut-off is used to indicate that in future rounds, the cutoff (or
closing rank) can actually become worse (i.e., larger) than the
min-cut-off.  Consider any virtual program $p$. We use the output $\wl(p)$ of $i$th round to compute $\mc(p)$ for
$(i+1)$th round as follows.  
%First note that it is entirely possible
%that $p$ may have unfilled seats. These unfilled seats, may or may not be de-reserved. 
%In the sequel, we assume that merit lists have been updated based on marks revision. 
We temporarily construct a \emph{reduced} waitlist for 
   $p$ by removing the following kinds of candidates:
     \begin{enumerate}
      \item Those who opted for reject or withdraw.
%      \item Those who were awarded a seat through the Defense Services
%        Priority Allocation (DS) (these candidates should be
%        considered for min-cut-off of the DS virtual program and not
%        for other virtual programs)
%      \item Those with CatChange=1
%      \item Those with CatChange=4
%      \item If $p$ is an engineering program then those with a
%        decrease in the marks in the engineering exam. Conversely, if
%        $p$ is an architecture program then those with decrease in
%        subjects related to the architecture program.
     \item Those whose seat got cancelled at the reporting center in $i$th round. 
     \end{enumerate}

     \noindent
     We then set $\mc(p)$ for the next round as follows:
     \begin {itemize}

\item   If the reduced waitlist is non-empty, $\mc(p)$ is set to
   $\rank(y,p)$, where $y$ is the last candidate in the reduced wait
   list. Note again that, $\rank$ here stores the extended merit list rank
     after revision\footnote{Recall Definition \ref{def:primary_merit_list}.
     {\bf Example}: If $p$ is an OBC-NCL virtual program, then we use
     the last rank in the standard OBC-NCL merit list. On the other
     hand, if $p$ is an SC virtual program then we use the last rank in
     the extended merit list, which, to recall is constructed by
     taking the SC rank list
     followed by the PC-SC rank list.}.
     
\item If the reduced waitlist is empty, $\mc(p)$ is set to 0. Thus, nobody applying to $p$ in the next round will get any min-cut-off benefit.
\end{itemize}

%   he reader is advised to refer to \ref{sec:mincutoff}
%   in Appendix for the method of computing \mc before any round. Note that \mc($p$) is 0 for the 1st round for each virtual program $p$.
%   
%   This was described briefly as the \mc\ benefit in
%   Section~\ref{sec:credchange}. 

\section{Summary}

We have described a multi-round deferred acceptance algorithm.  After
each round, candidates may reject, freeze, slide, or float with the
program offered.  As a result, the program allocated to a candidate in
subsequent rounds may differ.  In this case, it is guaranteed that the
new program allocated to the candidate will be preferred (by the
candidate) to the earlier one.
%,
%unless 
%{\color{red} experiences a downward revision of marks between
%those two rounds}
% {\color{blue} 
% the candidate becomes ineligible at the reporting center.
 %}.
%
Many other business rules have been incorporated using the concept of
virtual programs, virtual preference lists, and virtual merit lists.
To make all desirable effects happen between rounds, internally in the
data structures, the choices of the candidates need to be modified and
%their ranks may need to be changed, 
possibly their credentials may need to be changed.  
%{\color{red} Seats may become unavailable in the fourth
%round to the IITs.}  
%However, to maintain the performance guarantees,
%the pool of competing candidates does not change, and there is a
%predictability and trust in the system for these competing candidates.

\noindent {\bf Remark:}~ Section
\ref{sec:Implementation-details-of-the-algorithm} in the Appendix
presents the complete implementation details of the generic MRDA
algorithm along with all input-output formats.
%{\color{red} used for the year 2015}.  
The reader interested in the actual implementation of the DA algorithm
should refer to this section.

% chapter 6
\chapter{De-reservations}
\label{chapter:deReserve}

Recall that the DA algorithm as described in Section~\ref{sec:coreDA}
takes care of all business rule of categories as described in
Chapter~\ref{chapter:businessrules} except de-reservation.  When seats are
unfilled due to lack of candidates in a particular category, seats may
be de-reserved.

%\subsubsection*{De-reservation}
After a possible allocation to preparatory candidates, if seats in
one or more of OPEN-PwD, OBC-NCL-PwD, SC-PwD, ST-PwD seat categories are still vacant in any
program, they are de-reserved.  For example, unfilled SC-PwD
seats (that are also not filled with PC candidates) will be
de-reserved and treated as SC category seats for allocation in every
round of seat allotment. Similarly, unfilled OBC-NCL seats will be
de-reserved and treated as OPEN seats (there is no allocation to PC
candidates in this case).  However, unfilled SC and ST category seats
will NOT be de-reserved.

\begin{figure}[h]
\centering
\includegraphics[height=3.5in]{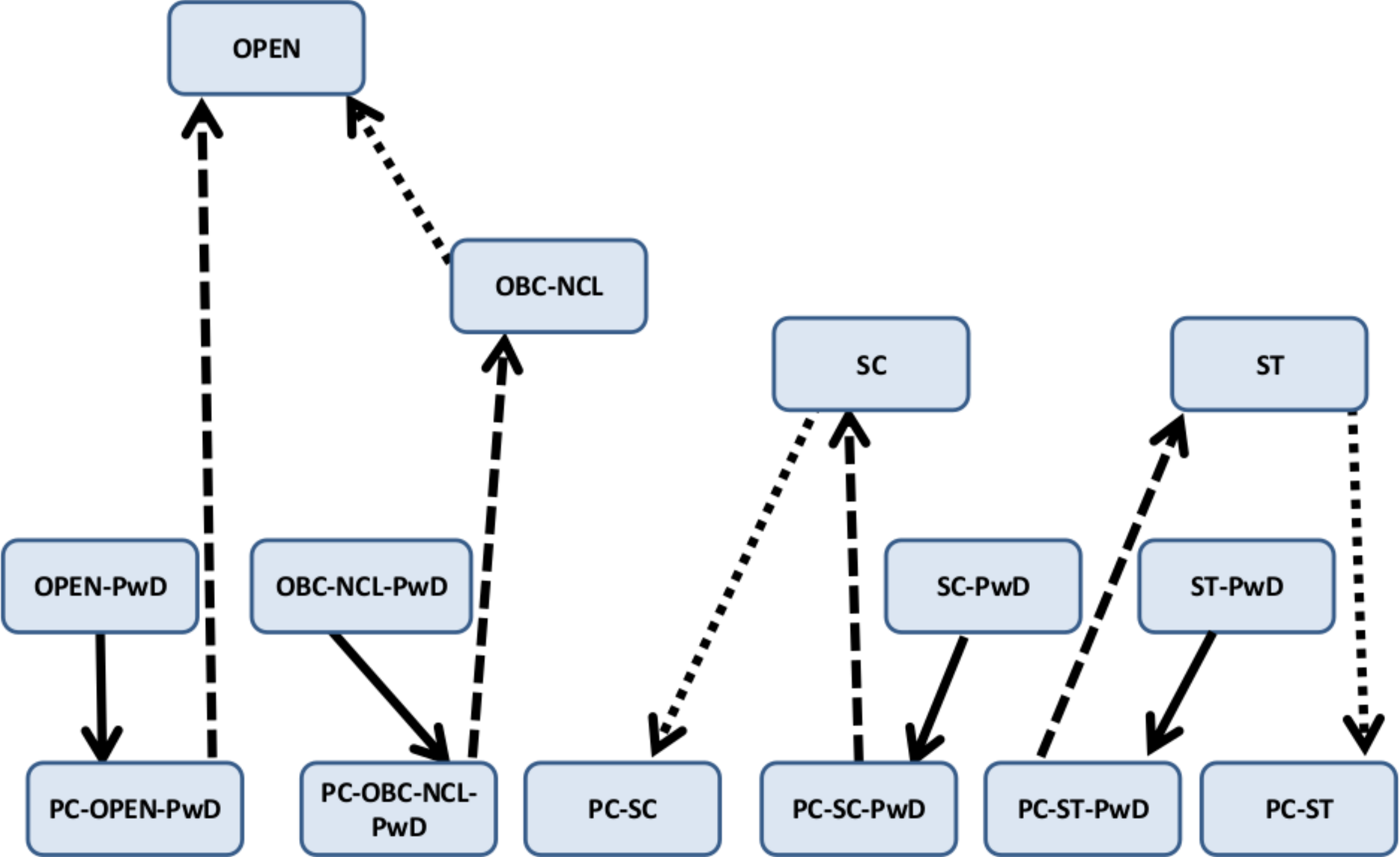}
\caption{De-reservation. Starting from the bold arrow and following
  dotted arrows, one arrives at a possible sequence in which seats are
  de-reserved.  For example, unfilled OPEN-PwD seats are considered to
  all eligible preparatory candidates (PC-OPEN-PwD); if at this stage,
  seats are still unfilled, they are offered to all eligible
  candidates (OPEN).}

\label{Figure:de-reservation-graph}
\end{figure}

The de-reservation rules stated above are succinctly depicted in
Figure~\ref{Figure:de-reservation-graph}.

There are two methods of incorporating  de-reservation rules.
% as described in
%Section~\ref{sec:singlePass} and
The first and a very simple method is to change the input seat
capacities available in a \emph{seat matrix}, and move unfilled seats
to the appropriate category, and run Algorithm~\ref{alg:DA} using the
revised seat matrix. This method involves multiple runs and has been
used for the joint seat allocation since the year 2015. This method is
described in the following section. The second method involves
modifying the virtual preference lists and the virtual merit lists.
This method, along with its advantages and disadvantages over the
first method, is described in the earlier technical
report~\cite{TechReport:2015}. %{\color{red} Section~\ref{sec:multiPass}}.

\section{Multi-Run DA with De-reservation}
\label{sec:multiPass}

Multi-run DA may be viewed as running the method in
Section~\ref{sec:coreDA} several times, each time updating the seat
matrix provided as input by appropriately de-reserving seats that were
not filled in the previous run.

%\sbas{I wish to comment the following two sentences since they are relevant if one knows the single-run DA that takes care of de-reservation using "long and wide" preference tables.}
%It is important to note the virtual
%preference lists are now very compact, as in
%Figure~\ref{Figure:NoDereserve} for the programs in IITs, and
%Figure~\ref{Figure:NIT-no-dereserve} for the programs  NITs.  The virtual merit
%lists are also simplified.

\subsection{Multi-Run DA Algorithm}
\label{sec:multiPassNIT}

%\yk{What is the meaning of ``for NITs'' and ``for IITs''?}

The Multi-Run DA algorithm works as follows.
\begin{enumerate}
\item We run the algorithm in Section~\ref{sec:coreDA}.\footnote{With the
  appropriate table for the virtual preference list depending on the
  program a candidate applies to, and the corresponding
  simplified virtual merit list.}

% We make use of
%   the primary merit lists for each virtual program, cf. Definition
%   \ref{def:primary_merit_list}.\footnote{Thus, we take care of PC seat
%     allocations in case of virtual programs that have PC versions
%     available.}

\item If there are unfilled seats that can be de-reserved, then we
  update the capacities by de-reserving unfilled seats to the parent
  category\footnote{We de-reserve any unfilled OPEN-PwD seats to OPEN
    seats, unfilled SC-PwD seats to SC seats, unfilled ST-PwD to ST, and
    unfilled OBC-NCL-PwD to OBC-NCL, and unfilled OBC-NCL seats to OP, and update
    the seat matrix appropriately. As an example, if there are two
    unfilled seats in an OPEN-PwD virtual program, then we de-reserve
    them by decreasing the capacity of that virtual program by 2, and
    increasing the capacity of the corresponding OP virtual program by
    2. It may be noted that the OPEN virtual program may simultaneously
    get some additional seats due to de-reservation from OBC-NCL during
    this step.}. We then re-run the algorithm as in Step~1 on all
  candidates.

    If there are no unfilled seats that can be de-reserved, i.e., all
    unfilled seats are in OPEN or SC or ST virtual programs, then we
    terminate.
%
%  \item If there are unfilled seats, and unsatisfied candidates, we
%    de-reserve OBC-PD, SC-PD, ST-PD to their respective parent
%    category. We then run DA as in Step 1.  Since no new OPen seat has
%    been created, it is not possible to have new OBC-PD, SC-PD or
%    ST-PD seats.
%
%  \item If there are unfilled vacant seats, and unsatisfied candidates
%    they will be in OBC, or SC, or ST.
%    Dereserve OBC to OPen.  Re-run DA as in Step 1.
%
%  \item It is possible that we still end up with unfilled SC-PD seat.
%    For example, a SC candidate moves to OPen, a SC-PD candidate moves
%    to SC.  Therefore we iterate on steps 2-4 till we reach a fixed
%    point.  At this point, there might be vacant SC and vacant ST
%    seats.
\end{enumerate}
This algorithm enjoys monotonicity across runs: The options available
to candidates are only enhanced in going from one run to the next.
More seats become available in parent programs, whereas the programs
that ``lose'' seats due to de-reservation do not hurt the allocation
since there is no demand for those seats anyway (in that run or in any
future run). Thus, candidates get the same or better allocation than
before.  Seats flow only upstream towards parent categories,
cf. Figure~\ref{Figure:de-reservation-graph}. On tests
consisting of various synthetic as well
as the actual data during the joint seat allocation,
the number of runs were always less than 5  in the last four years.

{\bf Additional Output:} The primary purpose of the seat allocation
algorithm is to allocate seats.
%There is no mention of any other requirement in the business rules
%\cite{BJan3}.
However, in addition to the allocation of seats, we can output the
opening and closing ranks in the Multi-Run Multi-Round algorithm after
the last round.  We read off the opening rank for each category if
there is any candidate in the virtual program. The closing rank is the
rank of the last admitted candidate in the corresponding virtual
program. However, if the corresponding candidate is from Preparatory
Course, we use the following convention. We use the actual rank of the
candidate in the corresponding Preparatory Course rank list and add
the suffix P to it (to indicate ``PC'').  (Note that the extended merit
list is for internal purpose only. Therefore, we must announce the
actual Preparatory Course rank instead of extended merit list rank).

%The closing rank is infinity if one or more of the following occur:

%\begin{itemize}
%\item Any de-reservation occurred from the corresponding virtual
%  program (at any stage).
%
%\item There are non-zero unfilled seats in the corresponding virtual
%  program when the multi-run algorithm terminates. (This can occur in
%  OPEN, SC and ST virtual programs. For other types of virtual programs,
%  such seats are de-reserved before the algorithm terminates.)
%
%\item If some of the seats are occupied by PC candidates.\footnote{In this
%  case, there is a closing rank for the corresponding PC course, which
%  is the rank of the last PC candidate to be admitted if there was no
%  de-reservation  and null/infinity if there was de-reservation.}
%\end{itemize}

%If none of the above occur, the closing rank is the rank of the last
%admitted candidate in the corresponding virtual program.

In addition, we also output the initial seat matrix, and the final
seat matrix for each program, with unfilled seats marked. The final
seat matrix is the matrix used in the final run of DA during which no
further de-reservation occurred. The following example shows how to
interpret the initial and final seat matrix together.
 \begin{example}
 Consider a program with:
 \begin{itemize}
 \item Initial seat matrix: 20 OPEN seats, 10 OBC-NCL seats, 2 OBC-NCL-PwD seats,
   and 1 OPEN-PwD seat. (Suppose there are no other seats.)
 \item Final seat matrix: 21 OPEN seats (18 filled), 11 OBC-NCL seats, 0
   OBC-NCL-PwD seats and 1 OPEN-PwD seat (filled by PC candidate).
     \end{itemize}
 \end{example}

 The immediate inference is that all 11 OBC-NCL seats in the final seat matrix
 are occupied, since if not, they would have been de-reserved. We also notice that
 the 1 OPEN-PwD seat is occupied by a PC candidate.

 We further interpret the above starting at the lowest level of Figure
 \ref{Figure:de-reservation-graph}. We infer that 2 OBC-NCL-PwD seats were
 both de-reserved to OBC-NCL, leading to 12 OBC-NCL seats. Of these OBC-NCL seats,
 $1$ was unfilled (hence the final seat matrix contained 11 OBC-NCL
 seats), and further de-reserved to OPEN, leading to 21 OPEN seats. Finally,
 we infer that 3 of these OPEN seats were unfilled.

% chapter 7
\chapter{Supernumerary seats for females}
\label{chapter:SupernumeraryFemales}
There are many institutes, such as the IITs, where the percentage of
female students in the undergraduate programs used to be (until 2018)
approximately 9\% of the entire student population.  Stakeholders in
the system sought an increase in the gender ratio.  The improvement in
the ratio was to be carried out in a phased manner --- at least 14\%
females in 2018, and thereafter, 17\% in 2019 and 20\% in 2020.

%It has also to be ensured that non-female candidates must not be adversely affected in achieving this target.
Until 2017, all seats in a virtual program were allocated in a
gender-oblivious manner.  One way to achieve the new objectives would
be to declare a 14\% reservation, similar to the reservations based on
birth category as described earlier.  This was not considered
acceptable since many practical constraints that had to be
satisfied. (See also Section~\ref{subsec:not-reservation}.) First,
some programs already had more than 20\% females already (even if the
Institute did not). It was not considered acceptable to burden the
institute with variable amount of new infrastructure requirements for
the females: seat matrices (i.e., capacities) had to be announced
before the process began. A related constraint was that the number of
non-female\footnote{In this chapter, a male or transgender candidate
  is considered as a non-female candidate.  } candidates could not
also be increased substantially, \footnote{The institutes were already
  finding it difficult to create the infrastructure for additional
  females, so it was difficult for them to create the infrastructure
  for any additional non-females.}  or decreased (so as to remove the
perceived ``injustice'') for that matter. One way to satisfy this
requirement is that the number of non-females admitted should be close
to their numbers in 2017 (to be more precise, average of the last few
years) by dividing the pool into females and non-females. We direct
the reader to Section~\ref{sec:wrong-approaches-females} for more
insights to some of the difficulties in reducing variability in the
number of females admitted while retaining a modicum of fairness.
% Secondly, While meeting these practical
% constraints and achieving the target minimum fraction of females in a
% program, we also have to ensure that the non-females should not be
% disadvantaged.

We describe how the MRDA algorithm, with suitable changes in the
virtual programs and the virtual preference lists, can ensure at
least 14\% seats for females in 2018 and fairness for all candidates
while satisfying the constraints stated above.
%Interestingly, just by suitably , this algorithm can be 
%We use the term {\em male} to denote a {\em non-female} candidate.

\section{The Method}
In order to implement the rule of supernumerary seats for females and
respecting all the constraints, the proposed algorithm will require
that each virtual program $p$ has two separate seat pools defined as
follows:
\begin{itemize}
\item Female-only($p$): This pool will be exclusively for females and
  they will be admitted on the basis of relative merit.
\item Gender-neutral($p$): As the name suggests, this pool will admit
  candidates through merit, but without any gender
  bias. 
\end{itemize}

The method of computing the number of seats in Female-only($p$) pool and
Gender-neutral($p$) pool for each virtual program $p$ is  described in
Section \ref{sec:seat-matrix-computation}. Now we describe the order
in which candidates are considered for a seat in these two pools of a
virtual program. It is this order that plays the key role in
establishing the fairness properties of the algorithm. 

\subsection{Order of seat allocation}
Consider a virtual program $p$ in the choice list of a candidate
$c$. $c$ will be considered for a seat in the virtual program $p$ based on
the gender as follows. 
\begin{itemize}
\item If $c$ is a female candidate: $c$ will first compete for a seat
  from the Female-only($p$) pool. Only after she fails to get a seat
  from this pool, will she compete for a seat from Gender-neutral($p$) 
  pool.  
\item If $c$ is a non-female candidate: $c$ will compete for a seat
  from the Gender-neutral($p$) pool only.
\end{itemize}

\section{Fairness properties}

The algorithm guarantees the following fairness properties.

\subsection*{Fairness for non-female candidates}
\begin{itemize}
\item {\em No reduction in the number of available seats.}\\ As shown
  later in Section \ref{sec:seat-matrix-computation}, the number of
  seats occupied by non-females in the gender neutral pool of an
  academic program $p$ (e.g. IIT Kanpur CSE) in 2018 will be equal to
  or greater than the number of seats that the non-female candidates
  got in that program in 2017. So potentially all the seats of
  the gender-neutral pool of $p$ may be allocated to non-females only.

\item
  {\em Guarantee of no-adverse effect.}\\
  The algorithm guarantees that at least one of the following
  properties will hold for each virtual program $p$ (e.g., IIT Kanpur
  CSE SC category):
\begin{enumerate}
\item The allocation of all seats (the union of the gender-neutral pool and
  the female-only pool) of $p$ is purely based on merit without any gender
  bias.
\item All gender-neutral seats of $p$ are occupied by non-females only.
\end{enumerate}
In order to better understand the second property stated above, we
provide an example. 
\begin{example}
Suppose the number of seats in the gender-neutral
pool and the female-only pool of a virtual program, say IITK CSE SC, are
declared as 10 and 2 respectively. 

The allocation produced by the algorithm will ensure the following. If
less than 10 seats from the gender-neutral pool are occupied by non-female
candidates, then all the 12 seats from IITK CSE SC are allocated on
pure merit basis among all the SC candidates who competed for a seat
in IITK CSE SC \emph{without any gender bias}.
\end{example}
\end{itemize}
  
\subsection*{Fairness for female candidates}
\begin{itemize}
\item The algorithm will guarantee that each academic program will
  have at least 14\% female candidates.
\item Under no circumstance will a female candidate  be denied a seat
  in a virtual program while allocating the same seat to a non-female
  candidate with (i) equal or worse rank than her, and, (ii)
  satisfying the same eligibility criteria.
%Under no circumstances a female candidate $c$ will be denied a seat in a virtual program but given to a non-female candidate (of the same category as that of $c$) with worse rank than $c$. 
(This property ensures that female candidates can still occupy more
than 14\% seats on the basis of merit). 
\end{itemize}
\subsection{Proof of the fairness properties}
We shall now establish that the fairness properties mentioned above
are indeed guaranteed by the algorithm.  The first fairness property
for non-females follows from the construction of seat matrices as
shown in Section \ref{sec:seat-matrix-computation}. We shall now
establish the second fairness property for non-female candidates.  Let
$p$ be any virtual program (for example, IIT Kanpur Civil Engg. SC or
NIT Surat Electrical Engg. Home State OBC\_NCL). Recall that there are
two seat pools of $p$ based on gender: Gender-neutral($p$) and
Female-only($p$). We begin with the following two facts which follow
immediately from the order in which female candidates are considered
for seat allocation. \\

\noindent
Fact 1: Among the females getting admitted to $p$, the top rankers
occupy Female-only($p$) pool only. That is, those in the female-only
pool will have higher ranks compared to those females in the
gender-neutral pool. \\

\noindent
Fact 2: Each female candidate applies to Gender-neutral($p$) only
after failing to get a seat from Female-only($p$).\\ 

If there is no female occupying a seat in Gender-neutral($p$), then
obviously all candidates occupying seats in Gender-neutral($p$) are
non-females only. So, in order to establish the second fairness property
for non-female candidates, we need to analyse the case in which at
least one seat of Gender-neutral($p$) pool is occupied by a female
candidate, say $c$. Recall that the seat allocation for
Gender-neutral($p$) pool is carried out purely on merit basis without
any gender bias/preference. This along with Fact 2 imply that rank of
each candidate (including $c$) occupying a seat of Gender-neutral($p$)
must be strictly better than the rank of each candidate who applied
for a seat in $p$ but got rejected. Fact 1 and Fact 2 also imply that
every female candidate occupying a seat from Female-only($p$) will
have strictly better rank than that of $c$. Therefore, rank of each
candidate who is allocated a seat from Gender-neutral($p$) pool or
Female-only($p$) pool is strictly better than the rank of each
candidate who applied but failed to get a seat in $p$. Hence, if we
had to carry out allocation of Gender-neutral($p$) $\cup$
Female-only($p$) seats on purely merit basis without any gender
bias/preference among all the candidates who applied to $p$, the set
of candidates getting allocated to $P$ will be identical to the
current set of candidates occupying seat from Gender-neutral($p$) and
Female-only($p$) pools. This implies the second fairness property for
non-females.

To establish the fairness property for females, we proceed as
follows. It follows from Fact 2 that every female candidate, after
failing to get a seat from Female-only($p$) pool, competes for a seat
from Gender-neutral($p$) pool. Further, each non-female candidate can
apply for a seat of program $p$ from Gender-neutral($p$) pool
only. Since seats are allocated from Gender-neutral($p$) pool purely
on merit basis and without any gender bias/preference, so it can never
happen that a female is denied a seat from $p$ but given to a
non-female candidate with inferior rank. 

The 14\% constraint (for 2018) fairness property is satisfied by the
seat capacities mentioned below. Hence we can conclude that all
fairness properties claimed by our algorithm are indeed satisfied by
it.

\noindent
{\bf Note:} In order to incorporate the rule of supernumerary seats
for females, there are a couple of simple and intuitively appealing
approaches, but each of them have one or more flaws. We illustrate these
approaches in Section~\ref{sec:wrong-approaches-females} of
Appendix. We encourage the reader to study them in order to appreciate
the non-triviality of the current algorithm.

\section{Constructing the seat matrices for 2018} 
\label{sec:seat-matrix-computation}

Consider an academic program $P$. Let us first define a few notations.
\begin{itemize}
\item $C$: Seat capacity of $P$ in 2017.
\item $f$: Number of female candidates who were allocated program $P$ in 2017.
\end{itemize}

We first describe the procedure for computing the number of seats for Female-only($P$) and Gender-neutral($P$).
\begin{itemize}
\item
If $f$ is less than 14\% of $C$, we create $x$ additional seats for females according to the formula: 
\[ f + x = 0.14(C + x) \]
This implies $x = (0.14C - f)/0.86$. Thus Female-only($P$) will have $f + x$ seats. In case $x$ is a fractional number, we need to consider its ceiling while rounding it. Gender-neutral($P$) will have $C - f$ seats. 
\item
If $f$ is more than 14\% of $C$ but less than or equal to 20\% of $C$, then Female-only($P$) will have $f$ seats and Gender-neutral($P$) will have $C - f$ seats. 
\item 
If $f$ is more than $20\%$ of $C$: 
In this case, Female-only($P$)  will have $0.2C$ seats and Gender-neutral($P$) will have $0.8C$ seats.
\end{itemize}
Once the number of seats for Female-only($P$) and Gender-neutral($P$) have been computed for an academic program $P$ as described above, we need to distribute them among its virtual programs according the prevailing business rules as follows.
For IITs which have only All India quota, there are eight virtual programs according to the respective categories - GEN, OBC, SC, ST, and their PwD counterparts. For non-IIT institutes, there are two types of state quotas: Home State and Other State (or All India). So there are 16 virtual programs defined by the quota and categories for these institutes.

\noindent
{\bf Note:} %The following two points provide additional details of the process of generating seat matrices for 2018.
\begin{enumerate}
\item
In case some JoSAA\footnote{Joint
Seat Allocation Authority (JoSAA) is the authorized body for
admissions ever since our algorithm was adopted.} institute wishes to achieve female percentage in the range [14\%,20\%], replace 14\% in the calculations above by the desired percentage. 
\item
In case of an academic program of capacity $C$ introduced for the first time in 2018, the number of seats in Female-only($P$) pool will be $14\%$ of $C$ and the remaining seats will be in the Gender-neutral($P$) pool. 
\end{enumerate}

\subsection{Supernumerary seats versus reservation for females}  
\label{subsec:not-reservation}

The creation and allocation of supernumerary seats for female
candidates is fundamentally different from reservation of seats for
female candidates and should NOT be mistaken for any kind of
reservation for females. 
\begin{example}
Suppose a program has 86 Gender-neutral seats and 14 Female-only
seats. Further, let us assume that the top 14 rankers opting for the
program are females and the next 86 rankers opting for the program are
non-females. 

In the current allocation scheme, the top ranking 14
females will first be allocated the 14 Female-only seats. The
non-females will be allocated the 86 Gender-neutral seats.  

In contrast, if the Female-only seats were following the allocation
similar or equivalent to ``reservations'', then the top ranking 14
females would first be allocated the top 14 Gender-neutral seats and
therefore the remaining 86-14 = 72 Gender-neutral seats would be
allocated to non-females. Further, 14 Female-only seats would be
allocated to lower ranked females, thereby depriving the
allocation of the program to the remaining (86-72 = 14) higher ranked
non-females.

In this example, with the ``reservations'' scenario, not only have we
increased the infrastructure requirements for females, some
non-females may perceive injustice when a superior scheme (such as the
one described in this chapter) is possible
\end{example}

% \section{Recommendation for building the seat matrices for 2018}
%The number of females admitted in IITs in the last 3 years have been respectively 904, 848, and 1006. So the maximum number of female candidates (or minimum number of male candidates) in the last 3 years were admitted in the year 2017. So male candidates may be averse to using 2017 data for calculating the seat capacity of the Gender-neutral pool of a virtual program. Ideally each IIT should consider the data of the last 3 years to compute the Gender-neutral seat matrix. This might lead to only a slight increase in the number of seats but is worth doing - This will go a long way in assuring the male candidates that they have the same or better chance in 2018 than the last 3 years.

\section{Implementation details}
The algorithm is implemented by an additional level of refinement for
each virtual program. Recall that till now an academic program was
split into various virtual programs on the basis of the category
(total eight) and state quota (All India, Home State, Other State). A
simple way to implement the rule of supernumerary seats for females is
to split each virtual program on the basis of gender. For example, an
earlier virtual program of IITs, say IIT Kanpur, CSE, OBC\_NCL will be
split into two virtual programs:
\begin{itemize}
\item
 IIT Kanpur, CSE, OBC\_NCL, Gender-neutral
\item
 IIT Kanpur, CSE, OBC\_NCL, Female-only
\end{itemize}
Likewise, an earlier virtual program of CSAB institutes, say NIT Surat Civil Engineering Home State ST, will be split into the following
two virtual programs:
\begin{itemize}
\item
NIT Surat, Civil Engineering, Home State, ST, Gender-neutral
\item
NIT Surat, Civil Engineering, Home State, ST, Female-only
\end{itemize}
An important issue that needs to be considered for the implementation of the above algorithm is whether we need to give preference to
category or gender while creating virtual preference list of a candidate. For example, if an OBC\_NCL female candidate with a valid GEN rank applies for an academic program $P$, what should be the order that should be followed? The candidate will be first considered for  a seat from the Female-only pool of GEN seats for program $P$ followed by Gender-neutral pool of GEN seats for program $P$. Thereafter, she will be considered for a seat from Female-only pool of OBC\_NCL seats for program $P$ followed by Gender-neutral pool of OBC\_NCL seats for program $P$. The justification for the above order is the following: The candidate is eligible for GEN seats. So attempts must be made to allocate her a GEN seat before considering her birth category or gender. 

Once we have created gender based virtual programs, all we need to do is to create the virtual preference lists of each candidate based on gender as recommended by the algorithm. To illustrate this, we provide examples.

\begin{example}
Suppose $c$ is a female candidate belonging to the SC\_PwD category with a
valid rank in SC merit list and CRL. Suppose one of her choices is IIT
Madras Chemical Engineering. Then we create the following virtual
programs, in order.\\ 

\noindent
1. IIT Madras, Chemical Engineering, GEN, Female-only\\
2. IIT Madras, Chemical Engineering, GEN, Gender-neutral  \\
3. IIT Madras, Chemical Engineering, PwD, Female-only\\
4. IIT Madras, Chemical Engineering, PwD, Gender-neutral  \\
5. IIT Madras, Chemical Engineering, SC, Female-only\\
6. IIT Madras, Chemical Engineering, SC, Gender-neutral  \\
7. IIT Madras, Chemical Engineering, SC\_PwD, Female-only\\
8. IIT Madras, Chemical Engineering, SC\_PwD, Gender-neutral\\
\end{example}
\begin{example}
Suppose $c$ is a non-female candidate belonging to OBC\_NCL category with a valid rank in CRL. Suppose one of his
choices is IIT Indore Electrical Engineering. Then we create the following virtual programs, in order.\\

\noindent
1. IIT Indore, Electrical Engineering, GEN, Gender-neutral\\
2. IIT Indore, Electrical Engineering, OBC\_NCL, Gender-neutral\\ 

\end{example}
\begin{example}
Suppose $c$ is a female candidate belonging to ST category. Suppose one of her choice is NIT Surat Civil Engineering. Suppose her home state is Gujarat. Then we create the following virtual programs, in order.\\

\noindent
1. NIT Surat, Civil Engineering, Home State, GEN, Female-only\\
2. NIT Surat, Civil Engineering, Home State, GEN, Gender-neutral\\
3. NIT Surat, Civil Engineering, Home State, ST, Female-only\\
4. NIT Surat, Civil Engineering, Home State, ST, Gender-neutral\\
\end{example}

The following additional points are very crucial in incorporating the rule of supernumerary seats for females.
\begin{itemize}
\item {\bf Vacant seats}\\
  Vacant seats, if any, from the female-only virtual program,
  is {\bf not} to be filled by any non-female candidate. The
  justification for this rule is the following: In this case, all
  seats from Gender-neutral($p$) will be occupied by non-female
  candidates, satisfying the fairness property for them. Seats from
  Female-only pool may have {\em supernumerary} seats created for
  improving gender ratio only. De-reserving these seats for
  non-females will defeat the purpose of supernumerary seats for
  females.
\item {\bf Handling foreign candidates} [for IITs only]\\
Foreign candidates will be admitted in {\bf gender-oblivious} manner only. Each foreign candidate will be considered for an IIT program if his/her rank is the same or better than the rank of the closing rank of the corresponding Gender-neutral OPEN virtual program. 
%The justification is the following. The total supernumerary seats for foreign candidates in each program is at most 10\% of its GEN %capacity. Creating 14\% female-only seats will be effectively creating 0 seats.    
\item {\bf Handling DS candidates} [for IITs only]\\
Recall that there are 2 supernumerary seats for DS candidates in each IIT. These seats will be gender-neutral only. 
%The justification is the same as that for the foreign candidates.
\item {\bf Handling Preparatory Course candidates} [for IITs only]\\
Preparatory course candidates will be handled {\bf based on the gender}. For example, any vacant seat from a Female-only virtual program will be offered only to Preparatory Course female candidates corresponding to that virtual program. Likewise, any vacant seat from a Gender-neutral virtual program will be offered to all Preparatory Course candidates eligible for that virtual program. 
\end{itemize}

%

%\chapter{Recommendations}
%\label{chapter:recommendations}
\chapter{Reduction of Vacancies}

As alluded in the introduction, joint seat allocation provides several
advantages to the student, including a single admission window, an
aligned academic calendar, and a fair allocation.  These beneficial
logistical advantages of having a single admission window is hard to
quantify in a nation as large as India.  

In this chapter, we discuss a subset --- the impact of the joint seat
allocation in terms of provable reduction in vacancies in the IITs
compared to 2014, and leading up to 2018. Since IIT seats are coveted
by a vast majority of engineering aspirants, any reduction is
substantially beneficial. We discuss the vacancies in IITs in
Section~\ref{sec:IIT-vacancies}, and the vacancies in NITs, IIITs and
other GFTIs (henceforth \emph{non-IITs}) in
Section~\ref{sec:non-IIT-vacancies}.

\section{Vacancies in the IITs}
\label{sec:IIT-vacancies}
Table~\ref{table:IIT vacancies} lists the vacancies over the
years\footnote{2018 vacancy data was not available at the time of
  writing this report.}, and the discernible trend is the reduction in
vacancies even when the number of seats have increased. An increase in
the number of seats should normally result in \emph{increase} in
vacancies, especially since the newer IITs\footnote{The Indian School
  of Mines, Dhanbad (ISM) was designated an IIT in 2016. In 2014, the
  number of vacancies at ISM resulted in an extra, special, local
  (secondary-market) spot round.  Nevertheless, ISM had 7 vacancies
  after this spot round. In 2015, ISM had 33 vacancies, and didn't
  feel necessary to conduct a spot round} are considered ``less
desirable''. The reduction clearly shows, and settles, the advantages
of the joint system.

\begin{table}[h]
\begin{tabular}{c c c c c c}
\toprule
\multirow{2}{*}{Year} & \multirow{2}{*}{Participating institutes} & \multirow{2}{*}{Total seats} & \multirow{2}{*}{Vacancies} & \multicolumn{2}{c}{2014 and prior institutes only} \\ \cmidrule{5-6}
& & & & Vacancies & Reduction from 2014 \\ \midrule
2014 & 16 IITs + ISM & 9,784 & 594 & 594 & \\
2015 & 18 IITs + ISM & 10,006 & 341 & 319 & 46\% \\
2016 & 23 IITs & 10,572 & 227 & 196 & 67\%\\
2017 & 23 IITs & 10,988 & 235 & 200 & 66\%\\ \bottomrule
\end{tabular}
\caption{The number of vacancies in the IITs over the years reduces
  substantially.}
\label{table:IIT vacancies}
\end{table}

Note that, throughout the years, not only did new institutes get added
in the system, but the seat matrices of individual institutes also
underwent modifications. Thus, the before-after comparison given here
may underestimate, or cloud, the benefits from the joint allocation.

To address this issue, we conducted a counterfactual experiment to
better assess how many vacancies were saved
\cite{Interfaces-paper:2018}. We simulated the earlier, separate
allocation process on the same input as joint allocation for the years
2015-2017. In these experiments, the candidates would apply (virtually
in the computer) only to the IITs first. Next, they apply only to the
non-IIT programs they preferred more than their IIT allocations.  As
output, we checked how many candidates got a better preference than
their IIT allocations, and presumed that these would represent the
vacancies in IITs (since these candidates are likely to desert their
IIT seats).
%\utk{Instead, how about ``since these candidates will desert
%  their IIT seats assuming their preferences remain constant
%  throughout the process". I think the reader might question why
%  ``likely'' is used here. This way, we can also explain our
%  assumptions}).  
We found that the joint process saved 371 (in 2015), 381 (in
  2016) and 629 (in 2017) seats in the IITs.

  As an aside, apart from saving vacancies, we found that many
  candidates obtained a better seat in the joint allocation scheme. In
  the earlier (separate) allocation process, a candidate A could
  occupy up to two seats: one in an IIT, and another in a
  non-IIT. Since the joint process allots her only one of those seats
  (the one preferred by her over the other), the second hoarded seat
  by A could now be offered to another candidate B, resulting in him
  getting a better preference. As a ripple effect, the seat vacated by
  B can now be offered to a third candidate C resulting in C getting a
  better preference as well. This cascading effect continues resulting
  in a number of benefited candidates, that is much greater than the
  number of vacancies!  The numbers of benefited candidates were
  1,866 (in 2015), 1,807 (in 2016), and 3,672 (in 2017).

\section{Vacancies in the non-IITs}
\label{sec:non-IIT-vacancies}
Since the non-IITs conducted their allocation after the IITs, we
didn't expect them to benefit from the joint process alone. We also
couldn't judge the impact to them from the counterfactual
experiment. Despite being unable to draw a clear conclusion on the
situation here, we report in Table~\ref{table:non-IIT vacancies} the
progression of vacancies in these institutions. We note the reduction,
and the downward trend in the reduction persists.
\begin{table}[h!]
\begin{tabular}{ccccccccc}
\toprule
\multirow{2}{*}{Year} & \multicolumn{3}{c}{Participating institutes} & \multirow{2}{*}{Total seats} & \multirow{2}{*}{Vacancies} & \multicolumn{2}{c}{2014 and prior institutes only} \\ \cmidrule{2-4} \cmidrule{7-8}
& NITs & IIITs & Other GFTIs & & & Vacancies & Reduction from 2014 \\ \midrule
2014 & 30 & 12 & 16 & 21,285 & 5,596 & 5,596 &  \\
2015 & 31 & 18 & 18 & 24,068 & 5,697 & 5,141 & 8\% \\
2016 & 31 & 20 & 18 & 24,323 & 4,901 & 4,379 & 22\% \\
2017 & 31 & 23 & 20 & 25,220 & 6,510 & 5,380 & 4\% \\ \bottomrule
\end{tabular}
\caption{Despite non-IITs admitting candidates after the IITs,
  vacancies in institutions reduce. Note the section on special round,
though. } 
\label{table:non-IIT vacancies}
\end{table}

\noindent {\bf Special Round} To mitigate the relatively high number
of vacancies, the non-IITs conducted a centralized special round in
2015 and 2017. In 2015, in the special round, 5,354 fresh allotments
were made, of which 2,683 candidates did not report, leading to a
final vacancy count of 2,883 at the non-IITs. In 2017, 5,352 seats
were allotted in the special round, of which 1,522 were rejected,
leading to a final vacancy count of 2,680.  Although the special round
is quite effective at reducing the vacancies after classes have
already begun, there is a need to understand the dynamics of the
vacancies in the main rounds in the first place as it indicates a
possible inefficiency in the joint allocation process.

We observe first that the non-IIT vacancies reduced in 2016 with the
introduction of the \emph{Withdraw} option, but, in absolute numbers
actually increased again in 2017.  We suspect that the reason for this
was due to a major change in the process. In 2017, school board marks
were no longer considered in determining JEE Main ranks, for admission
to the non-IITs. The school boards can be taken only once; however,
the IIT JEE Advanced can be taken twice. Further complicating the
matter, the JEE Mains can be taken three times.  Given this set of
situation, game theory based arguments suggest that it is beneficial
for more candidates to simply give up their non-IIT seat at the last
minute; they could still get confidently predict to regain these
seats again in subsequent years, if need be.

\begin{table}
\centering
\begin{tabular}{c r r}
\toprule
Round & 2016 & 2017 \\ \midrule
2 & 70 & 55 \\
3 & 293 & 129\\
4 & 805 & 365 \\
5 & 2,594 & 468\\
6 & N/A & 4,168 \\ \bottomrule
\end{tabular}
\caption{Withdrawals at the time of reporting for rounds 2 through 6
  for 2016 and 2017. Withdrawals were not possible at Round 1
  reporting, and they were not allowed at final round reporting (Round
  6 in 2016 and Round 7 in 2017). } 
\label{table:withdrawals}
\end{table}

Indeed, we observed a similar trend in withdrawals as can be seen in
Table~\ref{table:withdrawals} which shows the number of withdrawals
per round in 2016 and 2017. Also, most of the withdrawals happen in
the last possible round (the penultimate round in 2016, 2017 and
2018). We can literally visualize the candidate hoping against hope
that she gets a better preference in the next round, and, finally, she
ends up withdrawing in the end when this hope does not materialize.
%, and most of these result in vacancies because most fresh
%allotments are rejected in the last round.  

If we look at seats refused, as can be seen from
Table~\ref{table:SeatRJstats}, the number of seat rejections out of
fresh allotments goes on increasing as the rounds progress. In 2017
for example, in the 6th round, 77\% of the latest allotments were
rejected by the candidates.

\begin{table}[h!]
\centering
\begin{tabular}{c r r r}
\toprule
Round & 2015 (rejections/fresh) & 2016 (rejections/fresh) & 2017 (rejections/fresh) \\ \midrule
1 & 6,585 / 23,956 & 5,591 / 24,245 & 5,920 / 25,114 \\
2 & 3,571 / 6,954 & 2,373 / 5,977 & 2,695 / 6,262 \\
3 & 2,617 / 3,655 & 1,330 / 2,487 & 1,612 / 2,802 \\
4 & 2,166 / 2,624 & 1,010 / 1,698 & 1,180 / 1,905 \\
5 & N/A & 1,272 / 1,762 & 1,164 / 1,562 \\
6 & N/A & <No data> / 3,812 & 1,312 / 1,688 \\
7 & N/A & N/A & <No data> / 5,259 \\ \bottomrule
\end{tabular}
\caption[Seat rejections out of fresh allotments]{The number of
  seat rejections (when fresh allotments are made) for non-IIT allotments from
  2015 to 2017. %We do not have data for final round reporting in 2016
  %and 2017. 
  Note  how over $70\%$ of fresh allotments are rejected\footnotemark in
  later rounds.} 
\label{table:SeatRJstats}
\end{table}
\footnotetext{Note that seat cancellations (during document
  verification) are not counted towards rejections, but candidates
  with seat cancellations may be counted twice in fresh allotments
  (once at their first allotment, which got canceled and once again if
  they get allotted the second time). Also, withdrawals may be counted
  towards rejections only when they happen immediately after allotment
  (this happens very rarely in the data. The reason may be that the
  candidate reported first, and then withdrew in the same round). The
  number of seat cancellations and immediate withdrawals are generally
  small (less than a hundred).}
  
Based on the above observations, we believe the vacancy problem can be
mitigated by ensuring the following:
\begin{itemize}
\itemsep0pt
\item The number of withdrawals is reduced.
\item There is ample time after withdrawals for fresh allocations
\item The number of rejections in the final round(s) is reduced
\end{itemize}
In \cite{Interfaces-paper:2018}, we suggest a few simple
changes to the admissions process to obtain these improvements and
hence drastically reduce vacancies. The interested reader is
encouraged to go through them. 
%In 2018, our suggestions have not been
%implemented, and unsurprisingly, vacancies persist.

% chapter 8
\chapter{Conclusion}

In this report we have described the details of a multi-run
multi-round deferred acceptance (MRDA) algorithm used for allocating
seats in 100 institutes with approximately 39000 seats and with over a
million candidates. These institutes include the IITs (dictated by the
Joint Admission Board (JAB)), and other centrally funded technical
institutions (dictated by the Central Seat Allocation Board (CSAB)).

Prior to 2014, the allocations were decoupled leading to multiple
problems. There were fairness issues (see the example in the
introduction), and inefficiency issues such as seat vacancies, not to
speak of logistic nightmare for high school graduates criss-crossing
the country seeking admission towards the end of the admission process.

The method described in this report have been carried out by the Joint
Seat Allocation Authority (JoSAA) for the last 4 years. It is provably
fair and optimal in a formal sense. By providing a single window
centralized process, the logistic difficulties of admission from a
candidate perspective has reduced, even while keeping multiple merit
lists of the varied institutes as the criteria for admission, and with
no overbooking of seats allowed. It has led to a substantial reduction
in seat vacancies (with further recommendations for adoption for
JoSAA) as described here~\cite{Interfaces-paper:2018}.

From a policy perspective, authorities are urged to bring the entire
post high-school admission process (with multiple merit list criteria)
under this one unified process.

% chapter 9
\chapter{Appendix}

\section{Alternative algorithms to incorporate supernumerary seats for females}
\label{sec:wrong-approaches-females}
In this section, we present two simple algorithms which may appear to incorporate the rule of supernumerary seats for females correctly. However, each of them have one or more serious disadvantages.

\subsection{Algorithm 1}
This algorithm works as follows. Divide each virtual program into two
separate virtual programs for non-females and females respectively. In
the seat allocation process, non-females compete for non-female
virtual programs and females compete for female virtual programs
only. The capacities of these virtual programs may be fixed beforehand
similar to our current algorithm presented in
Chapter~\ref{chapter:SupernumeraryFemales}. The idea is that if the
applicant 
pool looks like that of last year, we will end up with as many
non-female candidates in each academic program as in 2017, whereas
extra seats may be created for female candidates in a program as per
the requirement.  

\noindent
{\bf Disadvantage of Algorithm 1:}\\
This algorithm has a serious flaw --- there may be potentially some
female candidates who get denied a seat by this algorithm which they
would have got otherwise by merit. This will happen, if for example,
there are many top-ranked females beyond the capacity of the reserved
sized female pool.  This violation would defeat the whole objective
--- in the pursuit of providing a guarantee of at least 14\% seats to
females, this algorithm may deprive some female candidates of their
right to compete for seats on the basis of merit. This serious problem
in the algorithm was not merely a theoretical possibility because our
simulations on the 2017 data revealed that the number of such female
candidates was non-zero.

\subsection{Algorithm 2}
This algorithm works as follows. First we carry out the usual
gender-neutral allocation like the previous years. We now freeze the
allocation of non-females. Let there be $i$ females getting admitted
to a virtual program $p$. If $i$ is less than 14\% of the number of
seats in $p$, we add additional seats to program $p$ in an appropriate
manner. Now we recompute the allocation for all females in these
female-only programs.
 
\noindent
{\bf Disadvantage 1 of Algorithm 2:}\\
This algorithm may lead to merit violation as follows. In the process
of satisfying the quota, there will be many programs where additional
seats for females are created. As a result, when we carry out the
allocation for females subsequent to the freeze, let's say a female
$x$ gets a more preferred program compared to the virtual program, say
$p$, allocated to her in the pre-freeze gender-neutral
allocation. Such a female will migrate to the more preferred program,
and her earlier seat (in the gender-neutral category) gets allocated
to some other female candidate $y$ with rank inferior to other
candidates in the gender-neutral pool. In other words, now a female
occupies a gender-neutral seat of $p$ with a rank which is worse than
a male candidate to whom virtual program $p$ was denied. So the final
allocation is unfair for non-females.

\noindent
{\bf Disadvantage 2 of algorithm 2:}\\
As mentioned in the beginning of
Chapter~\ref{chapter:SupernumeraryFemales}, one constraint that has to
be followed while incorporating the rule of supernumerary seats for
females is that program capacities (seat matrix) have to be frozen
prior to the joint seat allocation. This is because these capacities
are publicly announced. Moreover they should be determined in a
simple, transparent, and fair way. The allocation produced should not
violate the pre-announced capacities. Unfortunately, Algorithm~2
violates this constraint since the capacities for each round will
depend upon the outcome of the gender-neutral allocation for that
round. 
% Some may argue that this is a merit violation because of the way we define Gender-neutral seat in a static and rigid manner. 
In addition, there are non-trivial issues even in the generation of
the seat matrices dynamically. We highlight them below through
examples. 
\begin{example}
What if there are only 5 SC seats a program and 3 of them are occupied
by females in the gender-neutral allocation? Should we reserve the 3
seats for females? If so, it seems unjustified since the 3 females
for whom we reserved these seats, most likely, might have gone to more
preferred program after the female-only allocation that we carry out
after the gender-neutral allocation. 
%Moreover, it may lead to more than 14\% female seats.
\end{example}
\begin{example}
  What if there are only 2 ST seats and both of them are occupied by
  males in the pre-freeze allocation? Should we create 0.28
  supernumerary seats for females ? If we do not create any seat, it
  may lead to less than 14\% females violating the mandate of MHRD. If
  we create 1 seat for each such virtual program, we overshoot the
  14\% mark, violating the constraints of the institute. One could say
  that we may do it probabilistically: with probability 0.28 create 1
  seat and 0 otherwise. But, the institutes will not be ready for the
  fluctuation it may create in the number of female candidates.
\end{example}
\begin{example}
  What if nearly 40\% seats of a program are occupied by female
  candidates\footnote{There are some CSAB programs where it is indeed
    the case.}. Going with Algorithm~2, we will freeze 60\% seats for
  non-females and the remaining 40\% seats will be filled by females.
  One may argue that we should continue to keep 40\% seats for females
  in this program after the freeze, and before doing the dynamic
  allocation, since they were occupied by females on the basis of
  merit. But in this case, this argument is invalid, because the
  merit-worthy females who occupied these 40\% seats migrate to the
  programmes of their higher choices in the IITs (for example), and
  these 40\% seats will later (in the post-freeze phase) be occupied
  by females with much inferior rank. This begins to sound like 40\%
  seat reservation for females which is against the mandate (at most
  20\% supernumerary seats for females).
\end{example}

Note: The above examples are not hypothetical. These are real examples
taken from the data of the joint seat allocations of the years
2015--2017.

Based on the above examples, it is easy to observe that there are two fundamental problems in the dynamic seat matrix generation carried out by Algorithm~2.
\begin{enumerate}
\item Algorithm~2 entails creation of the supernumerary seats at
  virtual program level. This leads to a fluctuation and hence
  uncertainty about the number of female candidates admitted. It may
  potentially lead to fewer than 14\% females in an academic program
  (violating the mandate of MHRD) or much more females
  (infrastructural problems by institutes).  
The reader may verify that our algorithm in
Chapter~\ref{chapter:SupernumeraryFemales} does not have these two
shortcomings.
\item Handling of the programs where females are much more than 20\%
  is unjustified. As in the example, we may end up reserving a large
  number (e.g., 40\%) of seats for females.  Compared to this, our
  algorithm in Chapter~\ref{chapter:SupernumeraryFemales} handles it
  in a much more rational way: assigns 14\% - 20\% seats for
  female-only and the remaining to be Gender-neutral. It still allows
  40\% females if they come by merit, but not otherwise.
\end{enumerate}

\section{Summary of activities at reporting centers}
\label{sec:Reporting-center-activities}
In this section, we provide a summary of the activities that take place at reporting centers during a round. There are numerous activities carried out at a reporting center. However, we provide here the details of only those activities during a round that are essential from the perspective of seat allocation to be carried out for the next round.   First we mention the tables that get updated based
on these activities.
\begin{enumerate}
\item Allotment table.\\
For each candidate who is allotted a seat in a round, the allotment table provides the complete information of the seat allocated to her. This table is one of the output files of the seat allocation algorithm (see Section \ref{sec:Implementation-details-of-the-algorithm}). However, a candidate who is allocated a seat might not report at the reporting center. Even if she reports, the seat may get cancelled during document verification. Or the candidate willingly surrenders the seat in the same or later round, and thus withdraws from the joint seat allocation. This entire information is captured in two fields, namely, ``RStatus'' and ``Withdraw''. These two fields are appended to the allotment table of the current round and are used for the 
seat allocation of the next round.
\item Candidate table.\\
Candidate table is one of the input files of the seat allocation
algorithm (see Section
\ref{sec:Implementation-details-of-the-algorithm}). During document
verification, if the credentials of a candidate change, the candidate
table is updated accordingly. To mark that the credentials of a candidate
has  changed, the value of ``CatChange'' field in this table is updated
accordingly.  
\item Preference list (Choice table).\\
The preference list of a candidate may change during document verification. For example, if a candidate is found to be color-blind and the candidate had mentioned Mining program in the preference list, then the validity field of mining program in the preference list is marked N.
\end{enumerate}

\noindent
Now we provide a summary of the activities at the reporting centers followed by the way the tables mentioned above get updated based on these activities.

Once a candidate gets allocated a seat for the first time, the candidate has to report at the appropriate reporting center for document verification and seat confirmation. For this purpose, ``RStatus'' field of each such candidate is initialized as NR (acronym for Not Reported) and ``Withdraw'' field is initialized as N (acronym for Not) in the beginning of the round. Based on the activities at the reporting center, these fields are updated appropriately as described below.

If the candidate does not report during the round, the RStatus remains NR, and the candidate will not be considered for seat allocation in future rounds. If the candidate reports at the reporting center, then the following are the possible outcomes. 
\begin{enumerate}
\item
If the candidate is able to produce all the relevant documents and the documents are verified to be authentic, her seat gets confirmed. RStatus become RP (acronym for Reported).
If the candidate fails to produce some document or the documents produced have any discrepancy, the candidate's credentials have to be changed. For example, the candidate may be claiming to be OBC-NCL, but fails to produce valid OBC-NCL certificate at the reporting center. In this case, the candidate's category is changed to GENERAL. Another example is a candidate who gets Mining program but is found to be color-blind. There are many other examples that results in change in credentials. The change in credentials may lead to the following two possible outcomes.

The change in credentials may lead to cancellation of her seat. For example, if a candidate got seat in OBC-NCL virtual program but fails to produce valid OBC-NCL certificate, then her seat will get cancelled. Another case is a candidate who gets Mining program but is found to be color-blind. There are many other cases as well. In each such case that leads to the cancellation of a seat, the RStatus of the candidate is set to RC (acronym for Reported and seat Cancelled). The CatChange field of the candidate in the candidate table  will be set to 1 or 4. The candidate will be considered for seat allocation in future rounds based on the revised credentials. 

It is also possible that the credentials of the candidate change but the seat is not cancelled. For example, if an OBC-NCL candidate gets a seat from OPEN category but fails to produce valid OBC-NCL certificate at the reporting center. In such case, CatChange field of the candidate is set to 3. The RStatus of the candidate will be set to RP. The candidate will be considered for seat allocation in future rounds with revised credentials (the birth category will be GEN in this case). 

Note that a change in credentials may also result in making some choices (programs) in the preference list of the candidate invalid. For example, if the board marks of the candidate fail to satisfy the criteria of minimum board marks for admission into IITs (likewise CSAB), all IIT (likewise CSAB) programs in her preference list will become invalid. Another example is a candidate who opted for Mining program but is found to be color-blind. In this case, the Mining program will become invalid in the preference list  of the candidate. 
\item
During the verification of the documents, the candidate may be found to be ineligible for joint seat allocation. In such a case, the seat of the candidate is canceled and the candidate is removed from the list of eligible candidates (Candidate table) for future rounds. This candidate will not be considered for seat allocation in future rounds. 
\end{enumerate}

At the reporting center, if the seat of the candidate gets confirmed, the candidate may also exercise her options of Freeze/Float/Slide. 
Accordingly the ``Decision'' field of the candidate in the candidate table is updated to FR/FL/SL. If the
seat of the candidate gets canceled but the candidate is still eligible for seat allocation in future rounds with revised credentials, the Decision field continues to remain Float.  If a candidate does not report at the reporting center, Decision field is set to RJ. 
%In addition, RStatus field of the candidate continues to remain NR/DR as the case may be.

Note that there may be candidates in the current round of allotment who might have got a seat in previous rounds which they accepted as well. If the seat remains unchanged in the current round, then RStatus of such a candidate is set to RT (acronym for seat Retained). These candidates are not required to report at the reporting center again. But if the seat gets upgraded in the current round, then there are two possibilities. If the previous seat was in some CSAB institute and the new seat is in some IIT (or vice versa), then the candidate must report at the reporting center of some IIT (likewise CSAB institute for the reverse case). The RStatus of such candidate is set to DR (acronym for Dual Reporting). Processing of such a candidate is identical to the candidates with RStatus=NR as described above. If the upgraded seat is still from the same pool of institutes
(IITs or CSAB) as that of the previous seat, then the candidate is not required to report at the reporting center again. RStatus is set to RU (acronym for Retained and Upgraded) for such a candidate. Once a candidate gets a seat in a CSAB institute (likewise IIT), and her seat gets confirmed upon reporting at some reporting center of CSAB institutes (likewise IIT), she will not have to report again at any reporting center of CSAB institutes (likewise IIT). In other words, in normal circumstances, a candidate will report at a reporting center of CSAB institutes (likewise IIT) at most once though her seat may be upgraded multiple times.

In addition to the above activities at reporting center, the candidate may also visit reporting center for other activities. 
A candidate has to approach the reporting center if she decides, ever in future rounds, to exercise Freeze/Float/Slide.  
A candidate, who got a seat in some round, may also wish to surrender her seat and withdraw from joint seat allocation for that year. For this purpose, she has to report at the reporting center after initializing the withdraw process through his/her JoSAA login. The Withdraw field of such candidate in the Allotment file is set to Y after all the formalities of withdraw are completed at the reporting center.

Now we describe how the input for 2nd (or subsequent) round is computed based on the activities at reporting center described above.
%
%
%\subsection{Inputs for subsequent rounds}
%Apart from seat matrices which remains unchanged throughout all the rounds of seat allocation, the following tables are updated
%during the reporting center activities of each round. These updated tables are to be used for seat allocation in the next round.
\begin{enumerate}
\item Updating Candidate table.\\
The credentials of a candidate may change during the document verification at the reporting center during a round. 
%The following
%are a few examples.
%    \begin{enumerate}
%    \item Birth category change: A candidate allotted an SC seat is unable to
%      produce a valid SC certificate.
%    \item PwD status change: A candidate allotted a PwD seat is unable
%      to produce a valid PwD certificate.
%    \item Medically capability change: A candidate while reporting to
%      accept the offered seat is declared medically unfit for the
%      offered program.  Note that the candidate is unaware of this
%      situation a priori\footnote{Example: A seat in the mining
%        engineering program is offered to a student who turns out to
%        be color blind.  The mining program may not admit color
%        blind students, as per law.}
%    \end{enumerate}
The credentials will be updated accordingly in the candidate table for the next round. The CatChange field is updated accordingly based on the type of credentials that got changed. The ``Decision'' field is also changed based on the option (Freeze or Float or Slide)
chosen by the candidate at the reporting center. Here we would like to state some more information about Decision field.
Once a candidate reports at the reporting center, the following are the possibilities of the change in this field. 
\begin{enumerate}
\item
Her seat is confirmed. In that case, The Decision field may be set to Freeze or Float or Slide based on the option given by the candidate at the reporting center.
\item
Her seat is cancelled but she will still be considered for seat allocation in future rounds. Decision field will be set to Float for such candidates.
\item
She does not report at the reporting center. In this case, Decision field is set to RJ. 
\end{enumerate}
Note that the candidates who become ineligible for seat allocation after document verification are removed from the candidate table in subsequent rounds. The converse may also happen (though it is rare). The board marks of a candidate may increase after re-evaluation and a candidate, who was ineligible previously, may potentially become eligible for seat allocation in CSAB institutes and/or IITs (note that these two pools may have different eligibility criteria of board marks). 

\item  Augmenting the allotment table.\\
%Recall that for each round, allotment table is computed for each candidate. It has the complete details about the program assigned to the candidate. 
Based on the reporting center activities as described above, RStatus and Withdraw fields of the candidates is updated as described above. 
These two fields are appended to the allotment table of the current round. This augmented file will be used for computing the seat allocation for the next round (for computing Min-Cutoff for each virtual program). It is therefore given as input for the next round and called `allotment table of the previous round'.
\item Updating the preference list (Choice table).\\
Based on the credential changes, the preference list of the candidate will be updated. In particular, some choices may become invalid after document verification as described above. The ``Validity'' field of such choices is marked N. 
\end{enumerate}

Although we provided complete details of the reporting center activities above, a programmer needs to focus on the following
specific points while writing the code for seat allocation.
\begin{enumerate}
\item 
In order to determine if the seat of the candidate is cancelled, just check RStatus field and Withdraw field from the allotment table of the previous round. If RStatus=NR/DR/RC, or Withdraw=Y, the seat of the candidate is cancelled. 
\item
In order to determine the current preference list, the ``Decision'' field is required and the preference list needs to be pruned if needed. Also the validity field of programs in the preference list needs to be taken into account while reading - If validity field of a program is N, that program must be skipped while creating virtual preference list of the candidate. 
\item
For the first 3 years of the joint seat allocation, CatChange field in Candidate table used to be considered for seat allocation computation in each round.
But after a change in a business rule in 2018 (refer to Section \ref{sec:changes-in-business-rules-since-2015}), this field can be safely ignored as far as seat allocation in any round is concerned.
\end{enumerate}

%Based on the value of this field, the choice list of the candidate will be updated during the computation of seat allocation in the next round.

%\input{mincutoff}
%{\color{red} \input{choicefilling} }
\section{Survey Questions}
\label{sec:survey}
Candidates who wish to surrender their seat and withdraw from the joint seat allocation 
should be asked to fill out a survey before their fees are refunded. This survey should be analysed at the end of the joint seat allocation every year so that the efficiency of the joint seat allocation can be improved in the future years.
Here is a possible mini-survey design: 

{\small This survey is intended to obtain a detailed understanding of
  the performance of JoSAA 20xx, and possibly suggest areas for
  potential further improvements in efficiency of seat
  allocation. Complete privacy of your data is guaranteed. There will
  be no penalties of any kind based on information entered here. We
  appreciate your help in providing us this vital information.

Why did you choose to not take-up your allocation of program ??? in round ??? of JoSAA 20xx? Please select the appropriate option:\\[7pt]
\emph{\phantom{xx}} Want to write JEE again.\\
%\footnote{Keep this option only if this is allowed in the first place for people who accepted seat by paying fees and then changed their mind.}.\\
\emph{\phantom{xx}} Admitted to other Institute\\
\phantom{xsxxxxx}(i) Institute name \emph{\phantom{xxxxxxxx}} (ii) Date of admission notification \emph{\phantom{xxxxxxxx}}\\
\emph{\phantom{xx}} Other. (i) Please specify \emph{\phantom{xxxxxxxx}} (ii) Date of decision \emph{\phantom{xxxxxxxx}}(select on a calendar)

\vspace{7pt}
\noindent Thank you for your assistance!\\
}

\section{Validation Modules}
\label{sec:validation}
An important challenge is to verify whether the output of the DA algorithm for a large test case (nearly 13 lakh candidates) satisfies various business rules. Validation modules provide effective ways to achieve this goal. Each validation module will take the input and output files of the DA algorithm for a round and verify whether the output is indeed correct for the given input. There are two types of validation modules.
\subsection{Candidate specific validation modules}
There are 6 validation modules required to ensure that allotment meets the business rules for the candidates.
\subsubsection{Fairness}
Suppose a candidate $c$ is denied a virtual program $p$. Then the rank of each candidate getting $p$ must be superior to the rank of $c$. This module can be implemented efficiently with the help of the closing rank as follows. The closing rank of each of the virtual programs that rejected $c$ must be better than the rank of $c$. Note that this module seamlessly takes care of the current implementation of the supernumerary seats for females.

\subsubsection{Quota Eligibility}
NIT program has home state (HS) and other state (OS) quota. Similarly for other GFTIs, each program has home state (HS) and all India (AI) quota. Quota eligibility requires the following condition to be guaranteed for HS quota (similarly for OS and AI).  A candidate can be allotted a seat from HS quota of a program only if she is eligible for the HS quota of that program; conversely, if the candidate is not eligible for the HS quota, she should not be awarded the seat from HS quota. %(This description is different for AI).
\subsubsection{Category Eligibility}
The aim of this module is to verify that a candidate must be assigned seat from the category for which she is eligible. For example, a SC candidate who does not appear in CRL cannot be assigned a seat from any OPEN virtual program. In a similar manner, a GEN candidate cannot be assigned a seat from an SC virtual program.     
\subsubsection{Candidate willingness}
After a given round, the candidates who are allotted programs may opt for any of the following options: Freeze, Float, Slide, or Reject. These options must be respected while allocating programs to them in the following rounds.     
\subsubsection{Seat guarantee in later rounds (Min-Cutoff benefit)}
Each eligible candidate will get Min-Cutoff benefit. That is, if $c$ is a candidate applying to a virtual program $p$. If rank of $c$
is better than the Min-Cutoff($p$), then $c$ will surely be given a seat in $p$ irrespective of whether or not there is any vacancy in $p$.

\subsubsection{Fairness guarantee for non-female candidates} If the number of non-female candidates in a Gender-neutral virtual program $p$ turns out to be less than the seat capacity of $p$, then either there is a vacancy in $p$ or there is at least one female candidate getting a seat in $p$. 
\subsubsection{Fairness guarantee for female candidates} If a male candidate $c$ gets a seat in a virtual program $p$, then every female candidate eligible for the virtual program $p$, must be given a seat in $p$ if her rank is same or better than that of $c$.
%\subsubsection{Restricted fairness for category change}
%Let c be a candidate whose CatChange = 1. Both of the following must be true
%\begin{itemize}
%\item If rank of c is better than min-cut-off for program p but she does not get program p even after applying to p, then no candidate can get program p with rank worse than min-cut-off of p. 
%\item No candidate with CatChange = 1 can get p if his/her rank is worse than that of c.
%\end{itemize}
\subsection{Program specific validation modules}
There are two broad validation modules to verify whether the allotment meets the business rules of the programs.
\subsubsection{Validating de-reservation}
During multiple runs of DA in a round, the capacity of a virtual program may change due to de-reservation. Thus within a round, the seat matrix may change. However, the capacities of the sum of all virtual programs must remain unchanged. The checks we can do here are:
\begin{itemize}
\item Sum of capacities of all virtual programs is the same before and after de-reservation. In fact, the following equation should hold for each virtual program $p$
\[ \text{NewCap} = \text{InitCap} + \text{No. of seats de-reserved to }p - \text{No. of seats                                                                                                                    de-reserved from }p \]
\item A virtual program from which de-reservation happened cannot have any supernumerary seat. 
\item A virtual program from which de-reservation happened must not be denied to any candidate eligible for that virtual program. That is, there should be no eligible candidate who applied to this program and didn't get it.
\end{itemize}
\subsubsection{Validating the cause of supernumerary seats}
The number of candidates getting a virtual program may be more than the capacity of the virtual program. The surplus candidates are given supernumerary seats. The cause of supernumerary seats may be any one of following: 
\begin{itemize}
\item Multiple candidates at closing rank (EQ)
\item Min-Cutoff criteria (MC)
\item Foreign nationals (FR)
\item Foreign nationals with multiple candidates at closing rank (FE)
\item Foreign nationals getting seat by Min-Cutoff criteria (FM)
\item DS candidate (DS)
\item DS candidates with multiple candidates at closing rank (DE)
\item DS candidates getting seat by Min-Cutoff criteria (DM)
\end{itemize}
In order to verify the cause of supernumerary seats, additional information has been provided in the allotment table. The validation module for supernumerary seats should use this information.

In order to assist the people engaged in validating/testing the output of the seat allocation software, the allotment table will have 2 columns (Flag, SupNumReason). These columns will be used to compute Min-Cutoff of a virtual program for the next round, provide and verify the reason for the creation of supernumerary seats in a virtual program. In addition to the augmented allotment table, the following table will also be output by the DA implementation.
\subsubsection{Program Statistics}
      For each virtual program, the following information will be mentioned.
\begin{itemize}
\item Opening rank of the virtual program.
\item Closing rank of the virtual program.
\item MinCutOff of the virtual program.
\item Total candidates allotted to the virtual program.
\item Initial and final capacity of the virtual program
\item The number of seats that got dereserved to and from the virtual program\footnote{These are two different columns DereserveTo and DereserveFrom}
\item The number of supernumerary seats created in the virtual program.
\end{itemize}

\subsubsection{Allotment comparing module}
In order to compare two allotments (produced by two different implementations), a module that highlights the candidates whose allotment differs in the two allocations may prove to be very useful. This module will help testing/validating team and may help fixing of a bug, if any. Such a module was a part of the IITK implementation of the DA algorithm in 2015 and 2016, and it proved to be very helpful.

\section{Implementation Details of MRDA}
\label{sec:Implementation-details-of-the-algorithm}
In this section we present the implementation details of the algorithm. In particular,
we elaborate on the input, output, and the interface of the DA algorithm along with its interaction with the reporting center during any round.

\subsection{Algorithm: interface and interactions }
\begin{center}
\begin{figure}[h]
\begin{center}
  \includegraphics[width=0.9\textwidth]{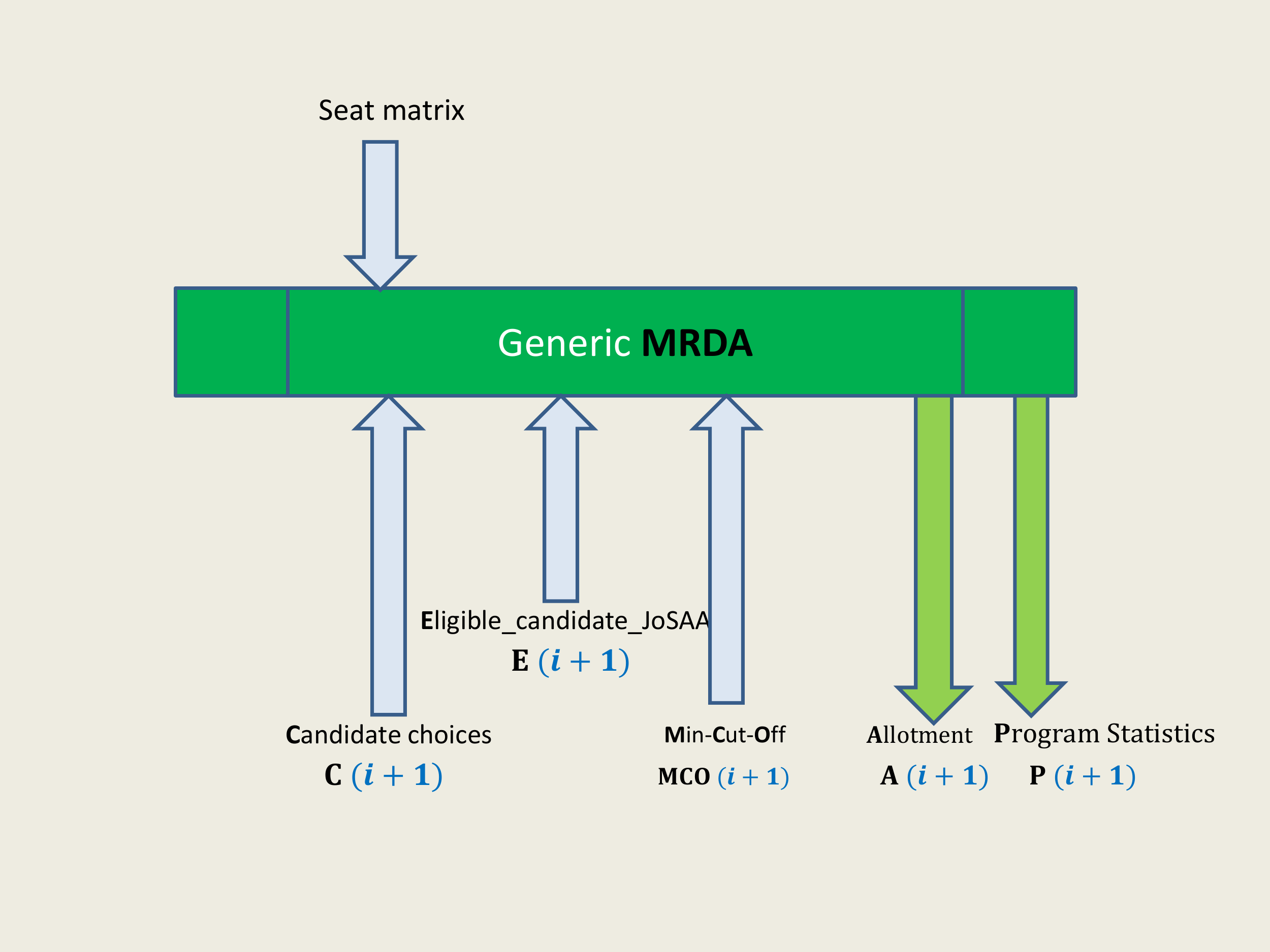}
\caption{The interface (input/output) of the generic DA algorithm}
\label{figure:interface}
\end{center}
\end{figure}
\end{center}

Seat allocations happen in multiple stages (or rounds).  The key difference between the first round and subsequent rounds is that some candidates have seats that the Joint Seat Allocation Authority (JoSAA) agrees to a guarantee - seats offered  in prior rounds (e.g. ``Freeze'') will continue to be available.  
Having a single implementation of the DA algorithm (with updated inputs) that works for every round is better than having multiple implementations. We have to validate only one implementation. We call such an implementation a ``generic'' implementation. Figure \ref{figure:interface} shows the interface (input/output) of the generic DA algorithm.

We provide below a summary of four  tables that will serve as input to the generic implementation of DA algorithm, and also two output tables.
\begin{itemize}
\item {\bf Seat\_Matrix}:
The table stores the list of programmes of various institutes that participate in the joint seat allocation along with their capacities . Seat matrix will remain the same for each round. (However, internally, during multiple runs, this matrix may change due to de-reservation.) Seat matrix is obtained from the institutes.

\item {\bf Choice list of Candidates}:
This table correspond to the preference list mentioned in our algorithm. This table stores the choices (programs) of each candidate. 
For each candidate the choices will appear in this table in the decreasing order of her preference.
The table for the (N+1)th round is defined by the table for the (N)th round, the candidate willingness option (freeze/float/slide/reject). The validity field of one ore more program may also be updated based on the activities at the reporting center. The willingness information is not a part of this table, but is available in the 63rd  column of the Eligible\_Candidate\_JoSAA table mentioned below.  The initial table is obtained after JEE Mains Rank List and JEE Advanced Rank List are declared. N is indexed starting from 1.

\item {\bf Eligible\_Candidate\_JoSAA} (Candidate table):
This table stores personal and other eligibility information of candidates. The information is collected at the time of the registration for the exam.  For JEE Mains, this information can be quite dated, being collected almost six months in advance. The information in this table is vulnerable, because, for example, the OBC\_NCL status of candidates can change, possibly due to change in income.
This table will be provided externally at the beginning of every round. For any round, the table will consist of all those eligible candidates
who have at least one valid choice in their respective choice lists.
For round (N+1), N $\ge$ 1, this table will have meaningful information in the following  fields.
\begin{itemize}
\item CredentialChange: This field will take value from \{1,2,3,4\}. For the 1st round, the value will be 2.
\item Decision:
This field will be one of Freeze/Float/Slide/Reject/Withdraw depending upon the option exercised by the candidate after a seat is allocated. This field will be used for computing the choice list of the candidate for the (N+1)th  round.
\end{itemize}

\item {\bf Min-Cutoff}.\\
This table will store min-cut-off for each virtual programme. For the first round, it will be set to 0 for all virtual programmes. Min-Cutoff for a virtual programme prior to the execution of round (N+1) will be computed using the allotment table of the (N)th round and the Decision field (column 63) of the Eligible\_Candidate\_JoSAA file mentioned above.
\end{itemize}
%Note: For rounds other than the first round, Eligibile\_Candidate\_JoSAA table and Candidate\_Choices table will be the only external input. 

\noindent
The output of the algorithm for each round will be the following two tables.
\begin{itemize}
\item {\bf Allotment}

This table will store the details of the program, if any, allotted to each candidate. The allotment table has two significant columns for purposes of multiround DA. Details appear below. 

\item {\bf Program Statistics}

This table will store the details of the program statistics (opening rank, closing rank, supernumerary information, and de-reservation information).
\end{itemize}

\begin{center}
\begin{figure}[h]
\begin{center}
\includegraphics[width=0.9\textwidth]{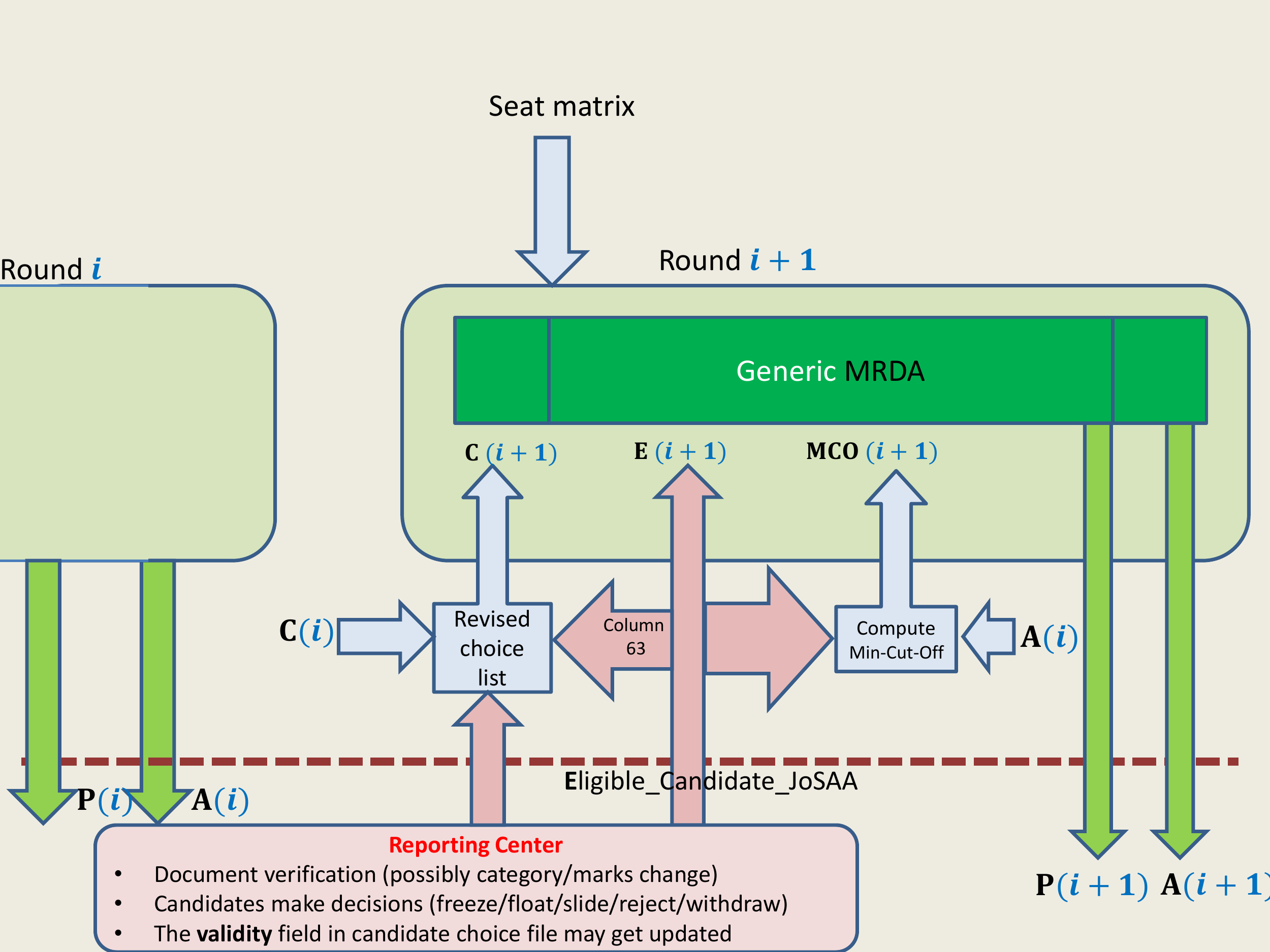}
\caption{The activities during $(i+1)$th round}
\label{figure:i-th-round}
\end{center}
\end{figure}
\end{center}

Please refer to Figure  \ref{figure:interface} for the interface of the generic DA algorithm and refer to Figure  \ref{figure:i-th-round} for various activities that take place during any round.

Figure \ref{figure:i-th-round} depicts the execution of the algorithm and activities that take place during $(i+1)$th round.
\pagebreak
%------------------------The Input format--------------------%
\subsection{Input format}
The complete details of the three input tables are given below.
\subsubsection{Seat Matrix}
\begin{center}
{\small
\begin{tabular}{|c|l|l|p{6cm}|}
\hline
{\bf S. No.} & {\bf Column Name} & {\bf Type} & {\bf Description}  \\ \hline
1. &  Quota  & char(2)  & AI/HS/OS \\ \hline
2. &  InstCd & char(3)     & Institute Code \\ \hline
3. &  BrCd   & char(4)     & Branch Code \\ \hline
4. &  GenderPool & char(6) & Gender pool with two possibilities:\\ 
   &         &         & {\color{blue} Neutral} or {\color{blue} Female} \\ \hline
5. &  OP     & int         & Open Seats \\ \hline
6. &  OP\_PwD & int         & Open-PwD Seats \\ \hline
7. &  SC     & int         & SC Seats\\ \hline
8. &  SC\_PwD & int         & SC-PwD Seats \\ \hline
9. &  ST     & int         & ST Seats  \\ \hline
10. & ST\_PwD  & int         & ST-PwD Seats \\ \hline
11.& OBC\_NCL & int         & OBC-NCL Seats \\ \hline
12.& OBC\_NCL\_PwD & int     & OBC-NCL-PwD Seats \\ \hline
13.& Total   & int         & Total Seats \\ \hline
14.& StCd1   & char(2)     & State code of Eligibility \\ \hline
15.& StCd2   & char(2)     & State code of Eligibility \\ \hline
16.& StCd3   & char(2)     & State code of Eligibility \\ \hline
17.& StCd4   & char(2)     & State code of Eligibility \\ \hline
\end{tabular}
}
\end{center}
\subsubsection{Choice list of a candidate}
\begin{center}
{\small
\begin{tabular}{|c|l|l|l|p{4cm}|} 
\hline
{\bf S. No.} & {\bf Column Name} & {\bf Type} & {\bf Description}  & {\bf Remarks} \\ \hline
1          & RollNo      & char(8) & JEE (Main) Roll number & \\ \hline
2          & OptNo       & int         & Option No  & \\ \hline
3          & Instcd      & char(3)     & Institute Code & \\ \hline
4          & BrCd        & char(4)     & Branch Code & \\ \hline
5          & Validity    & char(1)     & NULL (default)- if choice is valid & 
                                                       this field will be updated \\
           &             &             & N - Choice is not valid & at the time of Document verification \\ \hline
\end{tabular}
}
\end{center}
\subsubsection{Candidate Table}

\begin{center}
{\small
\begin{tabular}{|c|l|l|p{4cm}|p{2cm}|}
\hline
{\bf S. No}. & {\bf Column name} & {\bf Type and} & {\bf Description} & {\bf Remark} \\ 
             &                   & {\bf  Length } & & \\
\hline
1.     & RollNo      & char(8)       & JEE(Main) Roll Number & \\ \hline
2.     & AppNo       & char(8)      & JEE(Main) Application No & \\ \hline
3.     & NAME        & varchar(46)    & Candidate Name         & \\ \hline
4.     & MNAME       & varchar(46)    & Mother Name              & \\ \hline
5.     & FNAME       & varchar(46)    & Father Name              & \\ \hline
6.     & GName       & varchar(46)    & Guardian Name          & \\ \hline
7.     & SCode       & char(2)      & State Code of Eligibility & \\ \hline
8.     & Gender      & char(1)      & 1-Male, 2-Female,3-Transgender & \\ \hline
9.     & DOB         & char(10)     & Date of Birth (DD/MM/YYYY)   & \\ \hline
10.    & CAT         & char(2)      & GN/SC/ST/BC                  & \\ \hline
11.    & PwD         & char(1)      & 1-YES, 2-NO                  & \\ \hline
12.    & Nationality & char(1)       & 1-Indian, 2-OCI, 3-PIO, 4- Foreign (Other than OCI, PIO) & \\ \hline
13.    & AI\_Eng\_CRL\_Rank & float     & JEE(Main) All India Eng CRL Rank & \\ \hline
14.    & AI\_Eng\_OBC\_NCL\_Rank & float & JEE(Main) All India Eng OBC-NCL Rank & \\ \hline
15.    & AI\_Eng\_SC\_Rank      & float & JEE(Main) All India Eng SC Rank & \\ \hline
16.    & AI\_Eng\_ST\_Rank      & float & JEE(Main) All India Eng ST Rank & \\ \hline
17.    & AI\_Eng\_CRL\_PD\_Rank  & float & JEE(Main) All India Eng CRL-PwD Rank & \\ \hline
18.    & AI\_Eng\_OBC\_NCL\_PD\_Rank &float & JEE(Main) All India Eng OBC-PwD Rank & \\ \hline
19.    & AI\_Eng\_SC\_PD\_Rank   & float & JEE(Main) All India Eng SC-PwD Rank & \\ \hline
20.    & AI\_Eng\_ST\_PD\_Rank   & float & JEE(Main) ) All India Eng ST-PwD Rank & \\ \hline

\end{tabular}
}
\end{center}

\begin{center}
{\small
\begin{tabular}{|c|l|l|p{4cm}|p{2cm}|}
\hline
{\bf S. No}. & {\bf Column name} & {\bf Type and} & {\bf Description} & {\bf Remark} \\ 
             &                   & {\bf  Length } & & \\ \hline
21.    & Eng\_Rem\_Symb        & char(1) & JEE (Main) - BE/B.Tech Eligibility & \\ 
 & & &                                     Remark symbol & \\ 
 & & & `*' - Eligible under CRL  & \\ 
 & & & `=' - Eligible for OBC-NCL seats only & \\ 
 & & & `+' - Eligible for SC/ST/PwD seats only &\\ 
 & & & `\$' - Eligible for OBC-NCL-PwD seat only  & \\ 
 & & & `\%' - Eligible for GEN-PwD seat only  & \\ 
 & & & `N' - Non Eligible for Seat Allocation  & \\ \hline
22.    & AI\_Arc\_CRL\_Rank     & float   & JEE(Main) All India Arc CRL Rank & \\ \hline
23.    & AI\_Arc\_OBC\_NCL\_Rank & float   & JEE(Main) All India Arc OBC-NCL Rank & \\ \hline
24.    & AI\_Arc\_SC\_Rank      & float   & JEE(Main) All India Arc SC Rank & \\ \hline
25.    & AI\_Arc\_ST\_Rank      & float   & JEE(Main) All India Arc ST Rank & \\ \hline
26.    & AI\_Arc\_CRL\_PD\_Rank  & float   & JEE(Main) All India Arc CRL-PwD Rank & \\ \hline
27.    & AI\_Arc\_OBC\_NCL\_PD\_Rank & float & JEE(Main) All India Arc OBC-PwD Rank & \\ \hline

28.    & AI\_Arc\_SC\_PD\_Rank   & float   & JEE(Main) All India Arc SC-PwD Rank & \\ \hline
29.    & AI\_Arc\_ST\_PD\_Rank   & float   & JEE(Main) All India Arc ST-PwD Rank & \\ \hline

\end{tabular}
}
\end{center}

\begin{center}
{\small
\begin{tabular}{|c|l|l|p{4cm}|p{2cm}|}
\hline
{\bf S. No}. & {\bf Column name} & {\bf Type and} & {\bf Description} & {\bf Remark} \\ 
             &                   & {\bf  Length } & & \\ \hline

30.    & Arc\_Rem\_Symb        & char(1) & JEE (Main) - B Arch/B.Planning Eligibility & \\  
& & & Remark symbol & \\  
& & & `* - Eligible under CRL  & \\ 
& & & `=' - Eligible for OBC-NCL seats only & \\ 
& & & `+' - Eligible for SC/ST/PwD seats only &  \\ 
& & & `\$' - Eligible for OBC-NCL-PwD seat only  & \\ 
& & & `\%' - Eligible for GEN-PwD seat only & \\ 
& & & `N' - Non Eligible for Seat Allocation & \\ \hline
31.    & Adv\_RollNo         & char(7)  & JEE(Advanced) Roll Number & \\ \hline
32.    & Adv\_RegNo          & char(10) & JEE(Advanced) Registration Number & \\ \hline
33.    & Adv\_CRL\_Rank       & float    & JEE(Advanced) Common Rank List & \\ \hline
34.    & Adv\_OBC\_NCL\_Rank   & float    & JEE(Advanced) OBC-NCL Rank  & \\ \hline
35.    & Adv\_SC\_Rank        & float    & JEE(Advanced) SC Rank & \\ \hline
36.    & Adv\_ST\_Rank        & float    & JEE(Advanced) ST Rank & \\ \hline
37.    & Adv\_CRL\_PD\_Rank    & float    & JEE(Advanced) CRL-PwD Rank & \\ \hline
38.    & Adv\_OBC\_NCL\_PD\_Rank & float   & JEE(Advanced) OBC-NCL-PwD Rank & \\ \hline
39.    & Adv\_SC\_PD\_Rank     & float    & JEE(Advanced) SC-PwD Rank & \\ \hline
40.    & Adv\_ST\_PD\_Rank    & float     & JEE(Advanced) ST-PwD Rank & \\ \hline

\end{tabular}
}
\end{center}

\begin{center}
{\small
\begin{tabular}{|c|l|l|p{4cm}|p{2cm}|}
\hline
{\bf S. No}. & {\bf Column name} & {\bf Type and} & {\bf Description} & {\bf Remark} \\ 
             &                   & {\bf  Length } & & \\ \hline

41.    & Adv\_Rem\_Symb      & char (1)  & JEE (Advanced) - Eligibility  & \\
       &                   &           & Remark  symbol & \\
       &                   &           & `*' - Eligible under CRL  & \\
       &                   &           & `=' - Eligible for OBC-NCL seats only & \\
       &                   &           & `+' - Eligible for SC/ST/PwD seats only & \\
       &                   &           & `\$' - Eligible for OBC-NCL-PwD seat only  & \\
       &                   &           & `\%' - Eligible for GEN-PwD seat only & \\
       &                   &           & `P' - Eligible for Preparatory seat & \\
       &                   &           & `N' - Non Eligible for Seat Allocation & \\ \hline
42.    & Adv\_IsPrep        &  char(1)  &  Is eligible for Preparatory (1-Yes, 2-No)& \\ \hline
43.    & Adv\_Prep\_SC\_Rank  & float     & Preparatory Rank for SC candidates & \\ \hline
44.    & Adv\_Prep\_ST\_Rank  & float     & Preparatory Rank for ST candidates & \\ \hline
45.    & Adv\_Prep\_CRL\_PD\_Rank & float  & Preparatory Rank for all PwD candidates & \\  \hline
46.    & Adv\_Prep\_OBC-NCL\_PD\_Rank & float & Preparatory Rank for OBC-NCL  & \\ 
       &                              &       &        PwD candidates & \\ \hline
47.    & Adv\_Prep\_SC\_PD\_Rank & float  & Preparatory Rank for SC-PwD candidates & \\  \hline
48.    & Adv\_Prep\_ST\_PD\_Rank  & float  & Preparatory Rank for ST-PwD candidates & \\ \hline
49.    & Adv\_AAT\_Status    & char(1)   & 1-Qualified, 2-Not Qualified & \\ \hline
50.    & Adv\_DS            & char(1)   & DS Status (1-Yes, 2-No) & \\ \hline
51.    & Adv\_colour blind  & char(1)   & 1-Yes, 2-No & \\ \hline
52.    & Adv\_OneEyedVision & char(1)   & 1-Yes, 2-No & \\ \hline
53.    & Eng\_Top\_20        & char(1)   & NULL & \\ \hline
54.    & Arc\_Top\_20        & char(1)   & NULL & \\ \hline
55.    & Adv\_Top\_20        & char(1)   & NULL & \\ \hline

\end{tabular}
}
\end{center}

\begin{center}
{\small
\begin{tabular}{|c|l|l|p{4cm}|p{3.3cm}|}
\hline
{\bf S. No}. & {\bf Column name} & {\bf Type and} & {\bf Description} & {\bf Remark} \\ 
             &                   & {\bf  Length } & & \\ \hline

56.    & Board\_Mark\_Eng    & float     & NULL \\ \hline
57.    & Board\_Mark\_Arc    & float     & NULL \\ \hline
58.    & Board\_Mark\_Adv    & float     & NULL \\ \hline
59.    & Board\_RollNo      & varchar  & Board Roll No & \\  \hline
60.    & Board\_Year\_Passing & varchar & Year of passing of class 12th or equivalent & \\ \hline
61.    & Board\_Name        & varchar  & School Board Name & \\ \hline
62.    & CatChange**  & char(1)   & 1,2,3,4 (Default value is 2) & \\ \hline
%       &                   &           &  1-Yes, seat cancellation and penalty & \\
%       &                   &           &  2-No (Default) & \\ 
%       &                   &           &  3-Yes but no seat cancellation and no penalty & \\
%       &                   &           &  4-Yes and seat cancellation but no penalty & \\ \hline
%61.    & MarksChangeENG    & char(1)   &  Change in Marks & Satus of marks\\
%       &                   &           &  I: Increase,    & change in \\
%       &                           &           &  D: Decrease,    & Board\_marks\_ENG\\
%       &                   &           &  U:Unchanged(Default) & (from previous round)\\  \hline
%62.    & MarksChangeARC    & char(1)   & Change in Marks  & Satus of marks\\
%       &                   &           &  I: Increase,    & change in \\
%       &                           &           &  D: Decrease,    & Board\_marks\_ARC\\
%       &                   &           &  U:Unchanged(Default) & (from previous round)\\    \hline
63.    & Decision          & char(2)   & FR: Freeze  &  \\
       &                   &           & FL: float   & \\
       &                   &           & SL: Slide   & \\
       &                   &           & RJ: Reject  & \\ \hline
\end{tabular}
}
\end{center}

\subsubsection{Credential Change Description}
\label{sec:credential-change}
 The CatChange field flags the change in credentials of a candidate. The value of this field will be one of {\bf 1,2,3,} or {\bf 4}. The following is the description of different values of this field.
\begin{itemize}
\item[1:] If the credential changes are among the following:\\
1. Nationality change, 2. Home state change, 3. Birth category change\\
4. DS status change, 5. PwD status change, 6. Gender change\\
In this case, his/her seat is cancelled.
\item[2:] This is the default value. It means no change in candidate credentials.
\item[3:] If his/her credentials changed but seat was not cancelled. 
%In this case, the candidate will get the benefit of min-cut-off. 
Example: an OBC\_NCL candidate got a seat from GEN category but his/her OBC\_NCL certificate is 
found to be invalid at the reporting center.
\item[4:] If the credential changes are among the following:\\
1. OneEyedVision status change, 2. ColorBlindness status change\\
In this case, the seat of the candidate is cancelled.
\end{itemize}

%\noindent
%{\bf Important Note:} Round 2 onwards, we have to do the following. If a candidate gets CredentialChange = 1 or 4 in the beginning of the seat allocation, then we shall set CatChange=3 after the allotment so that in the subsequent rounds, the same candidate will be
%processed as a CatChange=3.

%------------------------The Output format--------------------%
\subsection{Output format}
\subsubsection{Allotment Table}
\begin{center}
{\small
\hspace*{-2cm}
\begin{tabular}{|c|l|l|l|l|}
\hline
{\bf S. No}. & {\bf Column name} & {\bf Type and} & {\bf Description} & {\bf Remark} \\ 
             &                   & {\bf  Length } & & \\ \hline
1. & RoundNo  & int   & Round No & \\ \hline
2. & RollNo   & char(8)   & JEE(Main) Roll Number & \\ \hline
3. & Birth\_Cat  & char(2) & Candidate birth category & \\ \hline
4. & Optno       & int & Option No & \\ \hline
5. & InstCd      & char (3) & Institute Code & \\ \hline
6. & BrCd        & char(4)  & Branch Code  & \\ \hline
7. & Rank        & float    & Rank used for seat allocation & \\ \hline
8. & AllottedCat & char(4)  & Allotted Category. 8 possibilities:  & First two characters  \\ 
   &             &          & OPNO, OPPH, BCNO, & correspond to the birth category \\ 
   &             &          & BCPH, SCNO, SCPH, & and the second two correspond \\ 
   &             &          & STNO, STPH        & to the PwD status of the candidates \\ 
   &             &          &                   & for which this program is reserved \\ \hline
9. & AllottedQuota & char(2) & Allotted Quota (AI/HS/OS/AP/GO) & \\ \hline
10.& GenderPool  & char(6)  & Allotted gender pool with two possibilities: & \\
   &             &          & {\color{blue} Neutral} or {\color{blue} Female} & \\ \hline  
11.& Flag        & char(1)   & Four possibilities: & D: DS seat is given to the candidate.\\ 
   &             &           & N: Normal &  A DS candidate can also receive \\ 
   &             &           & D: DS &  a "N" normal seat. \\ 
   &             &           & F: Foreign & \\ 
   &             &           & P: Preparatory & \\ 
   &             &           & (Default: N)  & \\ \hline
12.& SupNumReason & char(2)  & NA- Not Applicable (Default) & If 5 candidates are at the \\
   &             &           & EQ- Closing rank equality & same closing rank, any 4 \\ 
   &             &           & MC- Min Cutoff & of them can be marked as EQ.\\
   &             &           & FR- Foreign national & Similarly, if 5  supernumerary seats\\
   &             &           & FE- Foreign national with closing rank equality & are created due to min-cut-off,\\
   &             &           & FM- Foreign national with Min Cutoff & any 5 candidates clearing \\ 
   &             &           & DS- DS consideration & min-cut-off and whose \\ 
   &             &           & DE- DS with closing rank equality & category didn't change \\
   &             &           & DM- DS with Min Cutoff& can be marked as MC \\ \hline
13.& Withdraw    & char(1)   & Two possibilities: Y/N & To determine if \\
                                                & & & & candidate has withdrawn \\ \hline
14.& RStatus     & char(2)   & Six possibilities: & To determine the \\
   &             &           & NR,DR,RC,RP,RT,RU      & reporting status of candidate \\ \hline     
\end{tabular}
}
\end{center}
\noindent
Note: Allotment table output by the seat allocation algorithm has only the first 12 fields. The last 2 fields, namely ``RStatus'' and ``Withdraw'' are appended to the allotment table based on the activities at reporting center. The resulting table, called
allotment table of the previous round, is used for the seat allocation of the next round.
\subsubsection{Program Statistics}
\begin{center}
{\small
\hspace*{-2.5cm}
\begin{tabular}{|c|l|l|l|l|}
\hline
{\bf S. No}. & {\bf Column name} & {\bf Type and} & {\bf Description} & {\bf Remark} \\ 
             &                   & {\bf  Length } & & \\ \hline
1.   &  Quota      &     char(2)   & Quota & HS/OS/AI \\ \hline
2.   & InstCd      &     char(3)   & Institute Code  & \\ \hline
3.   & BrCd        &     char(4)   & Branch Code     & \\ \hline 
4.   & VCategory   &     char(4)   & Category of virtual & DSNO if the programme\\
     &             &               & programme. One among & is DS virtual programme.\\
     &             &               & OPNO, OPPH, BCNO, BCPH, & BrCd should be 0000 then\\
     &             &               &  SCNO,SCPH, STNO, STPH, & \\
     &             &               & and DSNO & \\ \hline  
5.   &  GenderPool &     char(6)   & Gender pool: & \\
     &             &               & {\color{blue} Neutral} or {\color{blue} Female} \\ \hline        
6.   & OpeningRank &     float     & Opening Rank & \\ \hline
7.   & ClosingRank &     float     & Closing Rank &  \\ \hline
8.   & MinCutOff   &     float     & The min cut off used & This field is computed in the \\ 
     &             &               & for allotment in the current round. &  beginning of the current round \\ 
     &             &               & & using the output of \\
     &             &               & & the previous round. \\ \hline
9.   & TotalAllotted &   int   & The number of candidates & \\ 
     &             &               & to whom the programme is allotted & \\ \hline
10.   & InitCap     & int       & Initial capacity  & \\  \hline
11.  & NewCap      & int       & Capacity after de-reservation  & Includes non-supernumerary  \\
     &             &               & & seats given to DS candidates \\ \hline
12.  & DeReserveFrom & int     & Number of seats that got & \\
     &             &               & de-reserved from this & \\
     &             &               & virtual programme & \\ \hline
13.  & DeReserveTo & int       & Number of seats that got & \\ 
     &             &               & de-reserved to this & \\ 
     &             &               & virtual programme & \\ \hline
14.  & SuperNum    & int       & Number of supernumerary seats & Includes supernumerary \\ 
     &             &               & created in this virtual programme & seats created due to\\ 
     &             &               &  & all possible reasons. \\ \hline
\end{tabular}
}
\end{center}

\section{Changes in Business rules since 2015}
\label{sec:changes-in-business-rules-since-2015}
We now state the changes in the business rules that have occurred in the last 4 years of joint seat allocation. 
Each of these changes were duly approved in respective JoSAA meetings after long deliberation, and in some cases justified through scientific analysis of data of one or more years of the joint seat allocation.
\begin{itemize}
\item Preferential allocation of seats for DS candidates. \\
  In 2015, the seats for DS candidates were allotted in preferential manner, and these seats had to be from the Open category. The reader may refer to Technical Report \cite{TechReport:2015} for the algorithm to handle DS candidates during Joint Seat Allocation 2015. We have reproduced this algorithm in Appendix \ref{sec:DS-complete-algorithm}.
However, from 2016 onwards, DS candidates are given seats in supernumerary fashion. 
\item Influence of board marks on merit list of CSAB institutes.\\
In 2015 and 2016, board marks of a candidate were considered in computing the final rank of a candidate in the merit list used for CSAB institutes. But from 2017 onwards, the board marks are used only as qualifying marks. The merit list for admission into CSAB institutes
is based only on the performance in the JEE main exam.
\item Provision to withdraw from joint seat allocation.\\
After accepting a JoSAA seat, a candidate may get a better (from the perspective of the candidate) program in some other institute outside JoSAA. Such a candidate should be allowed to surrender her
current seat and withdraw from joint seat allocation. But, there was no provision to withdraw in 2015. Without the provision of withdraw, the seat vacated by the candidate will never be filled. However, from 2016 onwards, option to withdraw was provided to the candidates. Each candidate who accepts a seat was allowed to withdraw till the end of the 6th round (total number of rounds are 7). 
\item Penalty of CatChange=1 candidates.\\
During document verification at the reporting center, certain changes in the credentials may result in setting CatChange field to 1.
Not only the seat of such a candidate gets canceled, as a penalty the candidate may compete only for vacant seats in subsequent rounds. Such candidate were derpived from Min-Cutoff benefit. The reader may refer to the Technical Report of 2015 for the details of the algorithm to implement this business rule. The objective underlying this business rule was to avoid creation of supernumerary seats.
However, it was found through simulations on the data of 2017 that the number of supernumerary seats avoided by this rule was less than 10. In addition, it was felt that the penalty imposed by this rule is too harsh for a candidate. So this rule was changed in 2018. According to the new rule, if a candidate's seat is canceled in a round because of failure to produce certain documents at a reporting center, the credentials of the candidate will be changed for future rounds, but the candidate will not have any kind of penalty in subsequent rounds.  %In other words, each candidate will get the Min-Cutoff benefit.  
\end{itemize}

\section{Detailed algorithm for handling DS candidates as per 2015 rule}
\label{sec:DS-complete-algorithm}
 In 2015, the business rule specified that DS candidates
 should be given the best program they prefer subject to the
 constraint that there are only two seats for DS candidates per
 institute.  Moreover, seats to DS category candidate needed to be
 allocated in a preferential manner from open category seats and
 creation of an additional seat (termed a ``supernumerary allocation'') needed to be avoided to the extent
 possible.\footnote{Supernumerary allocation could not be avoided, if
   for example, there are three candidates from DS with the same
   rank. There can be more complex cases.} 

   Our algorithm for 2015 can be divided in two parts. The first part is the same as the current algorithm described in Section~\ref{subsec:DS}. After executing DA as described in Algorithm~\ref{alg:DA}, while handling DS candidates as described in Section~\ref{subsec:DS}, we may have artificially increased the capacity
of some (open category) programs by maximum of two seats per institute.  Therefore, each seat which has
been allotted from a DS virtual program is to be mapped to a virtual
open program, and is thus processed one by one as follows. Let $s$ be
one such seat, say, in IITB, and let $c$ be the candidate who has been
given this seat. Suppose $c$ has been mapped to the EE program in
IITB.  Let $x$ be the candidate with the worst rank in the open
category waitlist of the virtual program EE in IITB. If there is one more
candidate in this waitlist with the same rank as the rank of $x$, then
we just assign program EE to candidate $c$ and this completes the
processing of seat $s$ (in a supernumerary manner).  Otherwise,  to
take care of the over allocation, $c$ replaces $x$. We now {\em run}
the DA algorithm with $x$ as the candidate not allotted any program
(and so $x$ applies to the next program in her preference list).  Note
that $x$ may displace another candidate, and this may lead to a {\em
  rejection chain}.
  
However, in very rare cases, there is a possibility of an undesired
condition arising out of this rejection chain. We now describe an example of a race condition that may arise while processing
some DS candidates. Thereafter, we present a method for detecting a race condition.
We conclude with a complete algorithm for handling DS candidates as per the 2015 rule.

\subsection{Example of Race Condition}
\begin{figure}[h]
\centering
\includegraphics[height=2.5in]{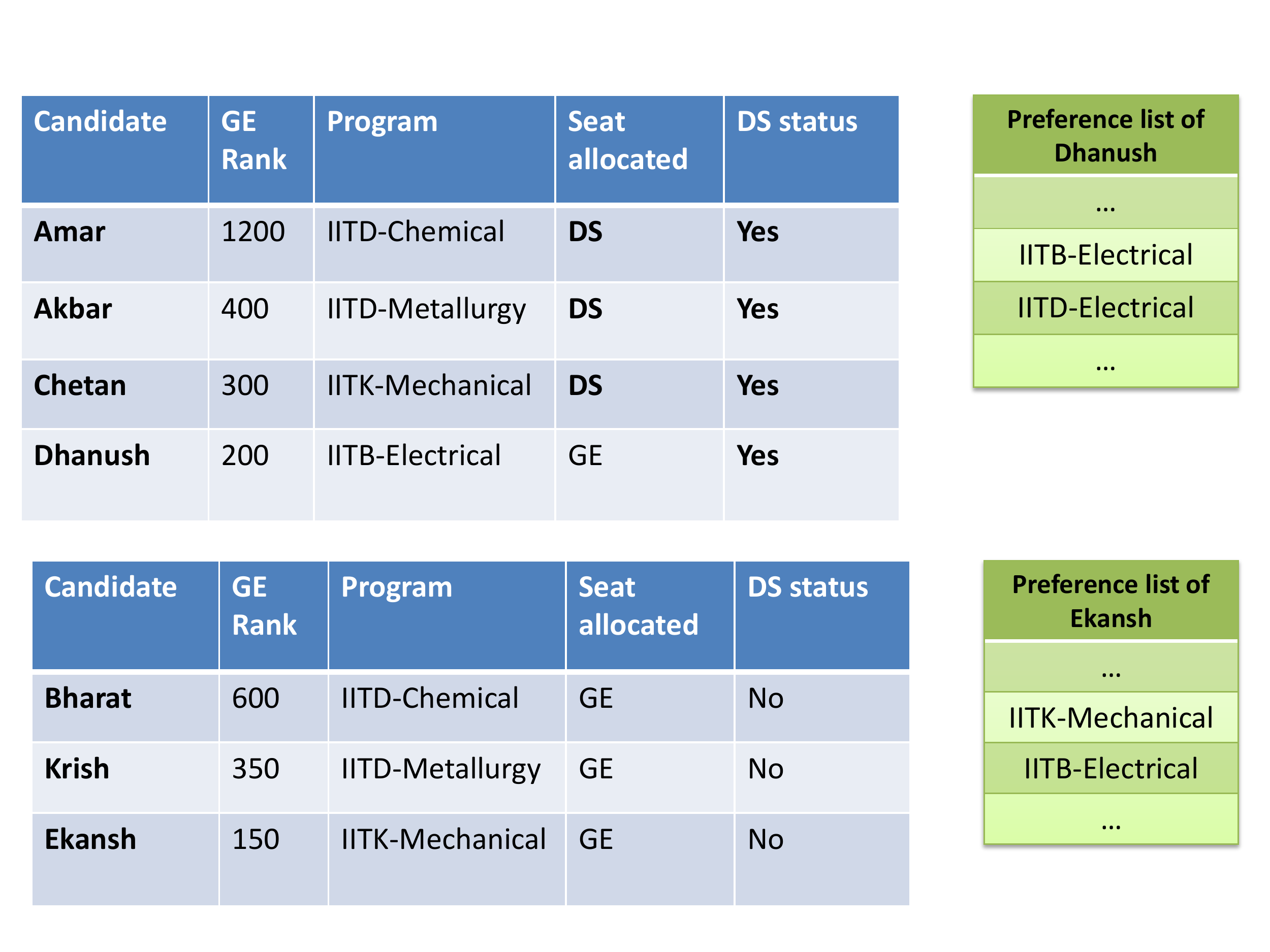}
\caption{Program allocation by the DA algorithm to 3 DS and 2 GE candidates.}
\label{figure:DS-example}
\end{figure}

Let Amar, Akbar, Chetan, and Dhanush be four DS candidates. 
At the end of the DA algorithm, Amar, Akbar, and Chetan get their program 
through a DS seat; But Dhanush gets his program IITB-Electrical through a 
GE seat because there were already two better ranked DS candidates who got 
DS seats in IIT Bombay. 
Moreover, let Dhanush happens to be the last ranked candidate getting 
IITB-Electrical. Let Bharat, Krish, and Ekansh be the last ranked GEN
candidates in IITD-Chemical, IITD-Metallurgy, and IITK-Mechanical 
respectively. The details of all these seven candidates with the programs 
allocated by the DA algorithm is shown in Figure \ref{figure:DS-example}. 

We now describe the processing of Amar, Akbar, and Chetan who got DS seat.
These candidates have to be given OPEN seats. 
In order to accommodate Amar, we need to remove Bharat and this leads 
to Bharat getting some other less preferred program in the rejection chain.
In a similar manner, Akbar gets IITD-Metallurgy after removal of Krish and 
Krish gets some other less preferred program in the rejection chain.
Let us process Chetan now. Since Chetan got seat IITK-Mechanical through DS 
quota we need to remove Ekansh from IITK-Mechanical. 
Next preferred program for Ekansh is IITB-Electrical. Recall that 
Dhanush is the last ranked candidate getting IITB-Electrical. Notice that 
though Dhanush is a DS candidate, he got OPEN seat in IITB-Electrical. 
So Ekansh will remove Dhanush from IITB-Electrical. So Dhanush will apply for 
his next preferred program which is IITD-Electrical. There are
already two DS candidates Amar and Akbar for IITD. 
Since Dhanush has better rank than Akbar, and Akbar has better rank than Amar, 
%Assuming that some other DS candidate with rank better than Amar has already taken a DS seat from IIT Delhi, 
so Dhanush will remove Amar from DS virtual program of IITD. 
So the processing of Amar in the past goes waste since he is not getting a 
DS seat in IITD (the seat vacated by Bharat for Amar goes waste, the seat remained vacant in the end and Bharat got a less preferred program).
A similar example can be constructed wherein the DS candidate who initiates
a rejection chain will have to be de-allocated.
In any such situation, we should revert the rejection chain
and allocate the DS candidate a supernumerary seat. In the current example,
Chetan has to be given a supernumerary seat in IITK-Electrical program. 

\subsection{Detecting Race Condition}
After processing each seat $s$, we need to check for race  condition.
Some key points to be noted about race condition are:
\begin{enumerate}
\item In the above example, if Dhanush had also applied for IITD-Chemical instead of IITD-Electrical, it would not have been a race condition. This is because, Dhanush will now simply occupy the space created by the processing of Amar in IITD-Chemical Open virtual program. Note that this doesn't result in any program being vacant and hence there is no race condition.\\
Thus, for race-condition-free allocation, candidates in a processed seat may change but their programs should not.

\item The number of candidates in any DS virtual program can be variable due to the rule of supernumerary allocation to same ranked candidates.
\end{enumerate}
We gather that for any candidate $x$ occupying a processed seat in old allocation (before processing of seat $s$), there must be a candidate $x'$ occupying a processed seat in new allocation such that $x$ and $x'$ have the same DS virtual program and $x.allotted\_program = x'.allotted\_program$. If this condition does not hold after processing of a seat $s$, we have a race condition and we will need to give the candidate occupying $s$ a supernumerary seat.

To summarize, we need to verify that the allotted programs to processed seats in Old allocation of a DS virtual program constitute a subset of allotted programs to processed seats in New allocation of the same DS virtual program. If this holds true for all DS virtual programs, then we can proceed with processing further seats, otherwise we need to revert to old allocation and give a supernumerary seat to the latest processed DS candidate.

To check the above condition, after we process a seat, we iterate over processed seats in old allocation (including the latest processed seat), for each allotted program of a processed seat, we find a matching processed seat in new allocation with the same allotted program. If at any point, no such match is found, we declare that there is a race condition and we need to revert to old allocation (before the processing of latest seat) and make one supernumerary seat.

\subsection{Pseudocode}
We now provide full details of the algorithm in this section.
\subsubsection{Details of the algorithm}

The algorithm maintains the set $D$ of seats in all DS programs. Since, due to ties in multiple DS candidates, number of seats in any program may change during the course of the algorithm, we keep $D$ as an instance of \textsc{allocation} object.
The algorithm processes the seats from $D$ sequentially.
At each stage, it also maintains a flag {\em Is\_processed}
(initialized to false) for each seat in $D$. After processing a seat, this
flag is set to {\em true}.
It picks an unprocessed seat $s$ from $D$
and does the following. Let $x$ be the candidate occupying the seat $s$ and
$p$ be the program opted by him/her. Let $w$ be the worst rank candidate
from OPEN category who has been assigned program $p$. $w$ is removed from
program $p$ and the capacity of $p$ is reduced by 1. We now start DA
algorithm with input $\{w\}$. Effectively, the candidate $w$ applies to the
next program in his/her preference list. This generates a rejection chain resulting in a different allocation than before. Let the allocation before processing be \textsc{old-allocation} and allocation after processing be \textsc{new-allocation}.
It is quite possible that the candidate corresponding to
any processed seat in $\textsc{new-allocation.}D$ is different from that in \textsc{old-allocation.}$D$.
However, if the {\em set} of programs associated to the
processed seats of $\textsc{new-allocation.}D$ differs from that in $\textsc{old-allocation.}D$, we revert to $\textsc{new-allocation}$,
create a supernumerary seat for $p$ and allocate it to $x$.

We need to consider each DS seat one-by-one, create their respective
rejection chain and revert it if it causes race condition. To detect race
condition after processing a seat $s$, all we need to do is to check if the multiset $S_1$ of allotted programs to processed seats in \textsc{old-allocation} is a subset of the multiset $S_2$ of allotted programs to processed seats in \textsc{new-allocation}. Note that, there cannot be a new processed seat created in \textsc{new-allocation} ($s$ is marked processed in both \textsc{old-allocation} and \textsc{new-allocation}). Hence,  $|S_2| = |S_1|$. Therefore, in order to detect race condition, all we need to do is to verify whether $S_1=S_2$. If $S_1\not=S_2$, then there is a race condition, else there is no race condition.  
%Therefore, for race-condition-free allocation, we must have $S_1 = S_2$. 
Notice that $S_1$ and $S_2$ are multi-sets and so  they should be matched element by element for equality. An element is a program (i.e. IITK\_CS, IITK\_CE etc. That is, consider branch names augmented with institute names).

Algorithm \ref{alg:DS-algo} presents the complete pseudocode of the
algorithm for handling the rule for admission of DS candidates.

{\em Notations} 
\begin{itemize}
\item[] ${\textsc{candidate}}(s)$: candidate to whom DS seat $s$ is allocated. Set to $\phi$ if the seat is unoccupied.
\item[] ${\textsc{program}}(s)$: OPEN virtual program corresponding to program allocated to candidate occupying DS seat $s$. Set to $\phi$ if the seat is unoccupied.
\end{itemize}
\begin{algorithm}[H]
\caption{Algorithm to incorporate the rule for admission under DS category}
\label{alg:DS-algo}

INPUTS:\\
The regular inputs for DA.\\
The Boolean \textsc{supernumerary-ok}, which indicates whether a supernumerary seat may be created if a DS related problem occurs. Initially we run with \textsc{supernumerary-ok}=False\\
Each seat has a Boolean flag \ip. We initialize it as \ip =False for all DS seats.\\
OUTPUTS:\\
The regular output of DA, as well as the waitlists for the DS virtual programs (these include the name of the actual program allotted to each candidate).\\
The list of stained programs $S$. If non-empty, this would be a trigger to get permission from the JAB Chairman and rerun with
\textsc{supernumerary-ok} = True.\\
% \KwData{Allocation produced by DA algorithm on all candidates. DA() takes list of candidates that don't have any allotted program as a parameter }
% \KwResult{Final allocation following DS reservation rule }
\begin{algorithmic}[1]
 \ForAll{Institutes}
 \State Create a DS virtual program $I$.
 \State $c(I) \leftarrow 2$
 \State $\merit(I) \leftarrow$ DS candidates in CML in CML order
 \State $\wl(I)$ will be an augmented waitlist, such that each entry will contain both a candidate as well as a program name.
 \EndFor
 \ForAll{candidates $x$ with DS tag}
    \State Create $\pref(x)$. For each program, first list virtual programs as usual as per birth category of $x$, then list corresponding institute DS virtual 
program (tag it with the relevant program).
 \EndFor

 \State Run the DA algorithm
 \State \textsc{Allocation}  $\leftarrow \wl(p)\ \forall \; p \in \cP$ and $\LP(x) \ \forall \; x \in \cA$ as per the output of the DA algorithm
 \State $\textsc{allocation.}D \leftarrow $ List of seats in all DS programs in \textsc{allocation}. A seat data-structure includes a candidate, a program, and a Boolean flag \ip. 
    %\State Add $x$ to set of candidates with $\LP(x) \leftarrow 1$
    %\State $Q \leftarrow $ Queue containing $x$ alone
    %\State Run DA with current $i(x')$, $\wl(p)$ and $Q$ as optional inputs

    %\If{$p_{x,\LP(x)}$ is a DS virtual program}
%        \State $p \leftarrow $ OP virtual program corresponding to tagged program in $\LP(x)$
%        \State Decrement $c(p)$
%        \State $L \leftarrow$ List of all candidates who must be rejected from $\wl(p)$ based on updated capacity, $\nrl(p)$ etc. \Comment{In case of no ties, $L$ will contain one candidate if the last candidate is not on $\nrl(p)$, and zero candidates otherwise.}
%        \State $Q \leftarrow $ Empty queue
%        \ForAll{y \in L}
%            \State Remove $y$ from $\wl(p)$
%            \State \Call{Reject}{$y$}
%        \EndFor
%        Run with current $i(x')$, $\wl(p)$ and $Q$ as optional inputs. During the run, if any DS candidate $x'$ is pushed out of a DS seat, add the corresponding OP virtual program to a list  $S$ of \emph{stained} virtual programs.
%    \EndIf
%
% \State Let \textsc{allocation} be the allocation made by the DA algorithm;

\ForAll{$s \in \textsc{allocation.}D$}
    \State $s.\ip \leftarrow $ false
\EndFor

\State $S \leftarrow $ Empty list  \Comment{List of stained programs $S$ is needed only if \textsc{supernumerary-ok} = False}
\While{$\exists s\in \textsc{allocation.}D$ with ([$s.\ip$]= false and \textsc{candidate}($s$) $\neq \phi$)}
%  \State Pick any unprocessed seat, say $s$, from $D$;
  \State \textsc{old-allocation} $\leftarrow$ \textsc{allocation};
  \State $L \leftarrow $\Call{processSeat}{$s$}
  \State $Q \leftarrow $ Empty queue
\algstore{myalg2}
\end{algorithmic}
\end{algorithm}

\begin{algorithm}
\begin{algorithmic}[1]
\algrestore{myalg2}
  \ForAll{$y \in L$}
      \State Remove $y$ from $\wl(p)$
      \State \Call{Reject}{$y$} \Comment{This adds $y$ to $Q$ as well}
  \EndFor
  \State Run DA with current $i(x')$, $\wl(p)$ and $Q$ as inputs (other inputs as usual). During the run, if a DS candidate $x''$ is pushed out of a DS seat and the set of programs associated with the processed seats in that DS virtual program changes, add the corresponding OPEN virtual program to a list  $S$ of \emph{stained} virtual programs.

  \State \textsc{allocation} $\leftarrow$ Output of DA \Comment{DA may maintain a list of affected DS virtual programs to make the loop below more efficient}
  \State \textsc{to-revert} $\leftarrow$ \Call{IsRaceCondition}{\textsc{allocation},\textsc{old-allocation}}

   \If{\textsc{to-revert}}
      \If{\textsc{supernumerary-ok}}
      	\State \textsc{allocation} $\leftarrow$ \textsc{old-allocation}
          \State Increment $c(p)$ by 1;
      \Else
        \State Do nothing \Comment{$S$ now includes both programs that lost a DS candidate after making room, as well as those that gained a DS candidate and may have a supernumerary seat (but will not be processed by the algorithm since that seat has been already processed).}
      \EndIf
   \EndIf
   \State $\textsc{allocation}.D \leftarrow $ List of seats in all DS programs according to \textsc{allocation}
%        	\State \textsc{allocation} $\leftarrow$ \textsc{new-allocation}
  \EndWhile
\State return \textsc{allocation} and $S$;  \Comment{Note that if a DS candidate has gotten a seat via DS in program $p$, he is not included in $\wl(p)$. This must be read separately by reading the augmented waitlist $\wl(I)$ (which specifies both the candidate and the program) where $I$ corresponds to the institute that hosts $p$.}

\algstore{myalg2}
\end{algorithmic}
\end{algorithm}

\begin{algorithm}
\begin{algorithmic}[1]
\algrestore{myalg2}

\Function{processSeat}{$s$}
  \State $s.\ip$ $\leftarrow$ true;
  \State $x \leftarrow $ {\textsc{candidate}}($s$);
  \State $p \leftarrow $ {\textsc{program}}($s$);
  \State Decrement $c(p)$ by 1;
  \State $L \leftarrow$ List of all candidates who must be rejected from $\wl(p)$ based on updated capacity, $\mc(p)$ etc. \Comment{In case of no ties, $L$ will contain one candidate if the last candidate does not clear $\mc(p)$, and zero candidates otherwise.}
  \State Return $L$
\EndFunction
\\
\Function{IsRaceCondition}{\textsc{new-allocation,old-allocation}}
\State $S_1 \leftarrow$  multi set of allotted programs to processed seats in $\textsc{new-allocation}.D$ \Comment{a program is institute+branch}
\State $S_2 \leftarrow$ multi set of allotted programs to processed seats in $\textsc{old-allocation}.D$ \Comment{a program is institute+branch}
\If{$S_1 = S_2$}
\State return False;
\Else
\State return True;
\EndIf
\EndFunction
\end{algorithmic}
\end{algorithm}

\section{History and background}
\label{sec:history-and-background}
%Before launching into the specifics of our project, we
This appendix provides some background regarding the institutions and examinations involved. %A casual reader may skip the details in this section.

Students in India must write a senior high school graduation exam
(known as the ``Board exam''); these are administered by the educational board to which their high school is affiliated. %There are state boards (roughly, each of 29 states in the country has its own board), national boards such as the Central Board of Secondary Education (CBSE) and the Indian School Certificate (ISC), and also some international boards.
However, these exams % subjects, the syllabi, and the exams for them
necessarily account for the wide heterogeneity in the preparation of students across schools and geographies. Hence,
scores in these exams are typically not considered appropriate for determining admissions to the country's most prestigious engineering colleges.

\paragraph{History of admissions to the IITs.} The first five IITs
(Kharagpur, Bombay, Kanpur, Madras and Delhi)
were founded during %the years
1951-61, %\sharat{verify.. Kharagpur was much later, not 51}\yk{I have verified 1951.}.
and almost immediately they created a countrywide Common Entrance
Examination for admissions purposes.  The examination was used to produce a single
ranking called a ``Merit List'' of candidates (more precisely, one
Merit List for each ``category'' of students, see Table \ref{table:categoryTag}). Next, in
a centralized process, candidates were then considered in the
increasing order of rank (starting with the top ranker), and allotted
their most desired program which was not already full based on the
preferences over programs that they submitted after ``counselling''
at the closest IIT. [This mechanism is known as Serial
dictatorship, e.g., see \cite{abdulkadirouglu1998random} The
%name of the
%examination changed to IIT Joint Entrance Exam
%(IIT-JEE) and then subsequently to simply
name of the examination subsequently changed to Joint Entrance Exam (JEE), and
the number of IITs has grown %(rapidly in recent times)
to
%\sharat{verify number}
23. %, but there was no fundamental change to the seat allocation process described above until 2015. %\footnote{Admissions to two institutions established in the early 1900s --- the Indian School of Mines Dhanbad (later IIT Dhanbad), and Institute of Technology, Benares Hindu University (later IIT-BHU), were also conducted as part of the IIT process, and recently these institutions were converted into IITs.}
As the number of candidates
grew, the IITs resorted to %, first, multiple choice questions to enable rapid grading, andlater also
a two stage examination process, with the first ``screening'' stage used
to select a subset of candidates who could
then write a more detailed second stage exam. % and be assigned an All India Rank (AIR) based on it.

\paragraph{History of admissions to the non-IIT CFTIs.}
%At the same time,
Starting in 1959, Regional Engineering Colleges (RECs) were
created in every major state to supplement the IITs. % and meet the country's need for engineers.
The admissions to these colleges was
conducted in a decentralized manner and many of them conducted their
own entrance exams, creating a logistical nightmare for high school students who were aspiring engineers. %, due to the huge multiplicity of examinations and admissions processes.
In 2002, the
RECs were renamed National Institutes of Techonology (NITs), and a
single All India Engineering Entrance Examination (AIEEE) was created
to centralize the examination and admissions process for the NITs, %under a body called the Central Counseling Board,
simplifying the logistics.
%greatly easing the burden on applicants and also the burden on
%institutes to conduct and grade examinations, and run
%admissions.
(There are now 31 NITs.)

%\yk{@Utkarsh/Sharat: Add two-four sentences about history of IIITs and  Other-GFTIs, especially their admissions processes.}

%Keeping up with the booming information technology sector, the
%government established several IIITs --- the
Several Indian Institutes of
Information Technology (IIITs) were established starting in 1997% (but especially in recent years)
; that exclusively offered programs allied to information technology. (There are now 23 IIITs.) Admissions to the IIITs as well as other engineering colleges funded by the central government were clubbed with admissions to the NITs. %, and women-only collegessuch as (the one in J\&K). \sharat{to add}. Admissions to these were
%clubbed either with the AIEEE process, or the IIT process (e.g.,
%Benares Hindu University Information Technology, IIT Dhanbad).

\paragraph{Merger of Examinations.}
Subsequently the AIEEE was merged with the first stage of the two
stage IIT exam process in 2012. The first examination is called the
JEE Main. The JEE Main score of candidates is used as follows:
\begin{enumerate}
  \item The non-IIT CFTIs use the JEE Main scores to construct their Merit
Lists and determine allocation of seats at the non-IIT CFTIs.  Until 2016, the Merit Lists were created by a combining the JEE Main score of a candidate with her Board exam score. Since 2017 they are exclusively based on the JEE Main score.
\item The IITs use the JEE Main scores to determine a subset of about 150,000 candidates (as of 2015) who qualify to be permitted to write the second stage ``JEE Advanced'' examination.
\end{enumerate}
The JEE Advanced examination is
conducted subsequently by the IITs for their own
admissions purposes, and typically consists of three separate exams
for Physics, Chemistry and Math. % conducted over 6 hours.
Subject cutoffs are set for each of the three, and a Merit List of
candidates who clear the cutoff is constructed based on their total
score across the three subjects for purposes of admission to IIT
programs. (Detailed tie-breaking rules ensure that ties in the Merit List play a negligible role, and similarly for the Merit Lists for the non-IITs umbrella.) %For the IITs (and since 2017, also the non-IIT CFTIs),
The Board exam score of candidates is not used
for ranking but purely to determine their eligibility based on a cutoff. %; candidates are required to obtain over $75\%$ marks in their Board exam or be in the top 20 percentile. %(English, for example, is tested in the Board exams.)

From 2012 to 2014, the seat allocation process for IITs (under IITs umbrella)
remained separate from that for the non-IIT CFTIs (under non-IITs umbrella). The
IITs conducted their admissions first, even before the Board exam
scores had come in (since almost all successful candidates obtained
the requisite Board exam score), and the non-IITs umbrella conducted its admissions
process subsequently.

%way CSAB constructs its merit lists: now the Board exam marks are no longer given weightage in determining rank, instead, they serve purely to determine eligibility (a candidate must get at least 75\% marks in the Board exam or be in the top 20th percentile, which is ``easy'' for strong candidates).

\paragraph{Merger of Seat Allocation processes.}
%With the merger of the AIEEE into the JEE in 2012, a joint seat allocation process was a natural next step. %However, the IITs, like any other elite institutions, enjoyed conducting their own independent admission process (with coordination only on the JEE Main exam which wasn't used to prepare the IIT Merit Lists anyway), and the CSAB similarly may not have taken the initiative to coordinate with the IITs on a joint seat allocation process.
% As it turned out, a
After the merger of examinations, an external nudge in the form of a public interest court case W.P.(C) 2275/2010 in the Delhi High Court (demanding coordination to reduce vacancies) caused the creation of a joint seat allocation process.
%\begin{DI}
%A directive from a higher authority can be instrumental in coaxing institutions to coordinate on a centralized allocation process.
%\end{DI}
%In June 2013, a Joint Seat Acceptance Committee (JSAC) 2013 was constituted, partly in response to ongoing court proceedings in W.P.(C) 2275/2010. However, the timeline was very short and no change was implemented in the seat allocation process conducted in July 2013.

Following a false start in 2013, a common seat allocation process for all the CFTIs including the IITs was launched in 2015, run by the Joint Seat Allocation Authority (JoSAA). This joint seat allocation process is the subject of the current paper. JoSAA provided candidates with a single window for admission to any of the over 80 CFTIs.

\begin{DI*}
Centralization can greatly reduce the logistical burden of
participation on both sides of the market (in addition to
improving the allocative efficiency and reducing market congestion if the
allocation is also done centrally).
\end{DI*}

Related efforts elsewhere include the Common Application for applying to hundreds of colleges worldwide
(in this case the allocation is not done centrally), centralized school
admissions, and centralized labor markets (see Section \ref{sec:prev-work}). In the case of the CFTIs, centralization has occurred
for the examination, application as well as the seat allocation
processes.

In 2015 and 2016, the JoSAA seat allocation process was conducted
after the JEE Main, JEE Advanced, and Board exam scores became available. Delays in announcement of Board marks was a major issue, indeed the joint process was very nearly called off
the very first time in 2015 due to such delays.  The IITs wanted to
proceed with their allocation, whereas the other institutes were
unable to rank candidates without Board marks being at hand. 
\begin{DI*}
  Aggregation of all relevant information and alignment of timelines
  of the concerned institutions can be bottleneck for centralized
  matching/allocation. If institutions construct their preferences
  based on the same information (and at the same time), this
  improves the chances of successful centralization.
\end{DI*}

Since 2017 the non-IIT CFTIs chose to stop using Board exam scores for constructing their Merit Lists, eliminating this issue, consistent with the trend of logistics getting simpler over time.

In Appendix \ref{app:discussion-of-JEE}, we briefly discuss the broader impact of the JEE.
We emphasize here that our mandate was restricted to designing an efficient and fair joint \underline{seat allocation} mechanism for the 80+ institutions involved (the CFTIs), that respects a set of business rules, treating the JEE as a given. %We summarize these business rules next.

\subsection{Broader view of the Joint Entrance Examination (JEE)}
\label{app:discussion-of-JEE}

The entire examination and seat allocation system for the CFTIs in
India under JoSAA based on the JEE is generally viewed as providing a good solution to the problem
of resource allocation in a supply constrained environment. It is
heartening that allegations of cheating in the examination
are highly atypical despite that 1.3 million candidates write the JEE
Main each year, and allegations
of corruption in terms leaking of exam questions or grading
malpractice are similarly atypical, despite the extremely high stakes. The exam is also viewed as being fairly successful at identifying talented candidates (however, many candidates may not really be interested in engineering; a large fraction of successful candidates see an engineering education as a stepping stone to lucrative careers in other fields).
%\sharat{controversial. Indeed Rajeev Kumar started the campaign   precisely because IIT KGP had backdoor admissions.}

The main questions that do arise about this system are around the
demands and incentives it generates for candidates and the JEE coaching
industry that has grown exponentially around it. Some of the concerns
that have been voiced are: (i) Wealthy candidates have an increasing
advantage due to coaching classes becoming increasingly adept at
systematically preparing candidates, and charging very high fees. (ii)
Candidates are ``burned out'' even before they start their studies at
these institutes due to at least two, many times three, and often six
years of extremely intense preparations merely in order to gain
admission. As such, they often do not invest in their education as
engineers, and a majority of them do not work in or around the area
for which they are trained. Instead, they go into consulting, finance
and information technology. (iii) Related to the above, candidates may
``lose'' years due to repeating the JEE after they have graduated from
senior high school. (Previously, candidates would commonly lose
multiple years and write the JEE three times, for instance. Since
2010, the rules prohibit writing the JEE Advanced more than one year after completing the 12th grade, whereas the JEE Main can still be attempted up to two years after.) %\sharat{not more than one year?? basically   two attempts. You could write the board focusing on the boards the   first year, and then write the exam twice.}
 There are no ready
solutions for these issues.  Reservation of seats for different
categories of students (see Chapter \ref{chapter:businessrules}) is obviously a hot button
topic which is heavily debated by stakeholders, observers and the
government alike. We remark here that changes to the examination
system and the reservation rules are serious issues that were fully
\emph{outside} the scope of the joint seat allocation project that
this paper describes; we are providing a description here merely to
provide the interested reader with some context. Our mandate was to
design a seat allocation process with high allocative efficiency that
while preserving good properties of the legacy mechanisms such as
\emph{fairness}, i.e., a candidate with a better rank in the relevant
Merit List should not be denied admission to a program if another
candidate with a lower rank was granted admission to that program.

\section{The challenge and the opportunity of overbooking}
\label{app:yield-prediction}
The fraction of Rejects from among fresh allocations is quite large, especially in later rounds. One may think of using yield prediction (i.e., admitting more candidates than the capacity of the program) as a way to improve the efficiency of allocation and to have less vacancies at the end of all the main rounds. This is harder than it would seem, despite the fraction of rejected fresh allocations being $60\%$ or more in later rounds.
Consider Round 3 in 2015. There were 7342 candidates whose allocation changed including 3720 fresh allocations. There were 2666 Rejects, almost all (2651 of them) from among the fresh allocations, meaning that $71.6\%$ of fresh allocations were rejected! It is tempting to think that one can use yield prediction to substantially take care of the issue of Rejects. Unfortunately, this does not work out as expected. Even for an individual program whose size may be 50 or 100 seats, a lot of virtual programs (split by category and further by Home State vs All India quota for NITs and many CFTIs) -- with the exception of the OPEN and sometimes the OBC virtual programs -- have a single digit number of seats, minimizing our ability to benefit from yield prediction when overage (i.e., admitting more candidates than the number of seats) is a problem. In fact, if we consider the subset of virtual programs with 10 or more rejected seats in that round, this accounts for only 740 of the 2666 Rejects. Thus, roughly, one could only hope to account for about $30\%$ of the Rejects (roughly those that occur in virtual programs with 10 or more Rejects, since with 5 Rejects, Poisson(5) has a reasonable likelihood of being 0 and even more chance of being 1, so there is minimum benefit from yield prediction), without risking significant overage. $60-80\%$ Rejects are present across categories, though the OPEN category has slightly higher fraction of Rejects closer to $80\%$ and accounts for 1647 of the Rejects, meaning more than $60\%$ of them. We do notice some patterns like if there are 4 or more fresh allocations then there is at least 1 Reject, if there are 6 or more fresh at least 2 Rejects, with 10 or more fresh at least $40\%$ are Reject and with 20 or more at least $70\%$ are rejected. Such observations may be the basis of refined business rules for conservative yield prediction to improve the efficiency of allocation. Use of an opaque/complex predictive model is not desirable due to lack of transparency and possibility of unfairness etc. In this context, it is imperative that the business rules must be transparently and completely specified, and be clearly fair to all concerned, in accordance with the prevailing laws. So far, this option has not been considered. However, it may be worth considering, based on the following reasoning: currently, about $70\%$ of vacancies in a given round persist until the next round. So over two rounds, the number of vacancies is reduced to about $70\%$ of $70\%$ = half. Instead, suppose we use some conservative yield prediction as above, and it reduces the vacancies in the resulting allocation by about $30\%$. This means that now, $70\%$ of $(100-30)\% = 50\%$ of vacancies in a given round persist until the next round. Over two rounds then, the number of vacancies is reduced to about $50\%$ of $50\%$ = $25\%$. The cost of doing this would be, in worst case, a handful of seats allocated in excess of capacity, maybe about 10 in total (though we would aim for 0 supernumerary based on data from the previous year). In a system with over 30,000 seats, this would appear to be a very small aberration. On the other hand, the benefit may be substantial: In 2017, 4168 candidates withdrew in the reporting following the Round 6 allocation. After Round 6 there was only one more main round (Round 7), so $70\%$ or more of these 4168 seats (i.e., over 3000 seats) remained unfilled at the end of all the main rounds.
Note that Round 5 happened only two days before Round 6. One would expect that if withdrawal was permitted only until Round 5, then most of these 4168 candidates (say, 4000 of them) would have withdrawn by then, and then with two more main rounds of seat allocation with conservative yield prediction as above, only $25\%$ or about 1000 of these vacancies would have remained at the end of the last main round (Round 7). Similarly, under such an approach, the vacancies resulting from rejected seats would also be more effectively dealt with.

Finally, we remark that steps should be taken to ensure that only serious candidates participate in the last rounds of admission after the Withdraw option is closed. For example, if a candidate wants to participate in such a round, she should be required to explicitly state that. Currently, all unallotted candidates are a part of future rounds by default. Making the candidate report physically, or pay the fee upfront could be a powerful tools to filter candidates who are not serious. 
%{\color{red} \input{statistics} }
%\input{simplified-version-of-handling-quotas}

\bibliographystyle{plain}
\bibliography{mypapers}

\end{document}